\documentclass[%
 reprint,
bibnotes,
 amsmath,amssymb,
 aps,
]{revtex4-1}

\usepackage[usenames, dvipsnames]{color}

\usepackage[export]{adjustbox}
\usepackage{subfigure}
\usepackage{graphicx}% Include figure files
\usepackage{dcolumn}% Align table columns on decimal point
\usepackage{bm}% bold math
\usepackage{hyperref}% add hypertext capabilities
\begin{document}

\preprint{APS/123-QED}

\title{Black Hole Echology: The Observer's Manual}% Force line breaks with \\

\author{Qingwen Wang}
\email{qwang@pitp.ca}
% \altaffiliation[Also at ]{Physics Department, University of Waterloo.}%Lines break automatically or can be forced with \\
\author{Niayesh Afshordi}%
 \email{nafshordi@pitp.ca}
\affiliation{%
Perimeter Institute for Theoretical Physics \\31 Caroline Street N, Waterloo, ON, N2L 2Y5, Canada
}%
\affiliation{%
Department of Physics and Astronomy,
University of Waterloo,\\ 200 University Avenue West, Waterloo, ON, N2L 3G1, Canada
}%


%\affiliation{
%\affiliation{
%\affiliation{%


\date{\today}% It is always \today, today,

\begin{abstract}
While recent detections of gravitational waves from the mergers of binary black holes match well with the predictions of General Relativity (GR), they cannot directly confirm the existence of event horizons. Exotic compact objects (ECOs) are motivated by quantum models of black holes, and can have exotic structure (or a ``wall'') just outside the (would-be) horizon. ECOs produce similar ringdown waveforms to the GR black holes, but they are followed by delayed ``echoes''. By solving linearized Einstein equations we can model these echoes and provide analytic templates that can be used to compare to observations. For concreteness, we consider GW150914 event, detected by the LIGO/Virgo collaboration, and study the model dependence of its echo properties.  We find that echoes are reasonably approximated by complex gaussians, with amplitudes that decay as a power law in time, while their width in time (frequency) grows (shrinks) over subsequent echoes. We also show that trapped modes between a perfectly reflecting wall and angular momentum barrier in Kerr metric can exhibit superradiant instability over long times, as expected.

% \begin{description}
% Secondary publications and information retrieval purposes.
% \item[PACS numbers]
% May be entered using the \verb+\pacs{#1}+ command.
% You may use the \texttt{description} environment to structure your abstract;
% use the optional argument of the \verb+\item+ command to give the category of each item. 
% \end{description}
\end{abstract}

\pacs{Valid PACS appear here}% PACS, the Physics and Astronomy
\maketitle

%\tableofcontents

\section{\label{sec1}Introduction}

Motivated by the black hole (BH) information paradox and cosmological constant problems, it has been suggested that non-perturbative quantum gravitational effects may lead to Planck-scale modifications of BH horizons.  Proposals to solve the BH information paradox include gravastars \cite{Mazur:2004fk}, fuzzballs \cite{Lunin:2001jy, Lunin:2002qf, Mathur:2005zp, Mathur:2008nj, Mathur:2012jk}, and firewalls \cite{Braunstein:2009my, Almheiri:2012rt}, amongst others \cite{Barcelo:2015noa, Kawai:2017txu}. %Fuzzball paradigm argues that infalling particle information is stored in the  horizonless ``microstates'' of spacetime and will eventually get out. Firewall proposal solves the paradox by assuming that ingoing and outgoing particles will lose the entanglement near horizon due to a ``firewall''. 
These Exotic Compact Objects (ECOs) all modify the standard structure of BH horizons, and should form by Page time $\sim M^3$, but can emerge as early as the ``scrambling time'' $\sim M \log M$ \cite{Hayden:2007cs, Sekino:2008he}.
    
%Attempting to solve the cosmological constant problem(s), \cite{Afshordi:2008xu,PrescodWeinstein:2009mp} build a gravitational aether theory which modifies the Einstein field equation by adding an incompressible fluid, called aether. 
Gravitational aether theory \cite{Afshordi:2008xu,PrescodWeinstein:2009mp}  which modifies the Einstein field equations by adding an incompressible fluid (aether) is a possible solution to the cosmological constant problem(s). BH solutions in this theory link the BH mass with the aether pressure at infinity, and yield a comparable pressure to the observed dark energy pressure for stellar BH masses of 10-100 $M_{\odot}$. The solution of the modified Einstein field equation deviates from the GR within the order of Planck length proper distance outside the (would-be) horizon. It is also suggested that replacing the horizon with a ``wall'' could be a source of  high energy astrophysical neutrino flux \cite{Afshordi:2015foa}  which is a possible source for the PeV neutrinos recently detected by IceCube observatory. %Besides, some semi-classical approximation \cite{Abedi:2013xua,Abedi:2015yga} also claim correction near horizon from GR BH.

A concrete physical model for replacing event horizon due to quantum gravitational effects  is provided in \cite{Saravani:2012is}. The spacetime ends at about the order of Planck length proper distance outside the (would-be) horizon with a wall containing a surface fluid. It is then shown that Israel junction conditions imply that the fluid has the thermodynamic entropy matching the Bekenstein-Hawking area law, for charged rotating BHs (Also see \cite{Holdom:2016nek} for a similar horizonless spacetime solution).

Recent detections of gravitational waves from binary BH mergers by the LIGO-Virgo collaboration \cite{TheLIGOScientific:2016agk, TheLIGOScientific:2016pea, Giddings:2016tla, Abbott:2016blz, Abbott:2016nmj, Abbott:2017vtc, Abbott:2017oio, TheLIGOScientific:2017qsa, Abbott:2017gyy} provide a way to test the structure around the horizon scale.
Shortly after LIGO's first detection, GW150914, \cite{Cardoso:2016rao, Cardoso:2016oxy} argued that introducing a wall to replace horizon might yield a similar ringdown waveform as GR BHs, but  produce delayed echoes (see \cite{Cardoso:2017njb, Cardoso:2017cqb} for a review) in the gravitational wave signal.  Using a phenomenological template by truncating the GR merger waveforms, \cite{abedi2016echoes} carried out the first search for echoes and claimed a 2.5$\sigma$ tentative evidence for them in the the first three (candidate) events in the LIGO public data (but see \cite{Ashton:2016xff,Westerweck:2017hus} and \cite{Abedi:2017isz} for a critique/rebuttal). 

An independent search \cite{Conklin:2017lwb},  using a  different methodology, has recently found evidence for echoes in each of LIGO's merger events (with the notable exception of GW150914) at $\sim 3\sigma$ significance level. However, we should note that the echoes reported in \cite{abedi2016echoes} and \cite{Conklin:2017lwb} are for different events, even though they are both broadly consistent with the hypothesis of near-horizon Planck-scale structure. In particular, \cite{Abedi:2017isz,Westerweck:2017hus} fail to find echoes in GW151226, which has the most significant evidence for echoes in \cite{Conklin:2017lwb}, suggesting that the two methods capture different parts of the echo waveform. 

Most recently, \cite{BNS} claim a tentative detection of (lower harmonics of) echoes, at $4.2\sigma$ level, from a ``black hole'' remnant in the aftermath GW170817 binary neutron star merger. %Others \cite{Maselli:2017tfq} also gives some phenomenological templates for parameter estimation of echoes.}

While one may consider other phenomenological echo templates (e.g., \cite{Maselli:2017tfq}), more realistic templates for fitting data may be found by solving (linearized) Einstein equations with modified boundary conditions near the horizon. Along this direction, most studies have so far focused on Schwarzschild BHs (e.g.,  \cite{Cardoso:2016rao,Cardoso:2016oxy, Price:2017cjr, Mark:2017dnq, Volkel:2018hwb, Volkel:2017kfj}). In this paper, we extend this to Kerr metric as realistic BHs have spin. \cite{Nakano:2017fvh} also presented echo templates by modelling the reflectivity of the angular momentum barrier in the Kerr spacetime. We, however, model the propagation in the full spacetime which provides a more realistic treatment at lower frequencies. 

Another related work is \cite{Bueno:2017hyj} which studies the echoes of scalar gaussian wavepackets in Kerr-like wormholes. In contrast, we study generic propagation in Kerr spacetime, with arbitrary boundary conditions, which can be applied not only to scalar fields (s=0), but also massless Dirac (s = $\pm$1/2), electromagnetic (s =  $\pm$1), or gravitational (s =  $\pm$2) fields. Interestingly (but not surprisingly), we come to some similar conclusions, e.g.,  {\it i)} Spinning ECOs give rise to unstable modes which, however, do not affect the echoes till very late times (depending on whether the initial frequency range is within the superradiance regime). {\it ii)}  It is hard to make a model-independent prediction for the first echo. 

A related phenomenological issue that arises when we replace the horizon with a wall is the emergence of superradiant instability for horizonless ergoregions \cite{1978CMaPh..63..243F, Cardoso:2007az,Cunha:2017qtt}. While this might suggest long-term instability of spinning ECOs,  which may be in conflict with astrophysical spin measurements for BHs \cite{Narayan:2013gca}, it was suggested by \cite{Maggio:2017ivp} that an absorption rate of the wall as small as 0.4\% is sufficient to quench the instability completely.%\cite{Cunha:2017qtt} proves the existence of nonlinear spacetime instabilities in ECOs under some conditions. We argue that our ECOs violate the null energy condition as one of their assumptions.} %Here, we can confirm the presence of this instability, in the form of growing echo amplitudes, at late times and for highly spinning ECOs with perfect walls. 

%Besides, we find that a perfectly reflective wall kills superradiant. That's intuitive because of energy conservation law. We can still have unstable modes around would-be horizon but since perfect wall doesn't absorb negative energy (it reflects everything) so we cannot extract energy from BH.  

We organize this paper as follows: Sec.\ref{sec2} provides the linear Einstein equations and boundary conditions used. Instead of normal boundary condition with no outgoing wave on the horizon, we put a wall standing just outside the would-be horizon. The reflection rate of the wall depends on the specific model of quantum BHs. Sec.\ref{sec3} presents echo solutions for different positions of a perfect wall and time-delays of a geometric formula given in \cite{abedi2016echoes}, while Sec. \ref{seca1} discusses how superradiance of Kerr geometry is manifested in echo templates. In Sec. \ref{sec5}, we provide an analytic fit to the echo templates, based on solutions in Sec. \ref{sec3}. We explore a soft wall with frequency-dependent reflection, as well as nonlinear corrections to initial conditions in Sec.\ref{sec4} for a more realistic picture. In Appendix \ref{a1}, we briefly discusses ergoregion instability developed in the presence of a perfect wall. While in principle the instability is significant at high spins, we show that these instabilities do not affect the first several echoes of typical binary merger events.  Finally, Sec.\ref{sec7} concludes our work. 

If not specified, we use units with $G=\hbar = k_B = c=1$. For concreteness, we use the best fit properties and waveforms resulting from the GW150914 merger event, provided by the LIGO-Virgo collaboration \cite{TheLIGOScientific:2016agk, TheLIGOScientific:2016pea} \footnote{https://losc.ligo.org/events/GW150914/}. In particular, the detector frame mass and reduced spin parameter of the remnant used for the echo calculation are $M_{\rm fin} = 67.6~ M_{\odot}$ and  $a = 0.67$.  Echo templates for other final masses can be found by rescaling our analytic templates, as long as the dimensionless binary properties are not too different from those of GW150914.

\section{\label{sec2}Propagation and Boundary Conditions in Kerr spacetime}

\begin{table}%The best place to locate the table environment is directly after its first reference in text
\caption{\label{master}%
Corresponding field $\psi$ for different spin weight $s$ in Master equation. Here $\rho^{-1}=-(r-i a \cos\theta)$
}
\begin{ruledtabular}
\begin{tabular}{c|cccc}
\textrm{s}&
\textrm{0}&
\textrm{-1/2, 1/2}&
\textrm{-1, 1}&-2, 2\\
\colrule
$\psi$ & $\Phi$ & $\chi_0, \rho^{-1} \chi_1$ & $\phi_0, \rho^{-2} \phi_2$ & $\Psi_0, \rho^{-4} \Psi_4$\\
\end{tabular}
\end{ruledtabular}
\end{table}

We study the propagation of gravitational waves using linearized Einstein equations in Kerr geometry which describes the spacetime of a spinning BH. In order to model an exotic compact object (ECO), we simply replace the Kerr event horizon with a wall, where boundary conditions for linear perturbations are modified. The initial condition here is an incoming wavepacket $h_{\rm in}$ from infinity, and we calculate the outgoing wavepacket $h_{\rm out}$ by solving the linear Einstein equations. 
As usual, we use the Newman-Penrose (NP) Formalism which greatly simplifies perturbation in Kerr metric, reducing to only a single master equation (known as the Teukolsky equation) which describes propagation of all scalar ($s=0$), massless Dirac ($s=\pm1/2$), electromagnetic ($s=\pm 1$) and gravitational ($s=\pm 2$) fields (see \citeauthor{teukolsky1973perturbations} \cite{teukolsky1973perturbations} for details):
\begin{widetext}
\begin{eqnarray}\label{eq:teuk} 
\left[\frac{(r^2+a^2)^2}{\Delta}-a^2 \sin ^2 \theta\right] \frac{\partial^2 \psi}{\partial t^2}+ \frac{4Mar}{\Delta}\frac{\partial^2 \psi}{\partial t \partial \varphi}+\left(\frac{a^2}{\Delta}-\frac{1}{\sin^2 \theta}\right)\frac{\partial^2 \psi}{\partial \varphi^2}-\Delta^{-s} \frac{\partial}{\partial r} \left(\Delta^{s+1} \frac{\partial \psi}{\partial r}\right)-\frac{1}{\sin \theta} \frac{\partial}{\partial \theta}\left(\sin \theta \frac{\partial \psi}{\partial \theta}\right)\nonumber\\-2s\left[\frac{a(r-M)}{\Delta}+\frac{i \cos \theta}{\sin^2\theta}\right]\frac{\partial\psi}{\partial \varphi}-2s\left[\frac{M(r^2-a^2)}{\Delta}-r-ia\cos\theta\right]\frac{\partial \psi}{\partial t}+(s^2\cos^2\theta-s)\psi= 0, ~ 
\end{eqnarray}
\end{widetext}
where the field $\psi$ for each spin weight $s$ corresponds to NP quantities presented in Table \ref{master}. 

The Teukolsky equation (\ref{eq:teuk}) is separable in coordinates in the frequency domain and can be decomposed into 4 ODEs. Furthermore, the symmetries in time and azimuth, allows for Fourier space decomposition in $t$ and $\varphi$:   
\begin{widetext}
\begin{eqnarray}
&&\psi=\frac{1}{2\pi} \int d\omega e^{i(-\omega t +m \varphi)} S[\theta] R[r],\label{R}\\
&&\Delta^{-s} \frac{d}{dr}	\left( \Delta^{s+1} \frac{dR}{dr} \right)+\left[ \frac{K^2-2is(r-M)K}{\Delta}+4is\omega r -\lambda \right]R=0,\label{r}\\
&&\frac{1}{\sin\theta}\frac{d}{d\theta}\left(\sin\frac{dS}{d\theta}\right)+\left(a^2 \omega^2 \cos^2\theta-\frac{m^2}{\sin^2\theta}-2a\omega s \cos\theta-\frac{2ms\cos\theta}{\sin^2\theta}-s^2 \cot^2\theta+s+A_{slm}\right)S=0,\label{s}
\end{eqnarray}
\end{widetext}
where $K=(r^2+a^2)\omega -am$ and $\lambda=A_{slm}+a^2 \omega ^2 -2am\omega$. The solution for the angular mode is spin-weighted spheroidal harmonic (full discussion can be found in \cite{Berti:2005gp}). We solve the radial mode numerically based on \citeauthor{Brito:2015oca} \cite{Brito:2015oca}, with publicly available Mathematica code, which was developed to study superradiance in Kerr metric \footnote{https://centra.tecnico.ulisboa.pt/network/grit/files/amplification-factors/}. Eq \ref{r} has  the following asymptotic solutions
\begin{eqnarray}
&&R=\mathcal{T} \Delta^{-s} e^{-ik_{\rm h}r^*}+\mathcal{O}e^{ik_{\rm h}r^*},  r \rightarrow r_+ , \\
&&R=\mathcal{I} \frac{e^{-i \omega r^*}}{r}+\mathcal{R}\frac{e^{i\omega r^*}}{r^{2s+1}},  r \rightarrow \infty, 
\end{eqnarray}
where $r^*$ is tortoise coordinate (defined as $r^*=\int \frac{r^2+a^2}{r^2-2Mr +a^2} dr$ that approaches -$\infty$ at horizon), $k_{\rm h}=\omega-\frac{a m}{2Mr_{+}}$ and $r_+ = M+\sqrt{M^2-a^2}$. 

In classical General Relativity, everything that reaches the horizon will fall into the BH, and thus theres is no outgoing wave at $ r \rightarrow r_+ $, i.e.  $\mathcal{O}=0$. However, for ECOs we assume that quantum gravity effects replace the horizon with (partially) reflective wall standing  the order of Planck length proper distance outside the (would-be) horizon. We shall assume that this modifies the boundary condition, so that the wall reflects the incoming energy flux  (see \cite{Nakano:2017fvh} for definition of energy near horizon) with a rate $R$ but does not change the phase:
\begin{widetext}
\begin{eqnarray}
&&|\mathcal{O}|^2=R_{\rm wall} \left | \frac{C}{D}\right| ^{s/2} |\mathcal{T}|^2, \qquad \arg[\mathcal{T} \Delta^{-s} e^{-ik_{\rm h}r^*}]=\arg[\mathcal{O}e^{ik_{\rm h}r^*}]\qquad \text{when }  r\rightarrow r_{\rm wall},\\
&&C=B \left\{-36 a^2 \omega ^2+36 a m \omega +[\lambda +(s+1) s-2]^2\right\}+\{ 2 [\lambda +(s+1) s]-1\} \left(96 a^2 \omega ^2-48 a m \omega \right)+144 \omega ^2 \left(M^2-a^2\right),\nonumber\\
&& \\
&&B = [\lambda +s (s + 1)]^2 + 4 m a \omega- 4 a^2 \omega^2,  \quad D= 256 k_{\rm h}^2 (2 M r_+)^8 [k_{\rm h}^2 + \frac{4(M^2-a^2)}{(4 M r_+)^2}]^2 [k_{\rm h}^2 + 
      \frac{16(M^2-a^2)}{(4 M r_+)^2}].
\end{eqnarray}
\end{widetext}

$R_{\rm wall}=1$ would correspond to a perfectly reflective wall, but the actual reflectivity and phase change depend on the specific quantum gravity model for ECOs. In the rest of the paper, we will present solutions to these equations with different choices of the reflectivity and discuss the important properties of solutions, such as echo templates, time-delays and superradiant instability.

\begin{figure}
\includegraphics[width=0.2\textwidth]{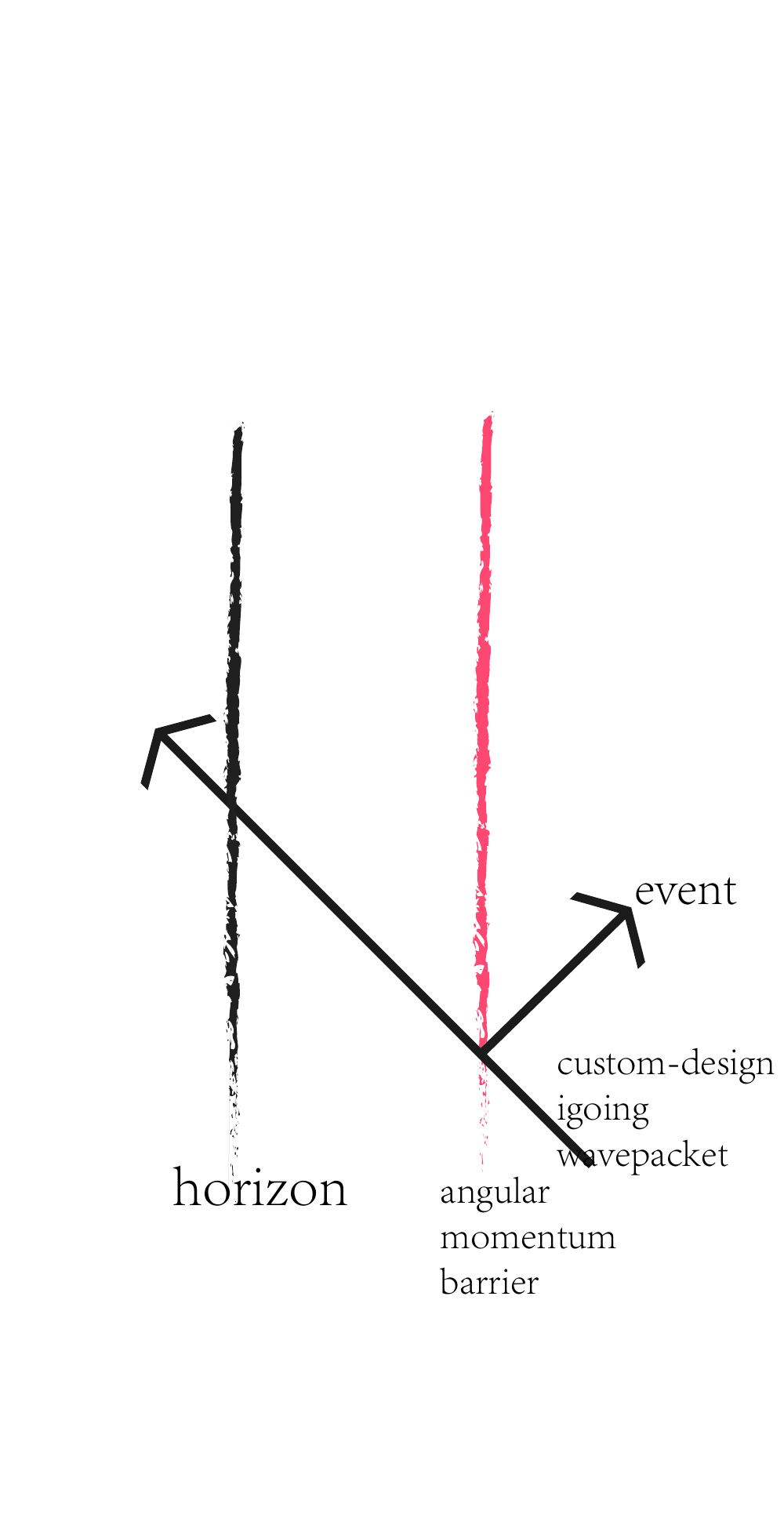}
\includegraphics[width=0.2\textwidth]{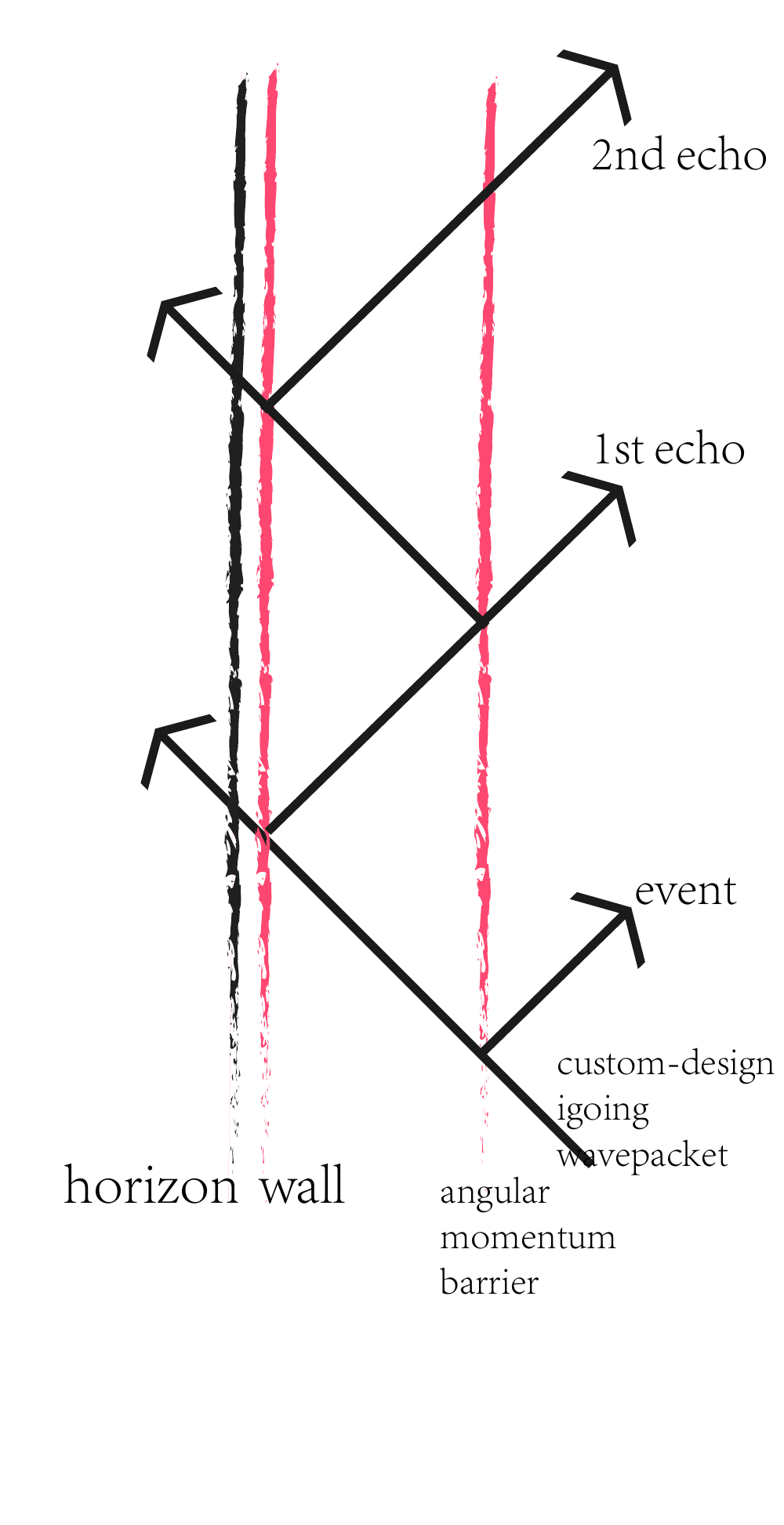}
\caption{\label{31} Black Holes (BHs) and Exotic Compact Objects (ECOs) with an ingoing wavepacket. For BHs, angular momentum barrier reflects low frequency modes but higher frequencies cross the barrier and fall through the horizon. For ECOs with a wall standing  the order of Planck length proper distance outside the (would-be) horizon, modes with intermediate frequencies can be trapped between the wall and the angular momentum barrier, slowly leaking out as repeating echoes.}
\end{figure}

\begin{figure*}
\raggedright
%\begin{minipage}[b]{0.2\textwidth}
\includegraphics[width=\textwidth]{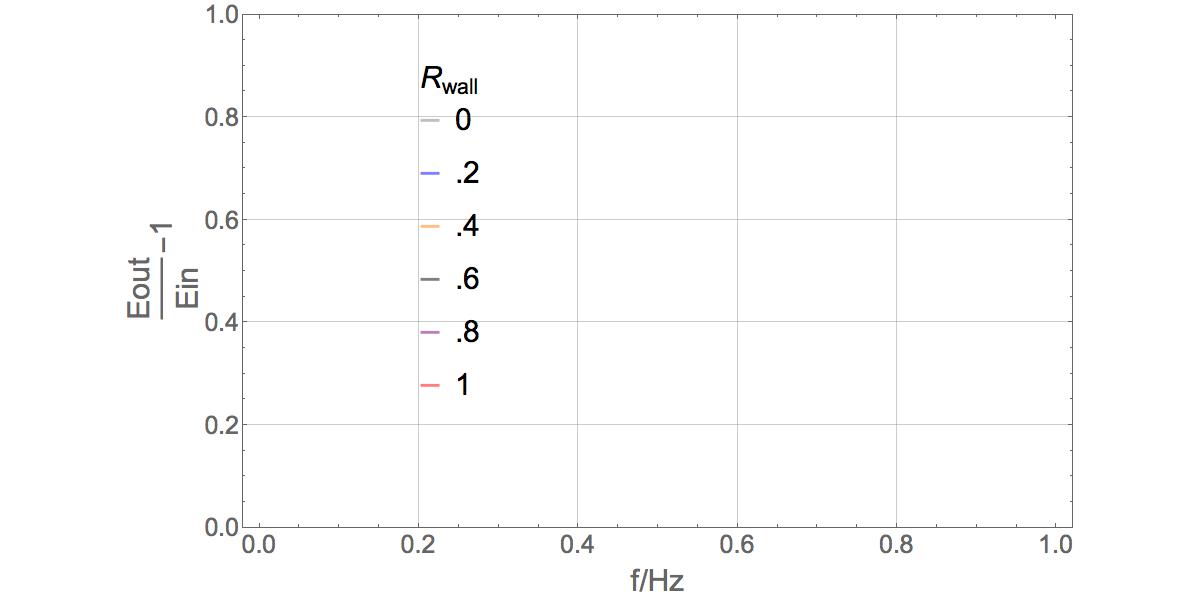} 
%\end{minipage}
\caption{\label{32} Echoes with different wall positions. Changing the positions of wall doesn't influence the shape of echoes a lot, but when putting wall closer to the would-be horizon and away from angular momentum barrier, the time-delay becomes bigger.}
\end{figure*}

\section{\label{sec3}Making Echoes}
Realistic predictions for Echo waveforms requires nonlinear simulations of the mergers of binary ECOs in full general relativity. As a consistent covariant formulation for dynamics of ECOs is yet non-existent, we have to rely on approximate methods to produce realistic echo templates. In order to do this using linear theory, we instead custom-design an ingoing wavepacket $\hat{h}_{\rm in}$ at infinity, so that the outgoing waveform matches the LIGO best-fit template $\hat{h}_{\rm LIGO}$ (without a wall). %In classical GR, $\hat{h}_{\rm in}$ is reflected by the angular momentum barrier leading to a ringdown waveform $\hat{h}_{\rm LIGO}$  confirmed by LIGO data. 
The higher frequencies will go across the barrier and fall into BH, as shown in Fig \ref{31} (left), while the lower frequencies are reflected. We thus assume 
\begin{eqnarray}
\hat{h}_{\rm LIGO}(\omega)= R_{\rm BH}(\omega) \hat{h}_{\rm in}(\omega),
\end{eqnarray}
where $R_{\rm BH}(\omega)$ is the reflectivity of the Kerr angular momentum barrier.
For an ECO, however, we have one more barrier near the would-be horizon as shown in Fig \ref{31} (right). Wavepackets with intermediate frequencies can now be trapped between two barriers and leak slowly every time when they hit the angular momentum barrier. Therefore, ECOs would have a similar ringdown waveform as classical BHs, but they are followed by delayed slowly decaying echoes.
\begin{eqnarray}\label{eq:hout}
\hat{h}_{\rm out}(\omega)= R_{\rm ECO}(\omega) \hat{h}_{\rm in}(\omega)=R_{\rm ECO}(\omega) \frac{\hat{h}_{\rm LIGO}(\omega)}{R_{\rm BH}(\omega)} f_{\text{cutoff}}(\omega), \nonumber\\
\end{eqnarray}
where $f_{\text{cutoff}}(\omega)$ is a low-pass filter introduced to suppress numerical noise at high frequencies, as 
the reflectivity of the Kerr angular momentum barrier $R_{\rm BH}(\omega)$, in the denominator, vanishes at high frequencies. Luckily, high frequencies leak out quickly in the first echo, and have small effect on the subsequent echoes. Our choice of $f_{\text{cutoff}}$ does not affect the second and later echoes, but it changes the first echo slightly by cutting the high frequency noise:
\begin{eqnarray}
&\hat{h}_{\rm out, fin}= \hat{h}_{\rm out} f_{\text{cutoff}},\\
&f_{\text{cutoff}}=\exp\left[ -\frac{1}{2}\left(\frac{2 \pi f({\rm Hz})-299.495}{1347.73}\right)^{16}\right], \label{cutoff}
\end{eqnarray}
where $\omega = 2\pi f$.

With the equations and boundary conditions given in the last section, we can numerically solve for  $R_{\rm BH}$ and $R_{\rm ECO}$ as a function of frequency. We use LIGO event GW150914 with $a=0.67$, $M=62~M_{\odot}$ and $z=0.09$. The mass is measured in the source frame and the finial mass used in our calculation is the mass in the detector frame $M_{\rm fin}=(1+z) M$. The waveform is dominated by the $(l,m) =(2,2)$ mode, which we shall focus on for the rest of the paper \footnote{Given the symmetries of Eqs. (\ref{r}-\ref{s}), we can easily extend the solution to $m=-2$ case using $R_{slm}[\omega]=R^*_{sl-m}[-\omega]$.}

The time dependence of the waveform can then be obtained by Fourier transforming $\hat{h}_{\rm out}(\omega)$, and is shown in Fig \ref{32}. We see that changing the position of the wall changes the time-delay between the echoes, but does not affect the individual echo waveforms significantly (as long as the wall is close the would-be horizon). %The ingoing wavepacket is partially reflected off the angular momentum barrier, part of it bounces back creates the same ringdown waveform as classical BH, and others go over barrier and arrive at the wall. Part of it goes back and across the angular potential barrier to produce first echo. The remained bounces back by barrier and repeat to produce second and later echoes. 
As we see in Fig. \ref{31}, in the geometric optics approximation, the time delay between echoes, $\Delta t_{\rm echo,geom}$ is given by the travel time from the angular momentum barrier to the wall and back \cite{abedi2016echoes}:
\begin{eqnarray}
&&\Delta t_{\rm echo,geom}=2r_*|^{r_{\rm barrier}}_{r_{\rm wall}}=2\int^{r_{\rm barrier}}_{r_{\rm wall}}dr \frac{r^2+a^2M^2}{r^2-2Mr+a^2M^2}\nonumber\\
&&=2r_{\rm barrier}-2r_{\rm wall}+2\frac{r_+^2+a^2M^2}{r_+-r_-}\ln\frac{r_{\rm barrier}-r_+}{r_{\rm wall}-r_+}\nonumber\\
&&-2\frac{r_-^2+a^2M^2}{r_+-r_-}\ln\frac{r_{\rm barrier}-r_+}{r_{\rm wall}-r_-}.
\end{eqnarray}
This can be well approximated by the following fitting function:
\begin{eqnarray}
&&\Delta t_{\rm echo,geom} 
=2\frac{r_+^2+a^2M^2}{r_+-r_-}\ln\frac{M}{r_{\rm wall}-r_+}+M G(a), \label{t_geom}\\
&&G(a)\simeq \frac{0.335}{a^2-1}+4.77+7.42 (a^2-1)+4.69(a^2-1)^2,\nonumber\\ &&\\
&&r_{\rm wall}-r_{+}= \frac{\sqrt{1-a^2} d_{\rm wall}^2}{4M(1+\sqrt{1-a^2})},
\end{eqnarray}
where we find the fit of $G(a)$ for the angular momentum barrier of $l=m=2$ mode, while $d_{\rm wall}$ is the proper distance from the wall to the would-be horizon.  The latter is expected to be comparable to Planck length for ECOs of quantum gravitational nature, but $\Delta t_{\rm echo}$ only depends on the exact value of  $d_{\rm wall}$ logarithmically (see Fig. \ref{32}). 

\begin{figure*}
\includegraphics[]{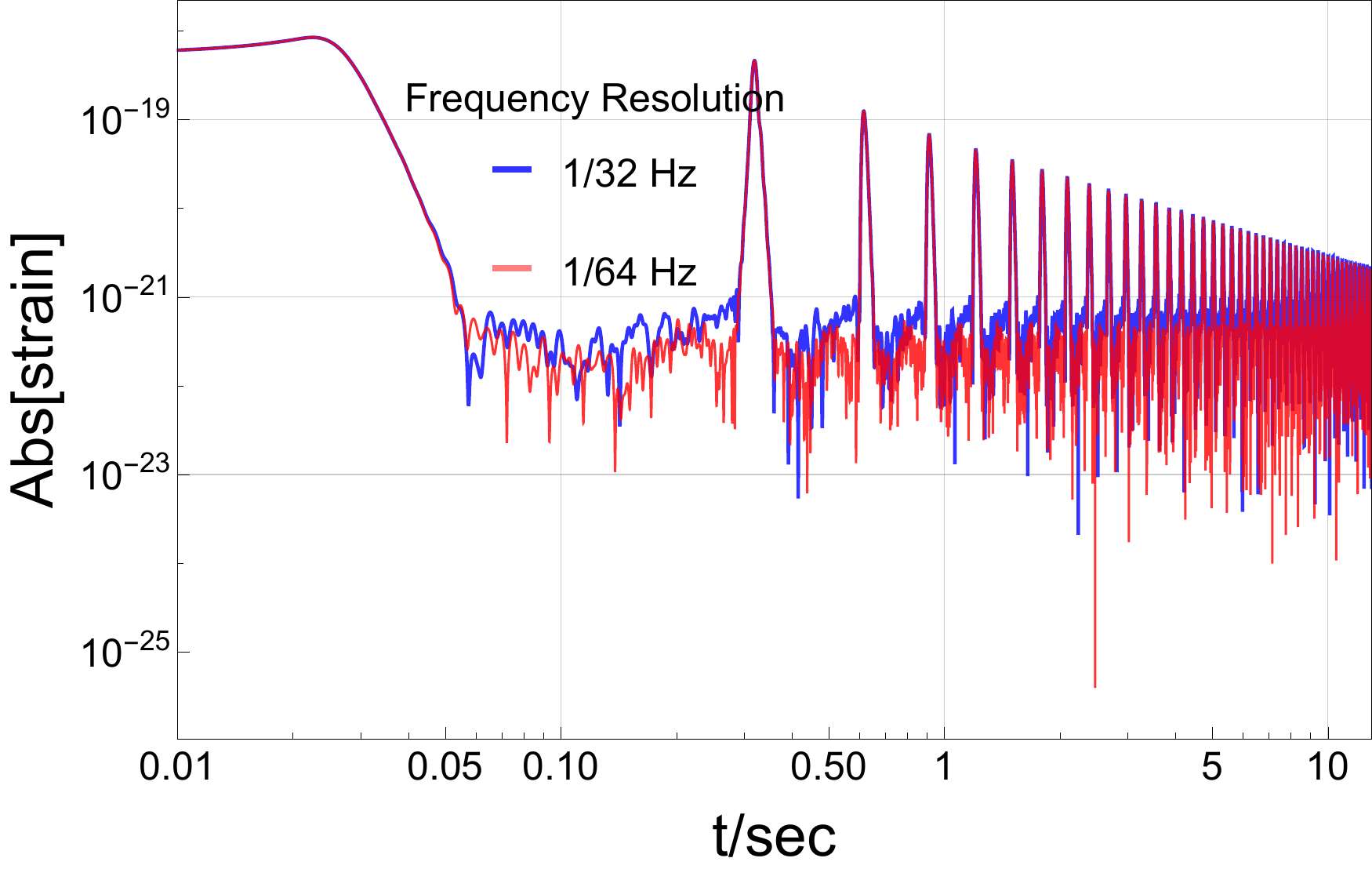}
\caption{\label{51} Predicted echoes for LIGO event GW150914 in the time domain with different resolution, assuming a prefect wall at a Planck length proper distance outside the horizon}.
\end{figure*}

\begin{figure}
\includegraphics[width=0.43\textwidth]{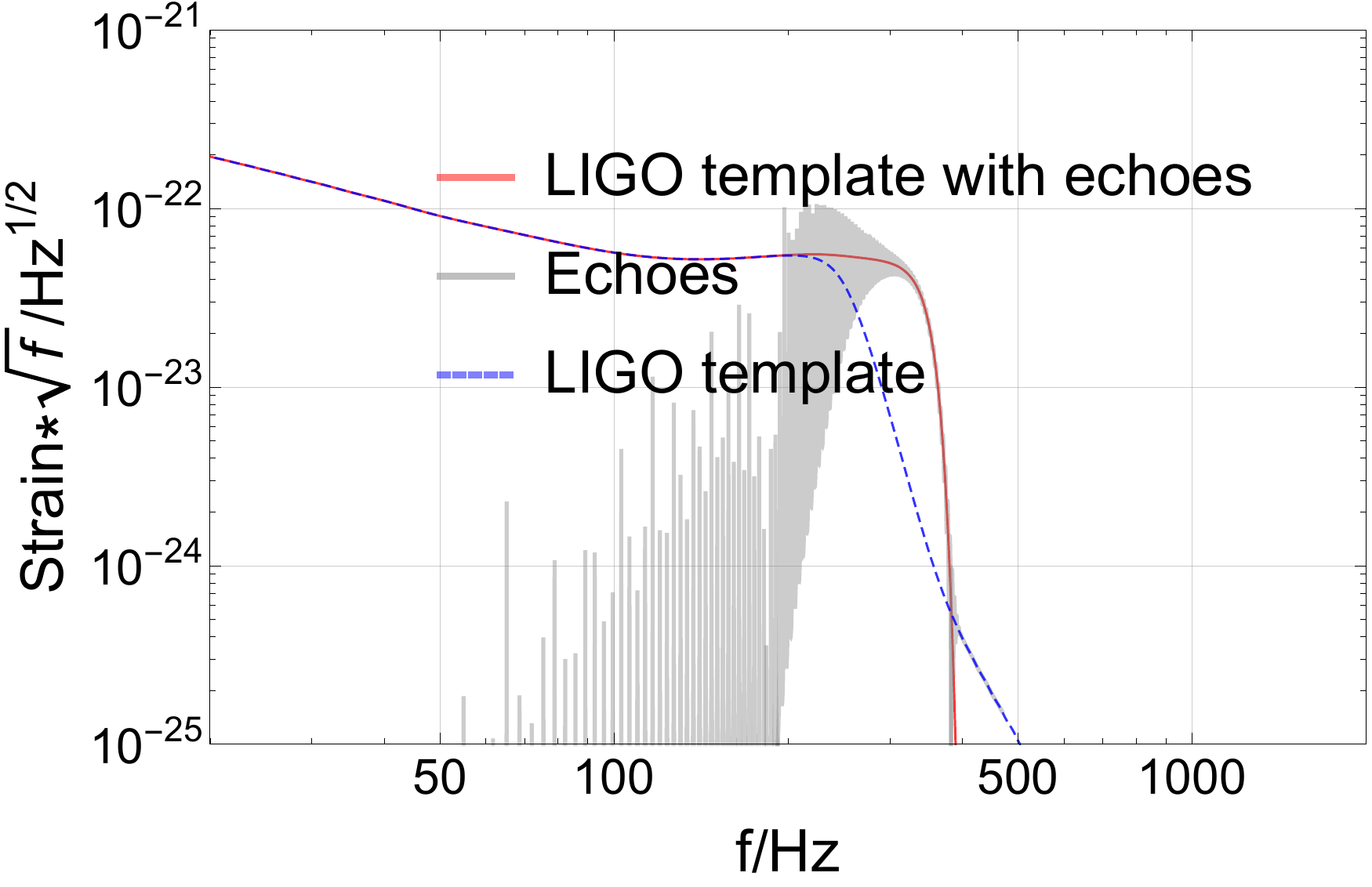}
\caption{\label{511} Predicted echoes for LIGO event GW150914 in the frequency domains, assuming a prefect wall at a Planck length proper distance outside the horizon}.
\end{figure}

The echoes in both time and frequency domain for the LIGO event GW150914 are shown in Fig. \ref{51} and \ref{511} with perfect wall standing  a Planck length proper distance outside the (would-be) horizon. Here, we show the Amplitude Spectral Density (ASD), which is the square root of the power spectral density. The latter is the average of the square of the fast Fourier transforms (FFTs) of the model. In the next section, we will study the structure of the echo in the frequency domain and present how superradiance affect the structure of echo.

\section{\label{seca1}Superradiance}

Scattering off Kerr BH can lead to superradiance of modes with frequency $0<\omega<m \Omega_{\rm H}$, which can extract energy from a spinning background \cite{1972Natur.238..211P}. Adding a (partially) reflective wall near horizon could turn this amplification to an instability, since modes trapped between the wall and the angular momentum barrier can extract the spin energy repeatedly \cite{1978CMaPh..63..243F, Cardoso:2007az}. In this section, we study this effect for the echoes in frequency domains. 

\begin{figure}
\includegraphics[width=0.43\textwidth]{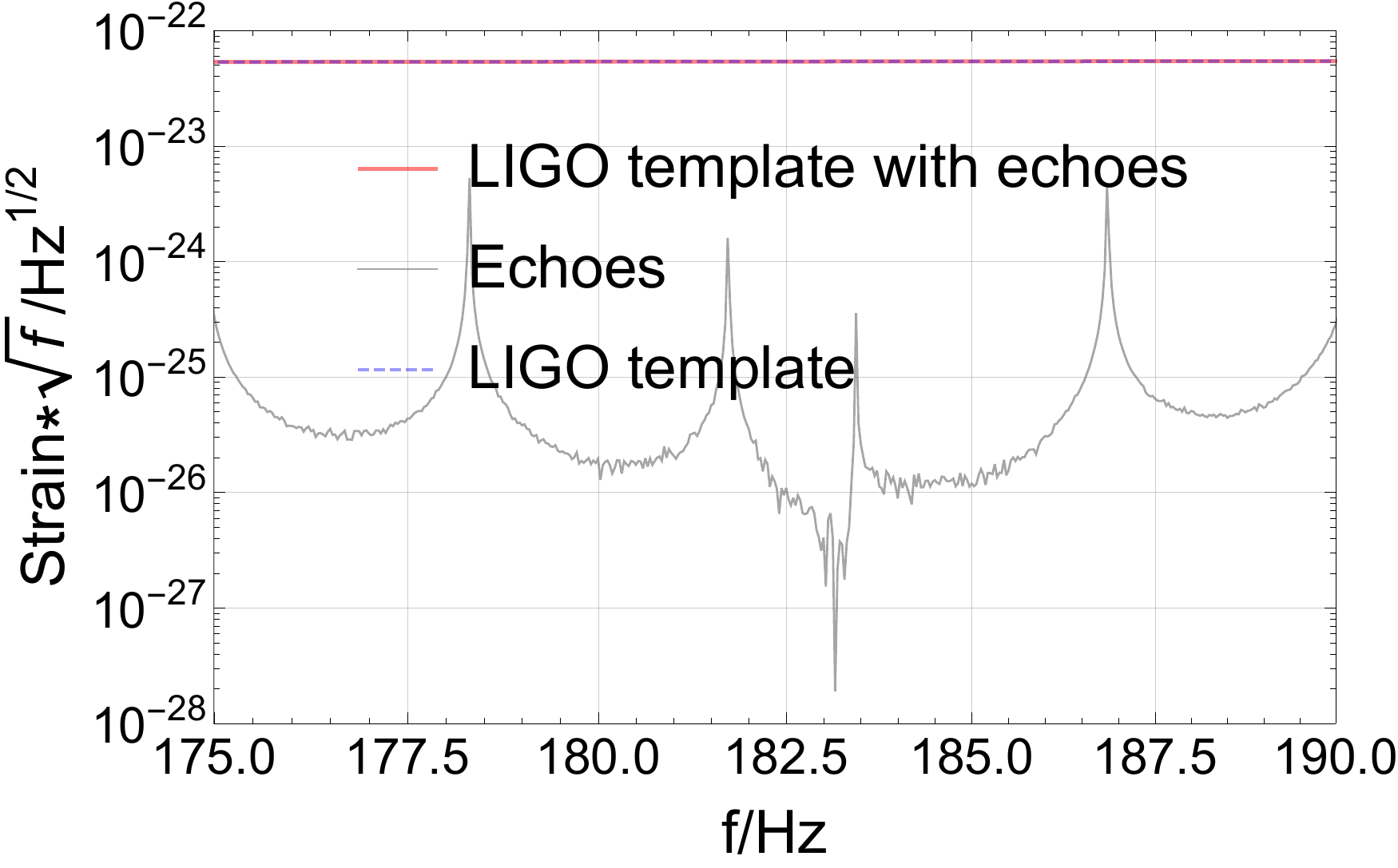}
\includegraphics[width=0.43\textwidth]{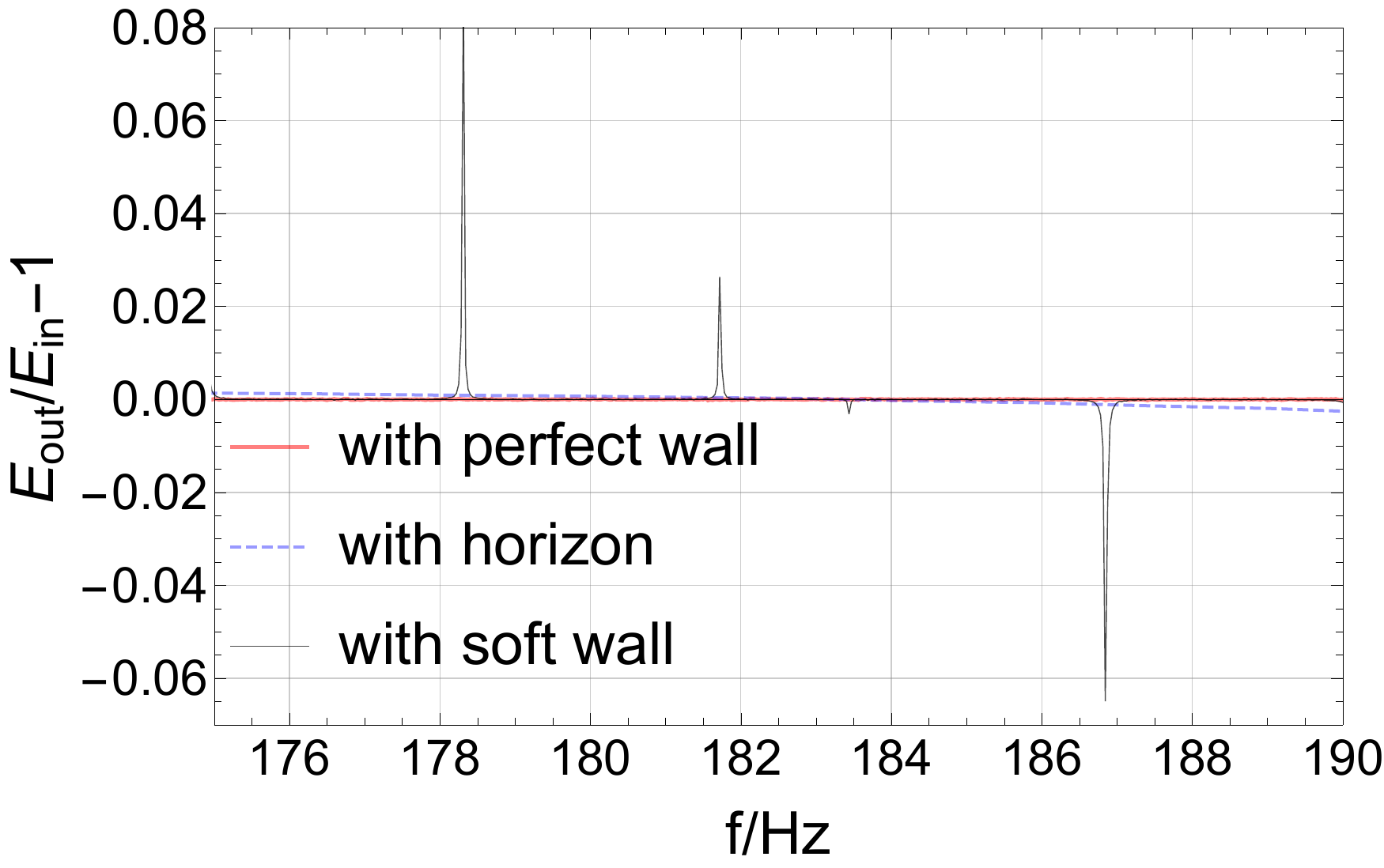}
\includegraphics[width=0.43\textwidth]{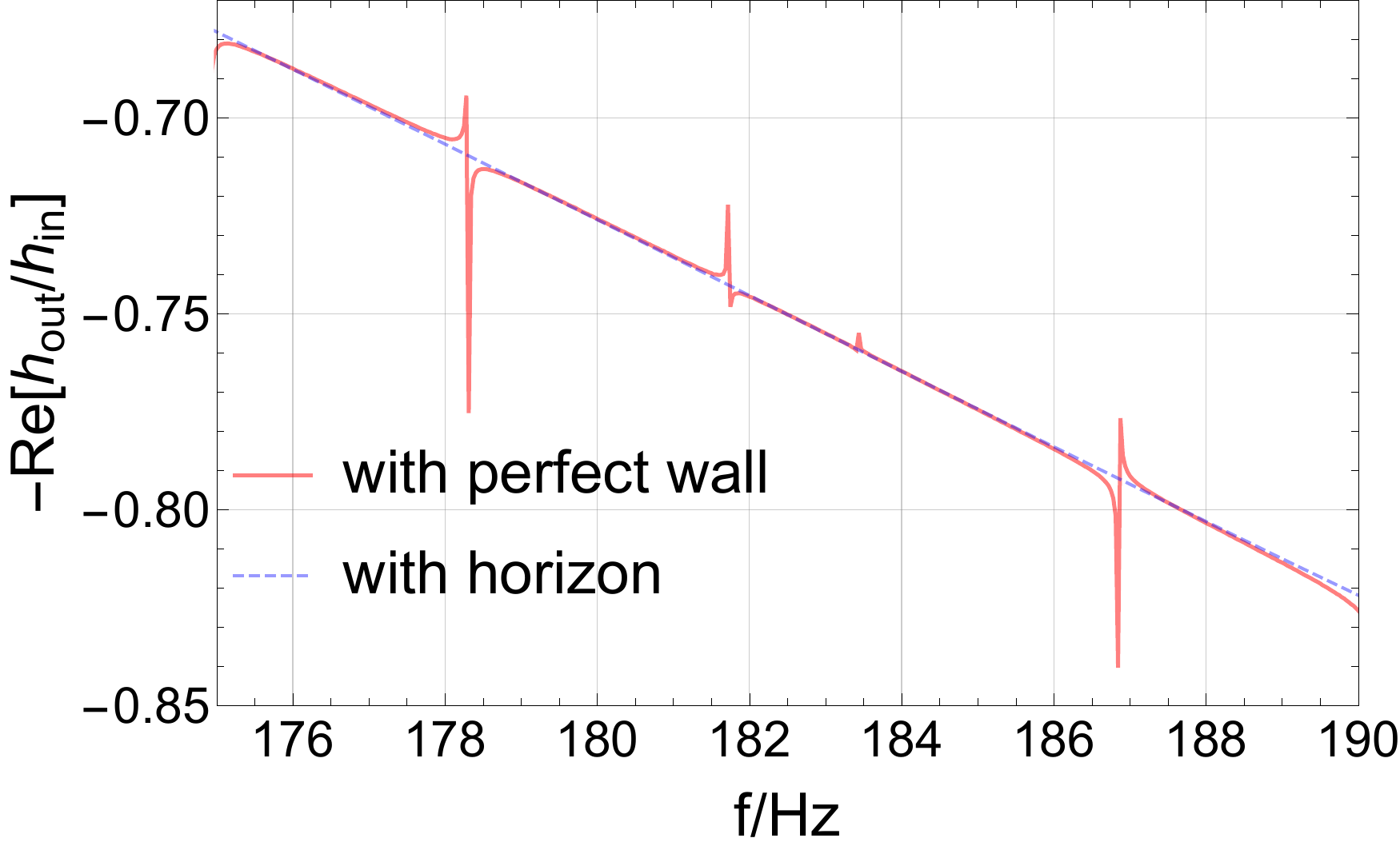}
\caption{\label{81} Superradiance in frequency domain for GR Kerr BH and ECO with a wall.}
\end{figure}

There is an odd looking spike in Fig. \ref{511} frequency domain around 183 Hz (see top panel in Fig. \ref{81} for a zoom-in). Indeed, this is exactly the threshold frequency for the superradiance. This is demonstaretd in the middle panel of Fig. \ref{81}, which shows the scattering amplification with the horizon, perfect wall and soft wall around that frequency. The vertical axis is the relative energy, extracted from around black hole by scattered gravitational waves. The blue dashed line shows superradiance slowly turning off with increasing the frequency, and we confirm that it ends exactly at frequency $f _{\rm max}=a m/[2 \pi(r_+^2+a^2)]=183~ \textrm{Hz}$, for $m=2$ as shown in the plot. In contrast, superradiance by soft wall (grey and thin curve) occurs at resonance peaks, corresponding to the ergoregion trapped mode (for more details, see Appendix \ref{a1} ). Since superradiance ends at 183Hz, the resonance peaks shift the direction, which is the reason we have an odd spike in the Fig. \ref{511} and \ref{81} top panel. 

The perfect wall (the red thick curve) in Fig. \ref{81} middle panel is a constant zero without any resonance peaks, since a perfect reflective wall kills superradiance, as all the energy that goes in, comes out eventually (see Appendix \ref{a1} for a subtlety in this argument). However, the odd spike structure remain in the amplitudes, as shown in Fig. \ref{81} bottom panel, where we change the  vertical axis to real part of outgoing to ingoing wave at infinity. We still see the sign flip in resonance structure at 183 Hz.

In the next section, we study the echo templates resulting from solving the linearized Einstein equations, which improves the simplistic geometric picture in Fig. \ref{31}. 

\section{\label{sec5}Minimal Echo templates}

%Template
% Based on last section, we develop a template for later echoes. As shown in fig(5.1), the n-th echo
% \begin{eqnarray}
% \hat{h}_n[\omega]&&= T \tilde{T} R^{n-1} \tilde{R}^n \hat{h}_{LIGO}[\omega]\\
% &&=R \tilde{R} \hat{h}_{n-1}[\omega]
% \end{eqnarray}
% Where R and T is reflection and transmission coefficient at infinity with out-coming wave from near horizon and $\tilde{T}$ is time-inverse of T with in-coming wave from infinity.

%Separate echo
%We separate the echoes in time-domain with a Gaussian cut off and Fourier-transform the cut off to frequency-domain, which is still a simple Gaussian ones.
%\begin{eqnarray}
% h_n[t]&&=g_{n}[t-t_n;\tau]h[t]\\
% \hat{h}_n[w]&&=\int \hat{g}_n[\omega '] h[\omega - \omega'] d\omega'
% \end{eqnarray}
% Where h is the whole template including initial burst and later echoes, $g_{n}$ is Gaussian cut-off function with $t_n$ as time of $n_th$ echo and $\tau$ as width of echo.We do it for first few echoes and get numerical reflection rate in frequency-domain shown in fig(5.2).

% {5.2 Rate picture}

% fitting function is  $\hat{h}_n[w]= R \hat{h}_{n-1}$

%The echoes in both frequency and time domain for the LIGO event GW150914 are shown as Fig \ref{51} with perfect wall standing a Planck length away from the would-be horizon. Here, we show the Amplitude Spectral Density (ASD), which is the square root of the power spectral density. The latter is the average of the square of the fast Fourier transforms (FFTs) of the model. 

Now that we have numerical predictions for echoes, we would like to provide simple fitting functions that could be used for quick visualization and data-fitting purposes. We call these fitting functions, templates. In order to find our templates, we define echoes in the time domain by the regions that surround the peaks of $|h(t)|$ and exceed a limit: $\ln\left[|h(t)|/|h|_{\rm max, n}\right] > -1, -1.5$ or $-2$. $|h|_{\rm max, n}$ is the height of the $n^{\rm th}$ peak of $|h(t)|$, which we call the $n^{\rm th}$ echo. Then we fit the $n^{\rm th}$ echo to a complex gaussian 
\begin{eqnarray}
h_n(t)&=& \exp [\Psi_n(t)+I \Phi_n(t)], \\ \label{gaussian}
 \Psi_n(t)&=&a_0+a_1t+a_2 t^2, \\
 \Phi_n(t)&=&b_0+b_1t,
 \end{eqnarray}
 where $a_0$, $a_1$, $a_2$, $b_0$ and $b_1$ are real numbers. This form is same as fitting the $n^{\rm th}$ echo to $ A \exp [ \frac{(t-t_0)^2}{2\sigma^2} ]$, where $A$ and $t_0$ are complex, while the width $\sigma$ is real.
 
\begin{figure}
\includegraphics[width=0.22\textwidth]{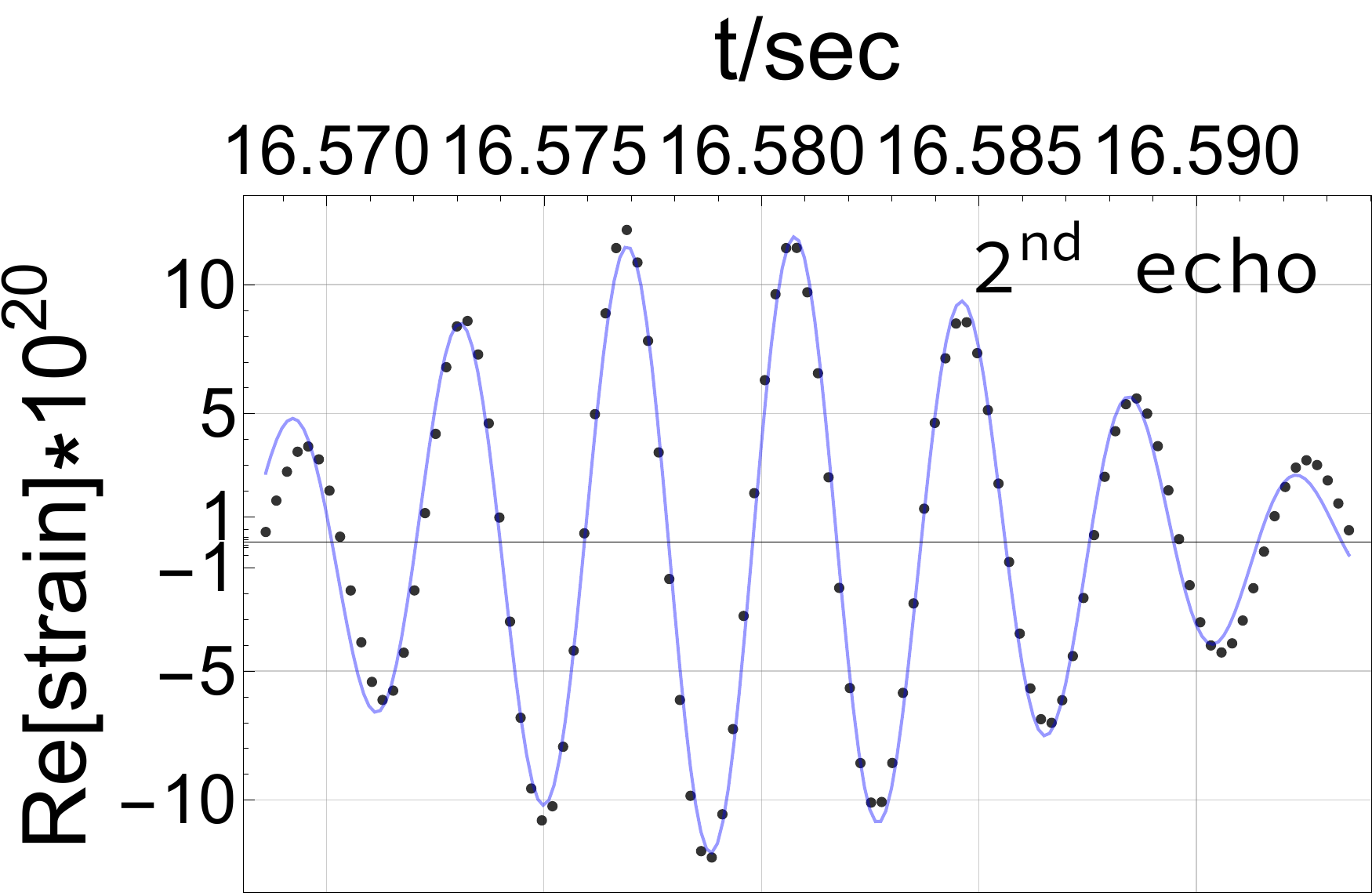}
\includegraphics[width=0.22\textwidth]{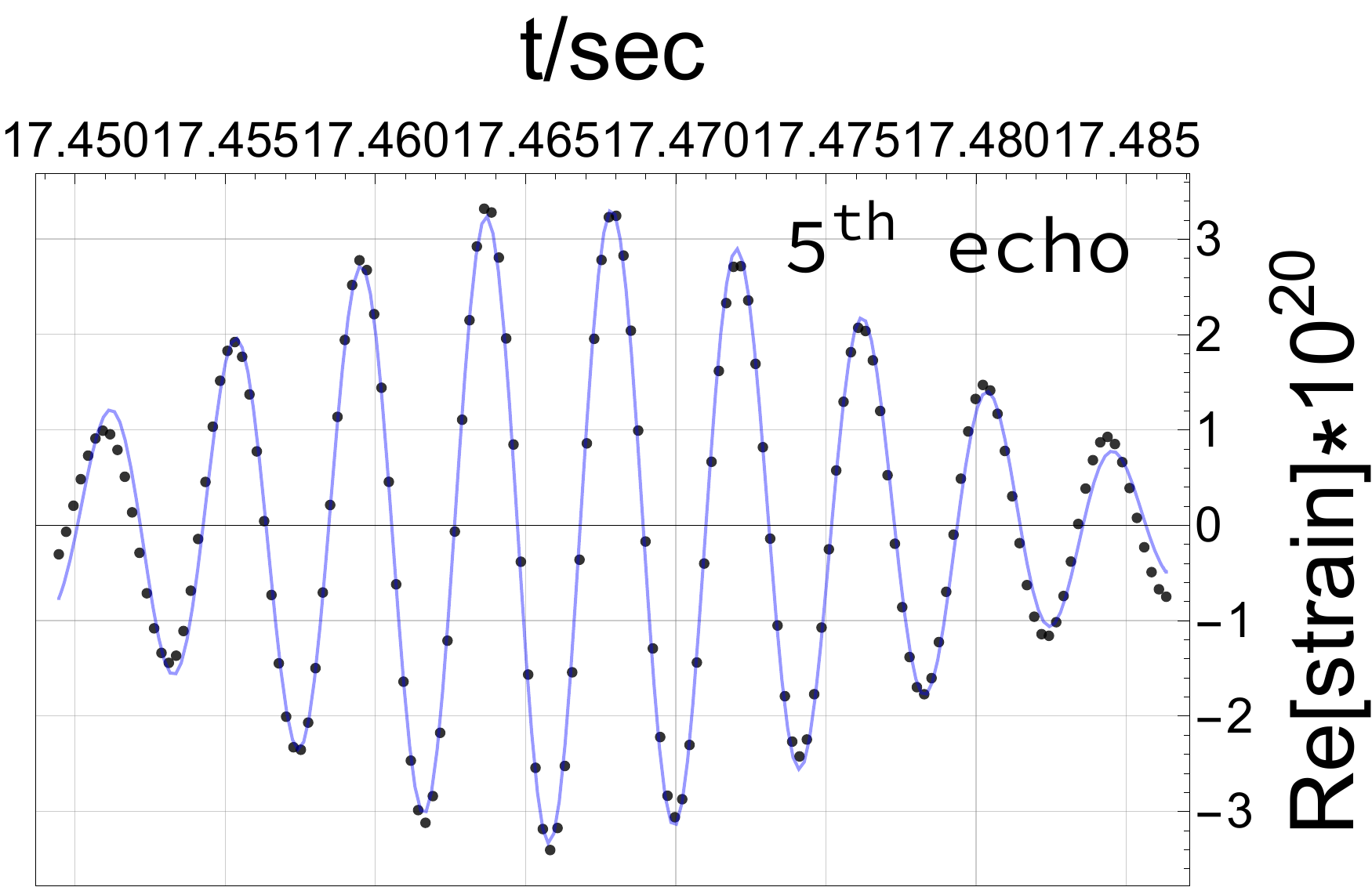}
\includegraphics[width=0.22\textwidth]{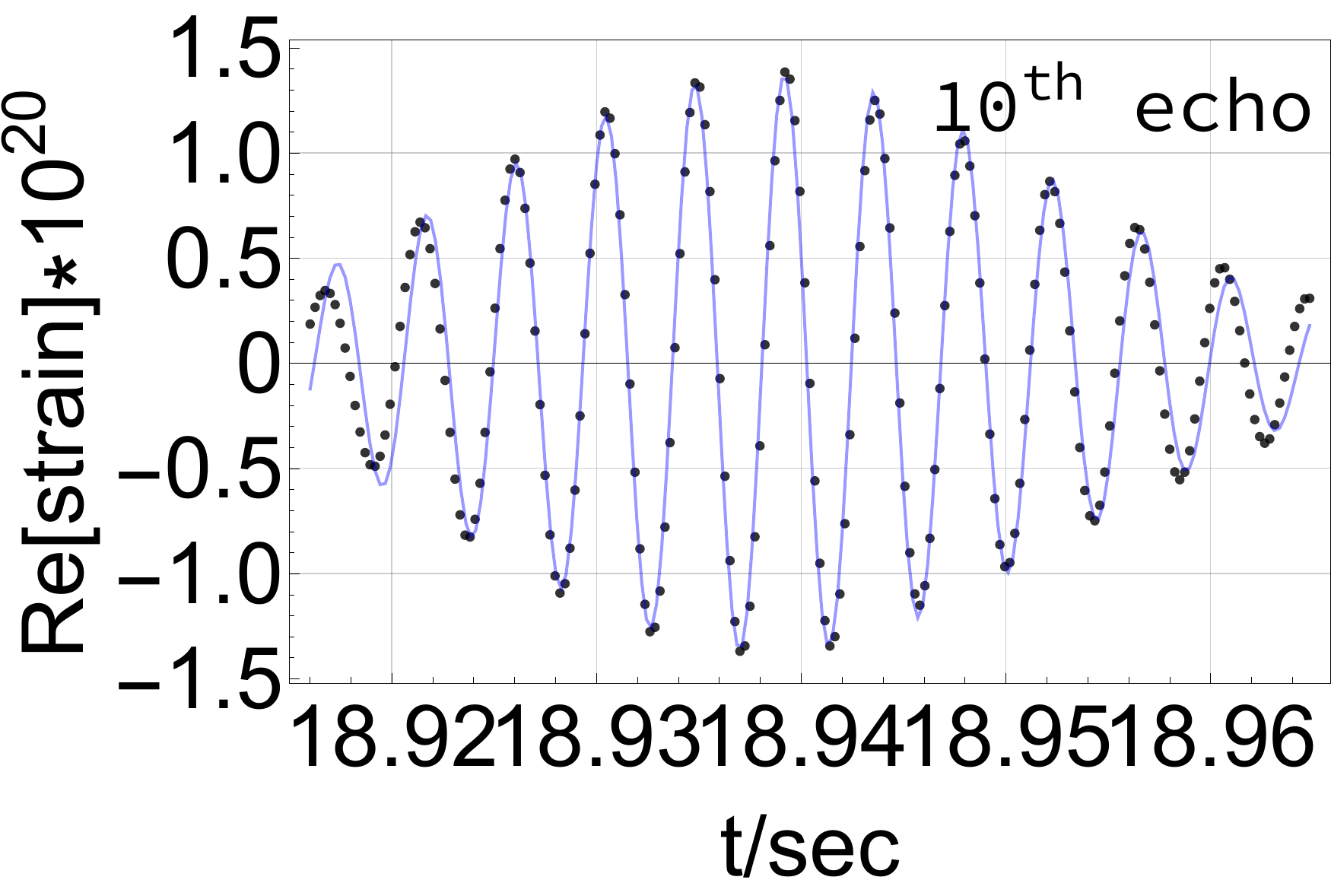}
\includegraphics[width=0.233\textwidth]{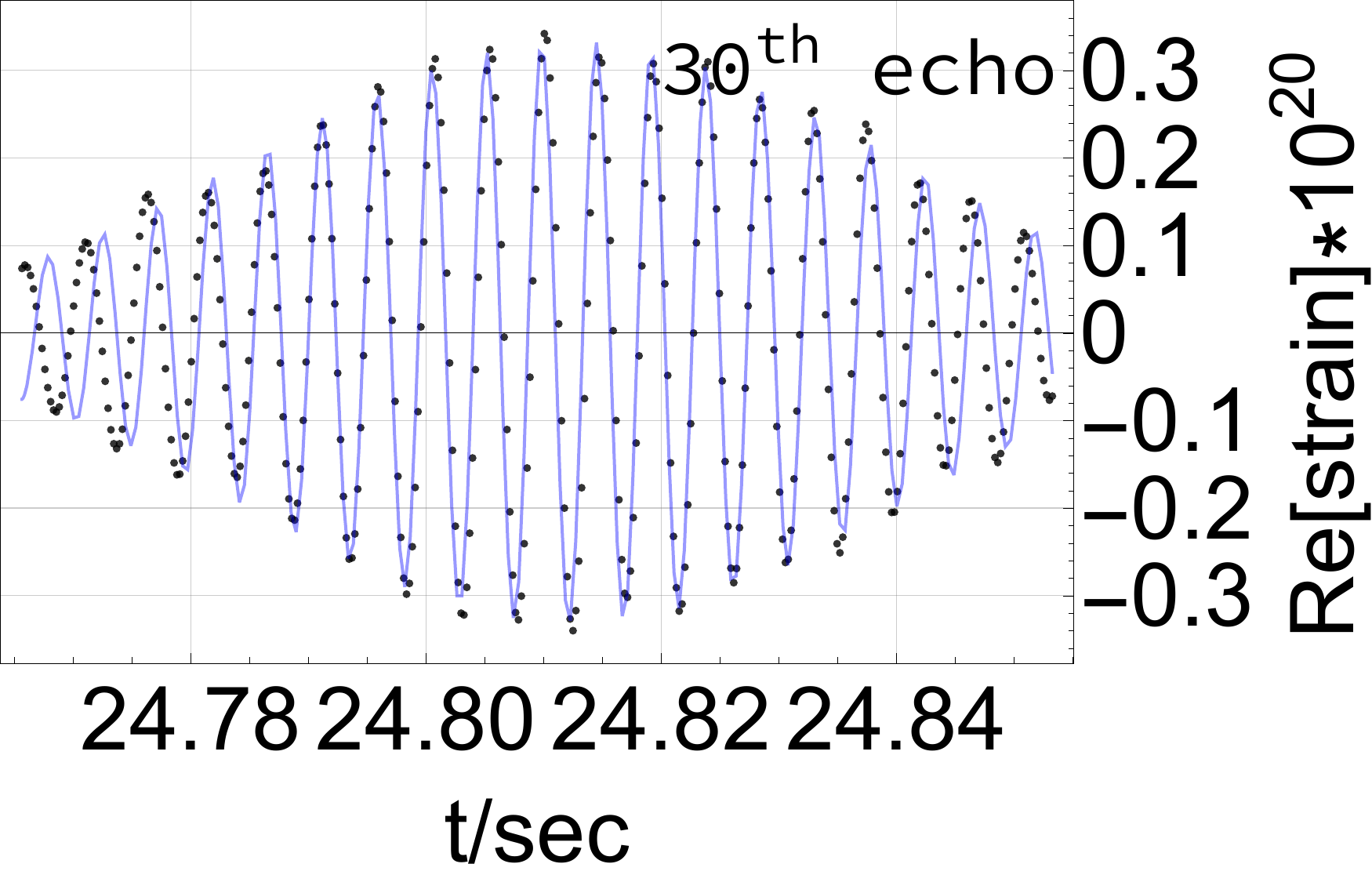}
\includegraphics[width=0.23\textwidth]{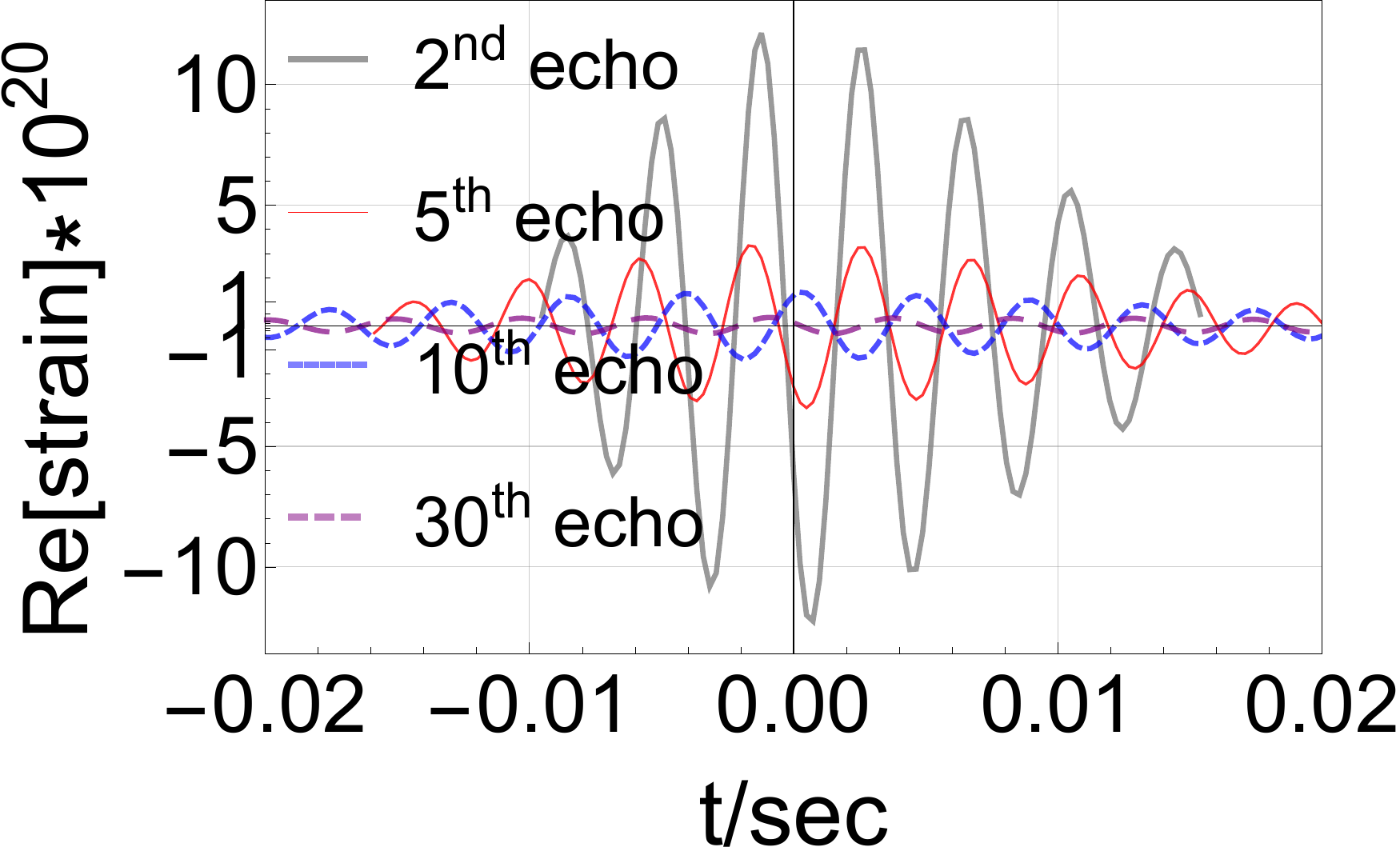}
\includegraphics[width=0.23\textwidth]{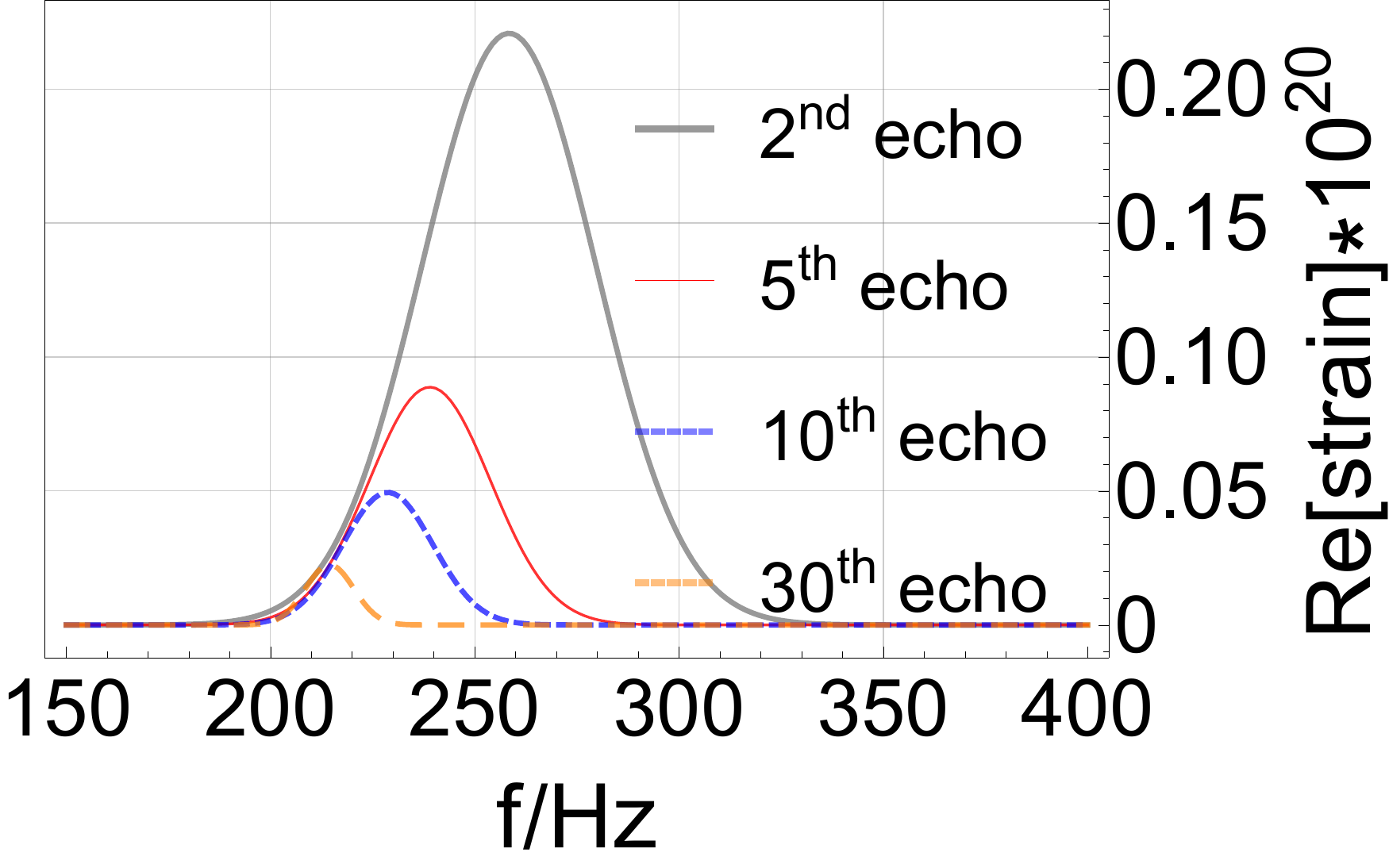}
\caption{\label{520} Best fit gaussians to the $2^{\rm nd}$, $5^{\rm th}$, $10^{\rm th}$ and $30^{\rm th}$ echoes within $\ln\left[|h(t)|/|h|_{\rm max, n}\right] > -1.5$. We see that as high frequency modes leak out faster, later echoes decay in amplitude and become wider in time domain, and high frequency is cuted in the frequency domain.}
\end{figure}

As an example, Fig \ref{520} compares the numerical solutions and gaussian fits for the 2$^{nd}$, $10^{th}$,  and $30^{th}$ echoes, with time origin shifted to center of each echo, and fitting the region with $\ln\left[|h(t)|/|h|_{\rm max, n}\right] > -1.5$.

Within this approximation, there are five real parameters for every echo that quantify its amplitude, width and center, both in time and frequency domain, as well as the overall phase at the center of the echo, as shown in Table \ref{t2}. 

\begin{table*}%The best place to locate the table environment is directly after its first reference in text
\caption{\label{t2} Some physical quantities of a single echo defined by the five parameters from the gaussian echo template (Eq. \ref{gaussian})
}
\begin{ruledtabular}
\begin{tabular}{c|ccc}
&\textrm{width}&\textrm{center}&\textrm{peak amplitude}\\
\colrule
\textrm{time}&
$\sqrt{-1/ (2a_2})$&
$-a_1/2 a_2$&
$\exp[a_0-a_1^2/4 a_2]$\\
\colrule
\textrm{frequency} & $ \sqrt{-2 a_2}/(2 \pi)$ & $b_1/(2\pi)$ & $\exp[a_0-a_1^2/4 a_2-1/2 \log[2 \sqrt{a_2^2}]]$\\
\colrule
\colrule
\textrm{overall phase} & $ b_0-b_1 a_1/(2 a_2)$ \\
\end{tabular}
\end{ruledtabular}
\end{table*}

\begin{table}%The best place to locate the table environment is directly after its first reference in text
\caption{\label{t3} Best fit gaussian echo template quantities (see Table \ref{t2} and Fig. \ref{52}) , for our minimal model of GW150914}
\begin{tabular}{|l|l|}
\cline{1-2}
%\textrm{parameter}&\textrm{average}\\
%\colrule
\textrm{peak amplitude in time / strain}&
$2.91 \times 10^{-19}/n^{1.32}$\\
\colrule
\textrm{width in time / msec} & $4.29+ 0.883 n$ \\
\colrule
\textrm{correction to} $\Delta t_{\rm echo,geom}$ \textrm{/ msec}& $1.52+1.71/(1+n)$ \\
\colrule
\textrm{peak frequency / Hz} & $177+102/ n^{0.3}$ \\
\colrule
\textrm{Overall phase}&
$-7.26+27.1 n^{0.945}+22.6 n$\\
\cline{1-2}
\end{tabular}
\end{table}

Table \ref{t3} provides the best fit parameters of our echo templates for all echoes, based on the LIGO event GW150914 and  averaging over the best fit functions with different echo domains $\ln\left[|h(t)|/|h|_{\rm max, n}\right] > -1, -1.5$ or $-2$. 
%For easy to understand and also simplicity, rather than presenting  $a_0$, $a_1$, $a_2$, $b_0$ and $b_1$, we choose other five physical quantities(defined in Table \ref{t2}, they are width, center and peak amplitude in time domain, center in frequency domain and the overall phase) to show the template. 

\begin{figure*}
\minipage{0.33\textwidth}
  \includegraphics[width=1.151\linewidth]{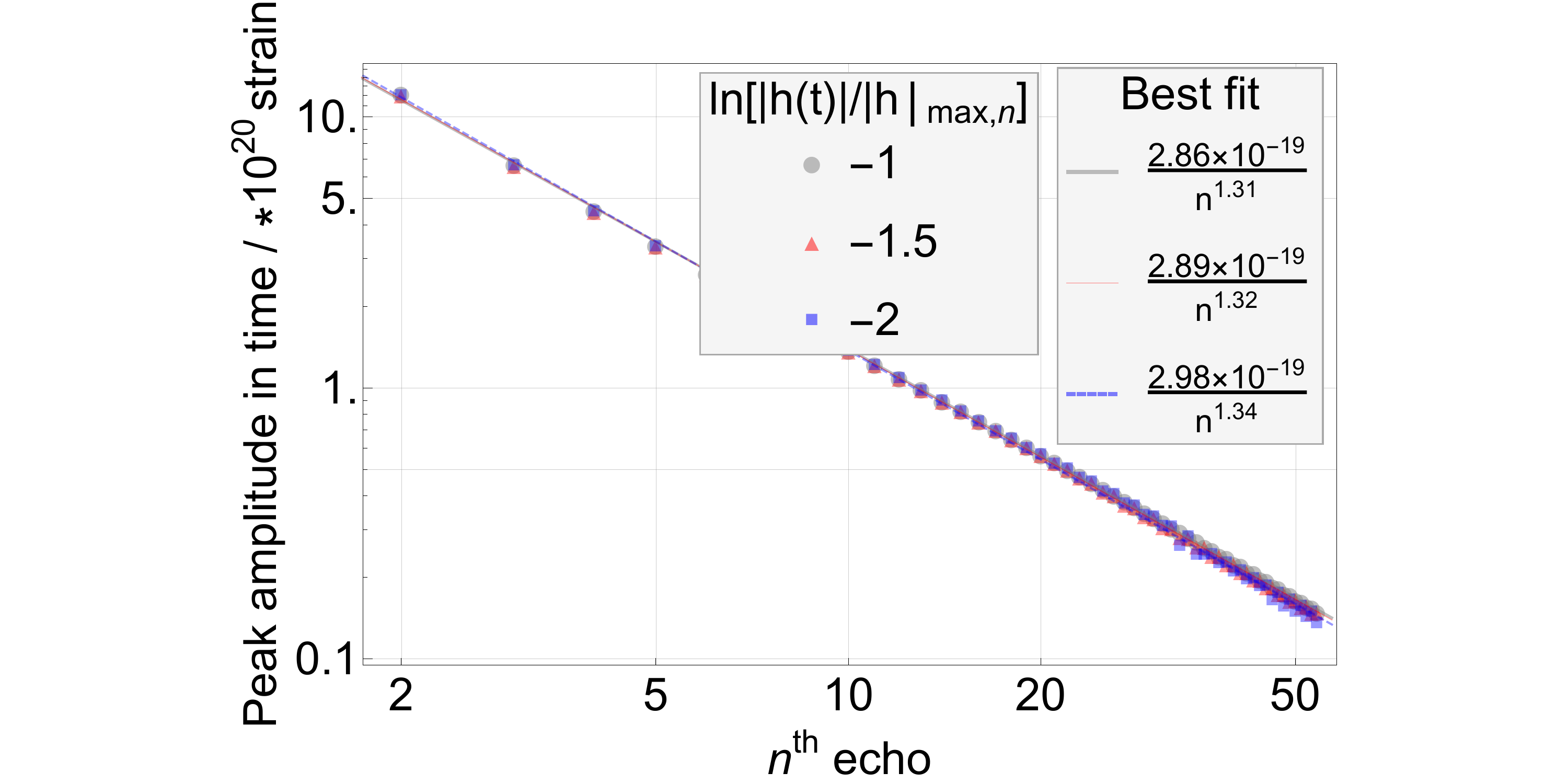}
    \includegraphics[width=1.154\linewidth]{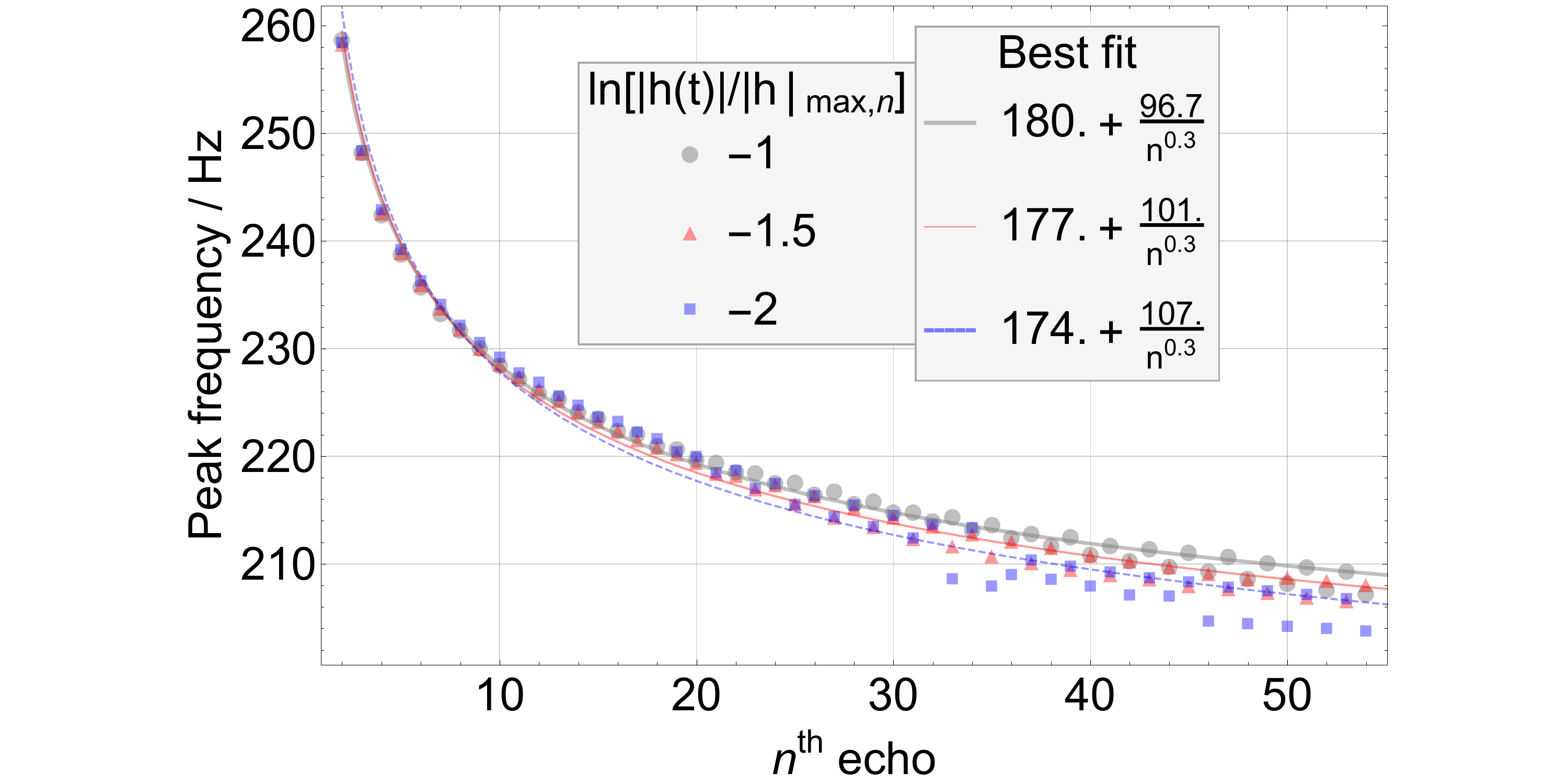}
\endminipage\hfill
\minipage{0.32\textwidth}
   \includegraphics[width=1.18\linewidth]{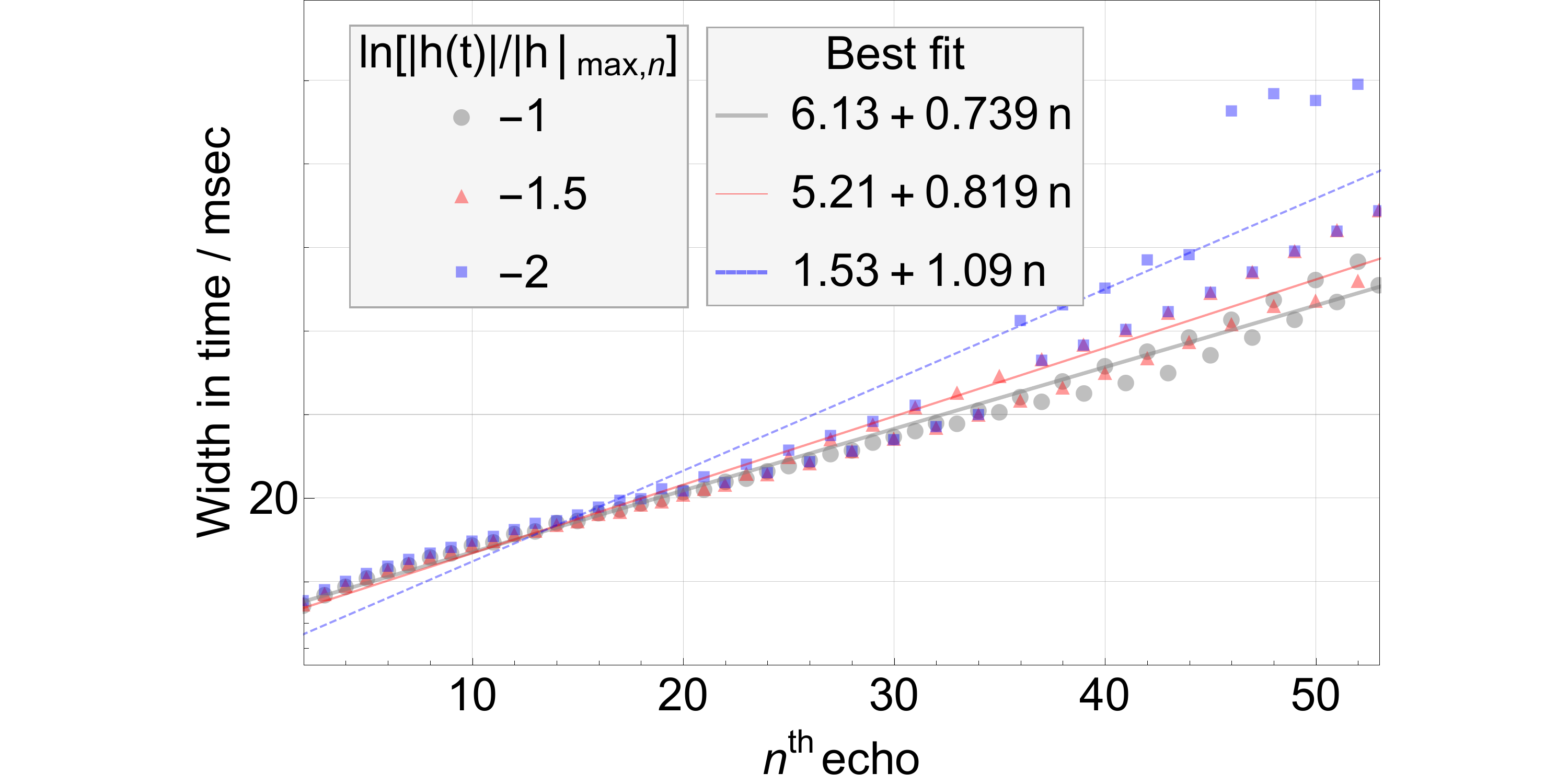}
    \includegraphics[width=1.175\linewidth]{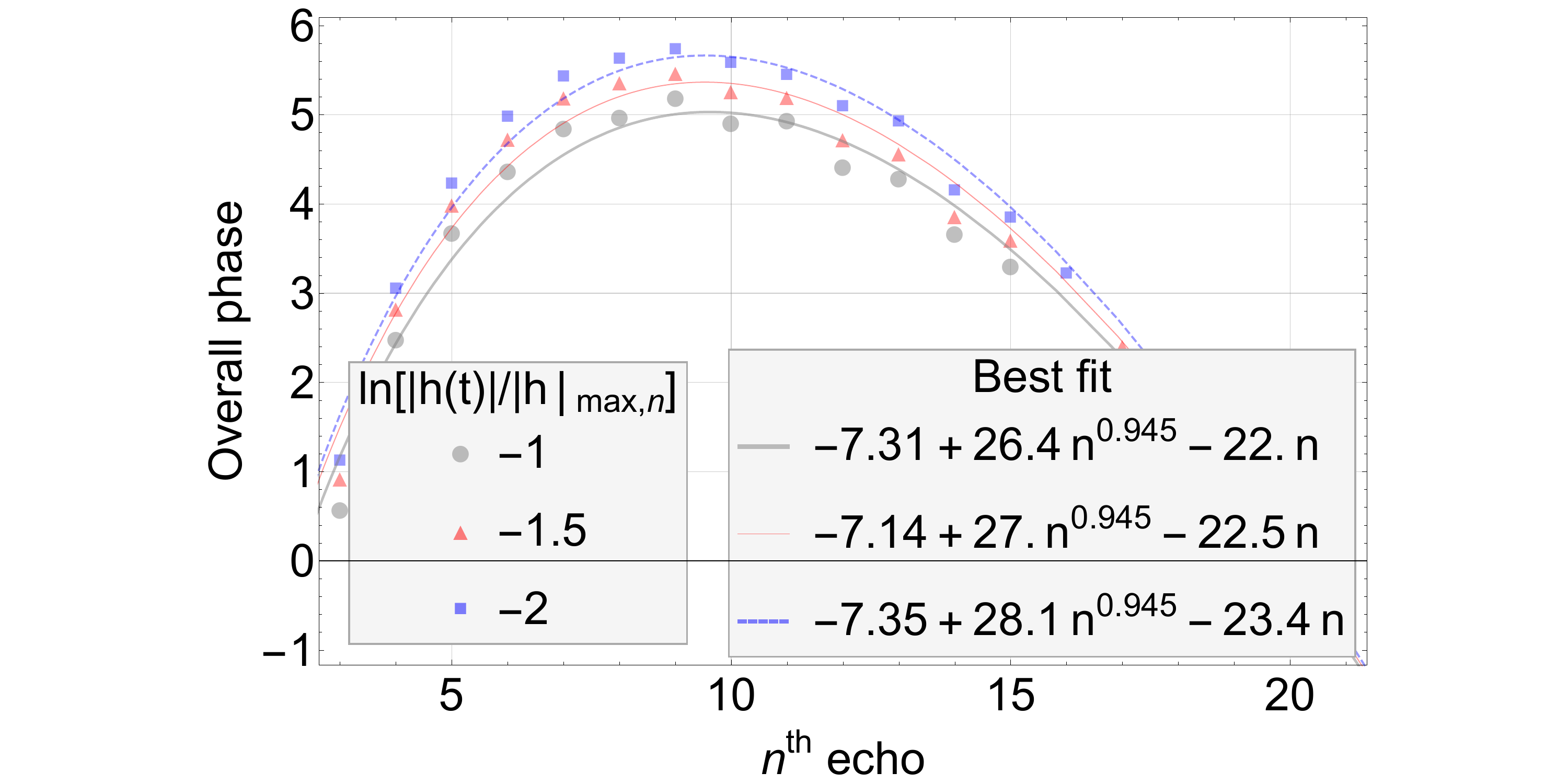}
\endminipage\hfill
\minipage{0.32\textwidth}%
   \includegraphics[width=1.18\linewidth]{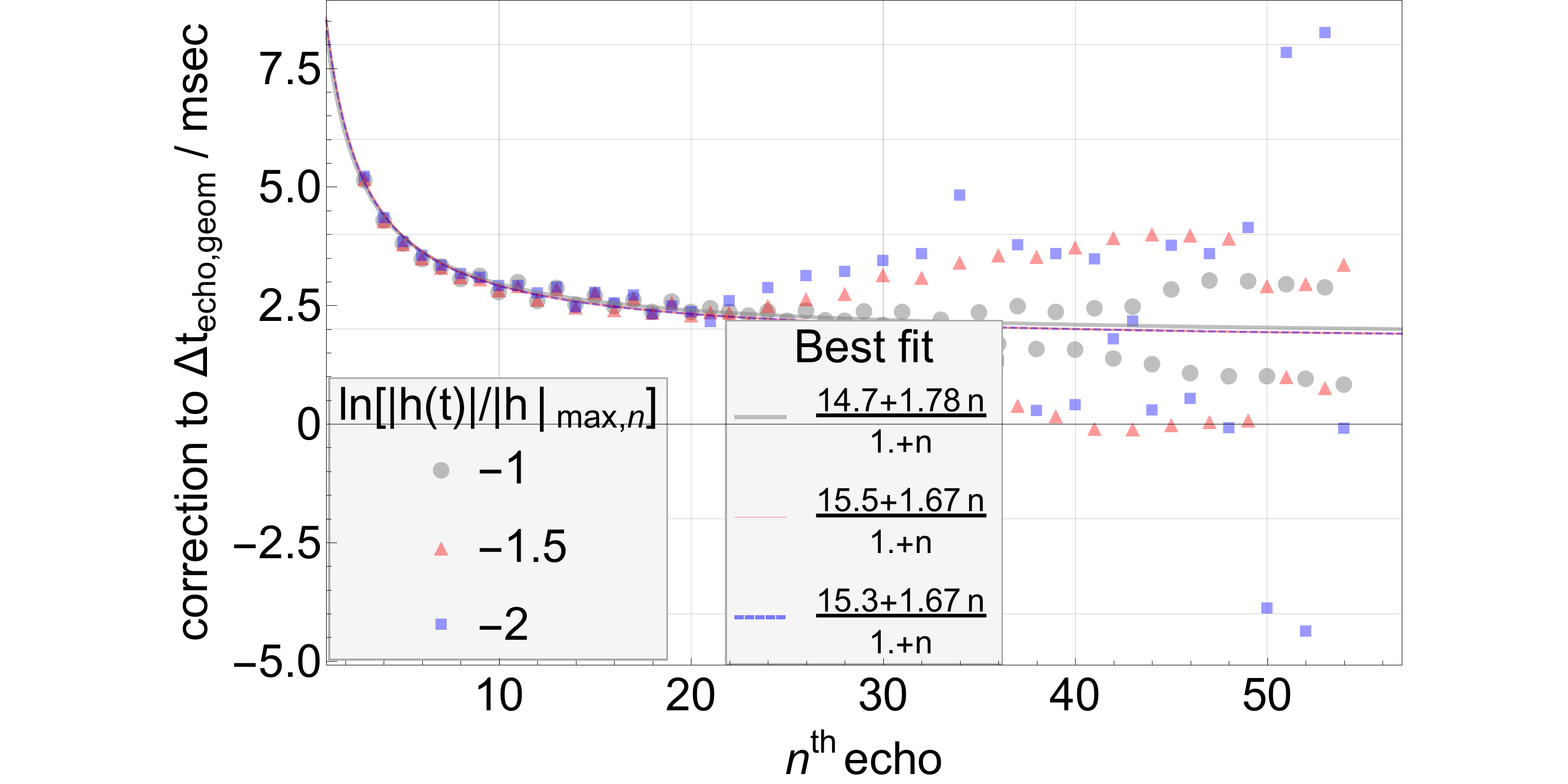}
    \includegraphics[width=1.18\linewidth]{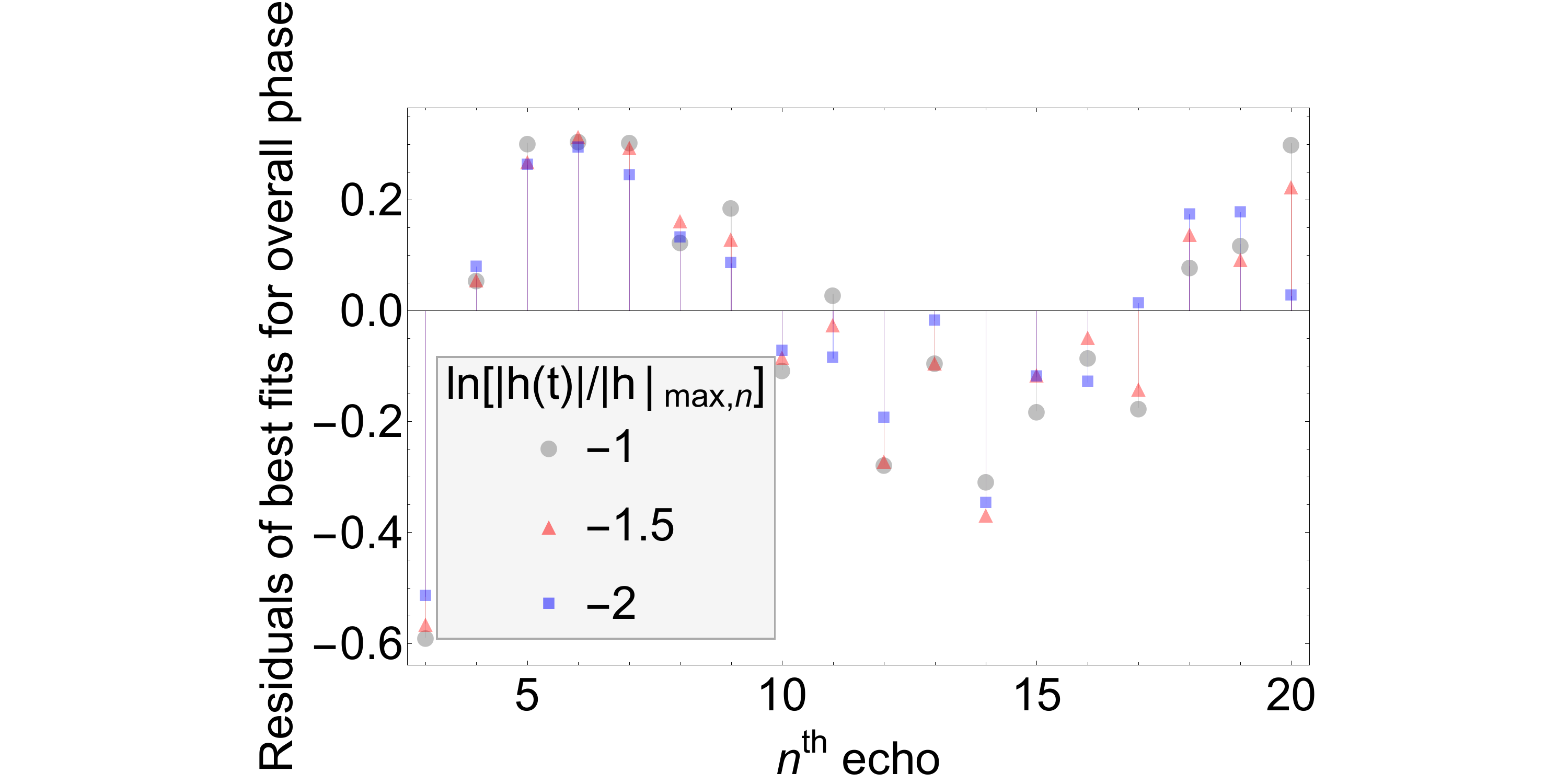}
\endminipage
\caption{\label{52} Best fit gaussian template parameters (for $\ln\left[|h(t)|/|h|_{\rm max, n}\right] > -1, -1.5$ or $-2$), in our minimal model of LIGO event GW150914, showing second and later echoes. The top three panels are in the time domains. Starting from left, peak amplitudes of echos in time are well fit by power laws. Middle panel is the width of the echoes, which become wider in time, as the high frequencies leak out more quickly. For the same reason, the peak frequency (bottom left) also decays with time. The top right panel gives corrections to $\Delta t_{\rm echo,geom}$ (Eq. \ref{t_geom}). Finally,  the bottom middle and right provide the overall phase at $t_{\rm center}$ of each echo and the residuals of the best fit. This is the only plots we show the residuals since the numerical error for the phase is relatively big. }
\end{figure*}

The best fits for each echo domain is also provided in Fig. \ref{52}. For correction to $\Delta t_{\rm echo, geom}$, we define time-delay as $\Delta t_{n}= t_{n} -t_{n-1} $. For all other plots, first echo is not included since it is very sensitive to the properties of the wall, as well as nonlinear effects from early stage of merger (see details in Sec. \ref{sec5}).  The top three panels in Fig. \ref{52} show the time domain properties as a function of the echo number. Starting from the left, peak echo amplitudes in time are all well fit by decaying power laws(\cite{Correia:2018apm} argue that the decay of echoes at early stages is polynomial). Middle are the width of the echoes, becoming wider for later echoes in the time domains, as the high frequency modes leak out more quickly. The top right panel gives correction to $\Delta t_{\rm echo,geom}$ (\ref{t_geom}), while the bottom left panel shows the decay of the mean echo frequency. The bottom middle and right provide overall phase at $t_{\rm center}$ and the residuals of the best fit for the phase. We only show the residuals for the phase, as the numerical error for the phase is relatively big.

\begin{figure}
\includegraphics[width=0.5\textwidth]{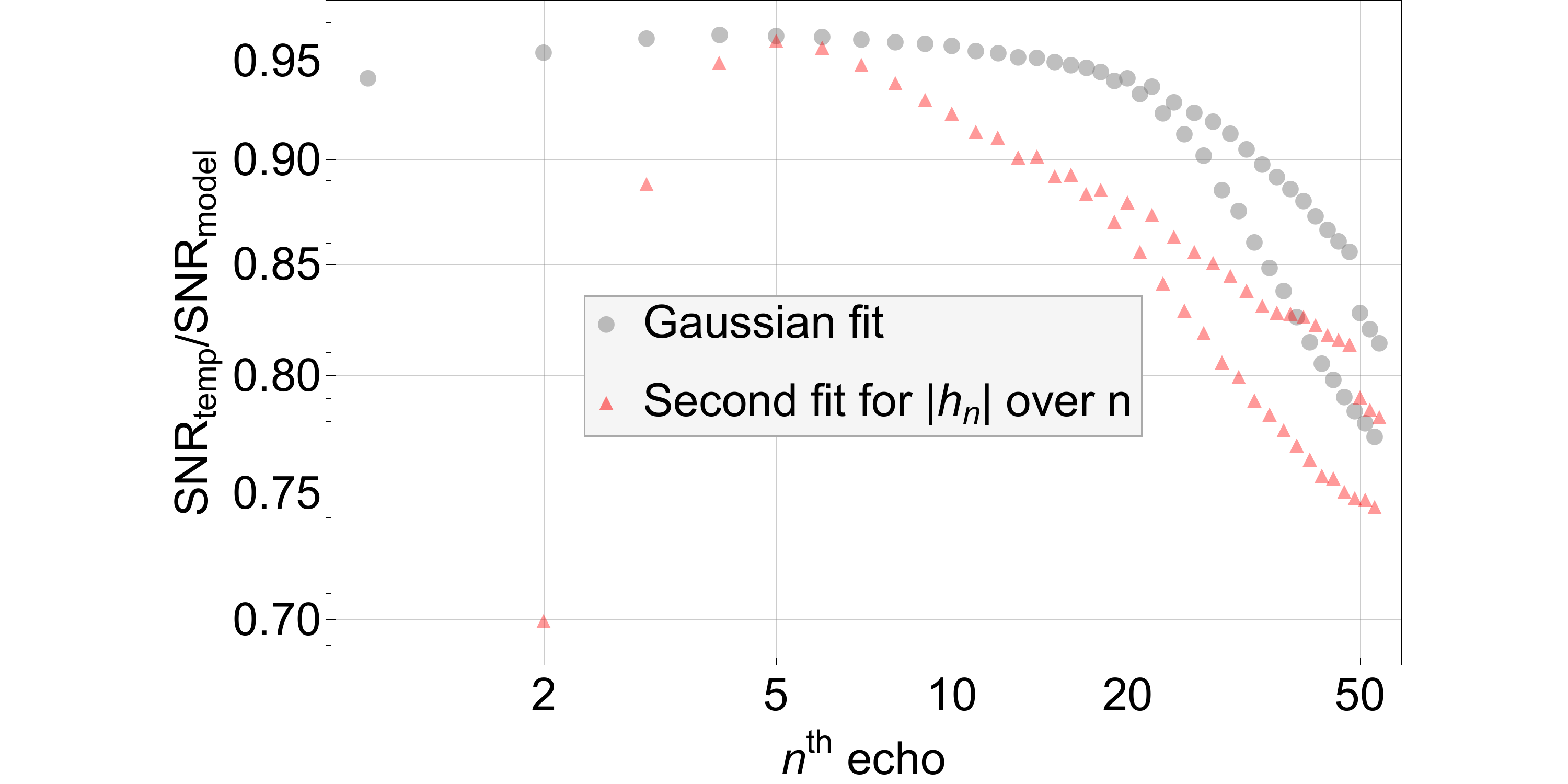}
\caption{\label{53} ${\rm SNR}_{\rm temp}$ compared to ${\rm SNR}_{\rm model}$, showing the quality of gaussian templates.}. 
\end{figure}

To visualize the quality of the template to fit data, Fig. \ref{53} shows the ${\rm SNR}_{\rm temp}/{\rm SNR}_{\rm model}$, where ${\rm SNR}_{\rm model}$ is the predicted signal-to-noise ratio for our numerical solution of echoes (assuming white noise), while ${\rm SNR}_{\rm temp}$ is a reduced value, if we use our Gaussian approximations of  Fig. \ref{520} (gray circles in Fig. \ref{53}). Using a second fit for how properties (i.e. width, center and amplitude) of $\Psi_{\rm n}(t)= \log |h_n(t)| $ depend on $n$ (Table \ref{t3}) further reduces ${\rm SNR}_{\rm temp}$ (red triangles in Fig. \ref{53}). We notice that the quality of Gaussian fit drops for later echoes, which could be either due to build-up of numerical error or systematic deviations from a single gaussian fit.  The secondary fit for $\Psi_{\rm n}$ vs $n$ further reduces SNR as the width in time and time delay, shown as Fig. \ref{52}, do not have a simple behavior. However, the power law fit to the peak amplitude in time $\propto n^{-4/3}$ is surprisingly good. Also, as we discussed before, since the shapes of first few echoes are much more dependent on the initial conditions, it might be better to use independent Gaussians to fit them in data. Finding a reasonable fit for phase information $\Phi_{\rm n}$ vs $n$ proves even more challenging, as a small change in phase leads to a significant change in echo profiles. Fortunately, model-agnostic searches (e.g., \cite{Conklin:2017lwb}) based on cross-correlating different detectors can be done independent of the phase information. 

\section{\label{sec4}Beyond the minimal model}
%As shown in Fig \ref{32}, amplitude of first echoes seems comparable to the ringdown waveform but much bigger than the later echoes. However, this big echo seems missed in the LIGO public data. Also, the first echo doesn't seem to appear as power law with later ones when doing template fit. Several reasons might lead to big first echo:  first, for boundary condition, we only assume wavepacket from infinity, real case should have wave outcoming from near horizon. Actually the reality is more complicated starting with two BHs with nonlinear process, not the linear perturbation of a single finial BH. Second, soft wall might be more realistic than a perfect wall. Some quantum gravity models show similar behavior: for instance frequency bigger than Hawking temperature can excite microstates in Fuzzball.

While our minimal model for echoes has only one free parameter (wall distance to the horizon, $d_{\rm wall}$) in addition to those of GR, the reality can be more complicated. Here, we explore the two main deviations expected from the minimal model due to nonlinear effects in GR and quantum gravity. 

\subsection{Nonlinear Mergers Effects}

Our assumption of a custom-designed incoming wavepacket, as a placeholder for black hole binary merger, is almost certainly too naive to provide a realistic echo template, as it misses the nonlinear nature of the merger. While numerical simulations can now provide realistic waveforms for black hole mergers in GR, a covariant formulation of ECOs that could produce realistic echo waveforms is currently missing. However, we can get an idea about the extent of nonlinear corrections to linear results by noticing that the Kerr background for Teukolsky equation (\ref{eq:teuk}) is dynamical during the merger event, and thus the frequencies can be shifted by ${\cal O}(30\%)$, between the ingoing and outgoing waves at merger \footnote{Fort example, the best-fit for the dominant quasinormal mode frequency for GW150914 is 10-20\% offset from the linear theory predictions for the best-fit Kerr metric (Fig. 5 in \cite{TheLIGOScientific:2016src}). } . We shall explore the extent of this effect on echoes by introducing a blueshift parameter $s$, in the ingoing linear initial conditions:    
%Nonlinear simulation of full merger picture is complicated but we introduce nonlinear correction into our template. We redshift(blueshift) the LIGO event template, hence our initial input to denote the correction from nonlinear process, such as mass increase, wall(or modified horizon) position change etc. We encode the redshift parameter as s and shift the LIGO template in frequency domain:
\begin{eqnarray}\label{s_shift}
\hat{h}_{\rm LIGO, shifted} [f]=\hat{h}_{\rm LIGO} [ f/s ].
\end{eqnarray}

\begin{figure*}
\includegraphics[width=0.3\linewidth]{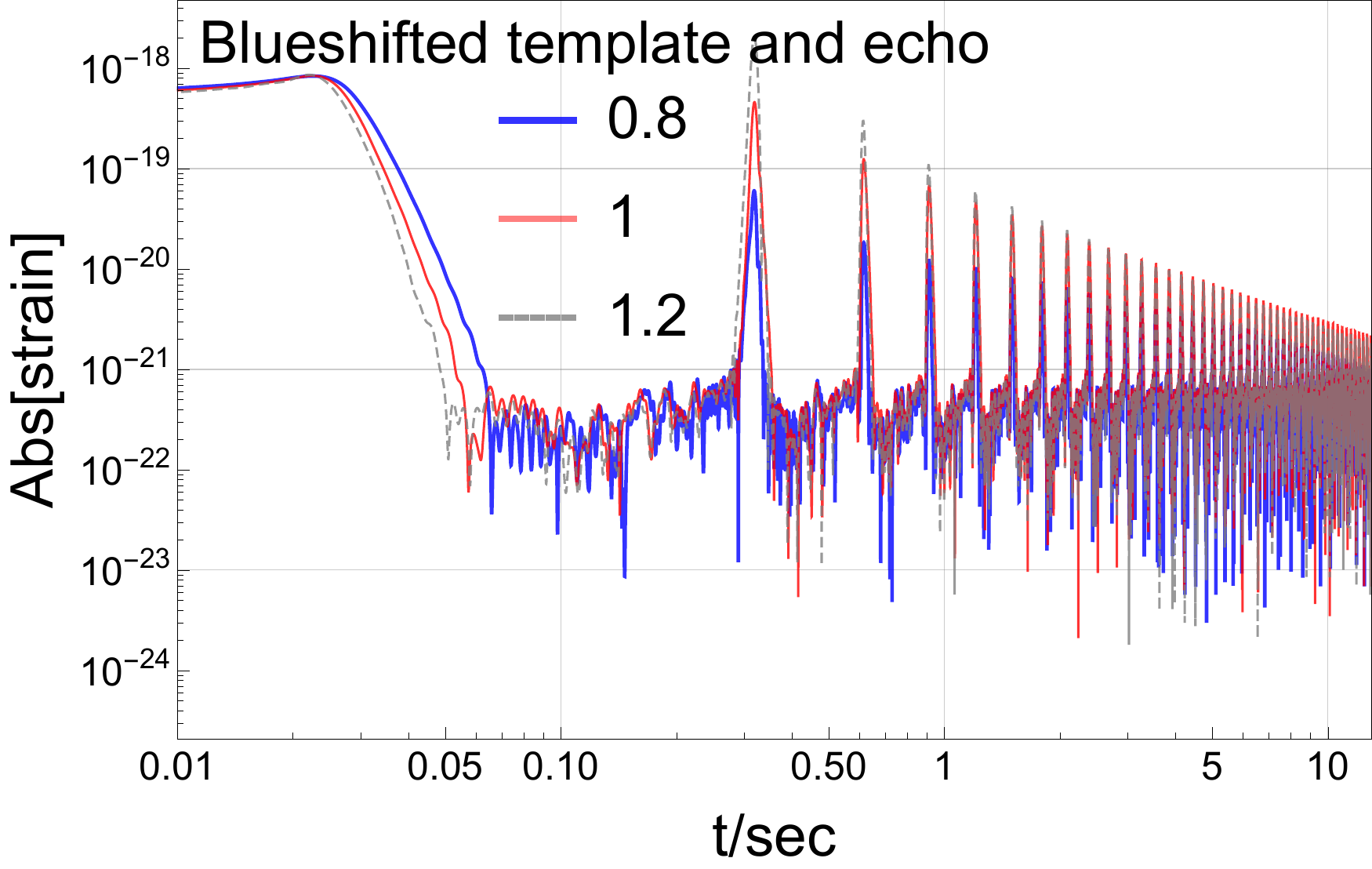}
\includegraphics[width=0.3\linewidth]{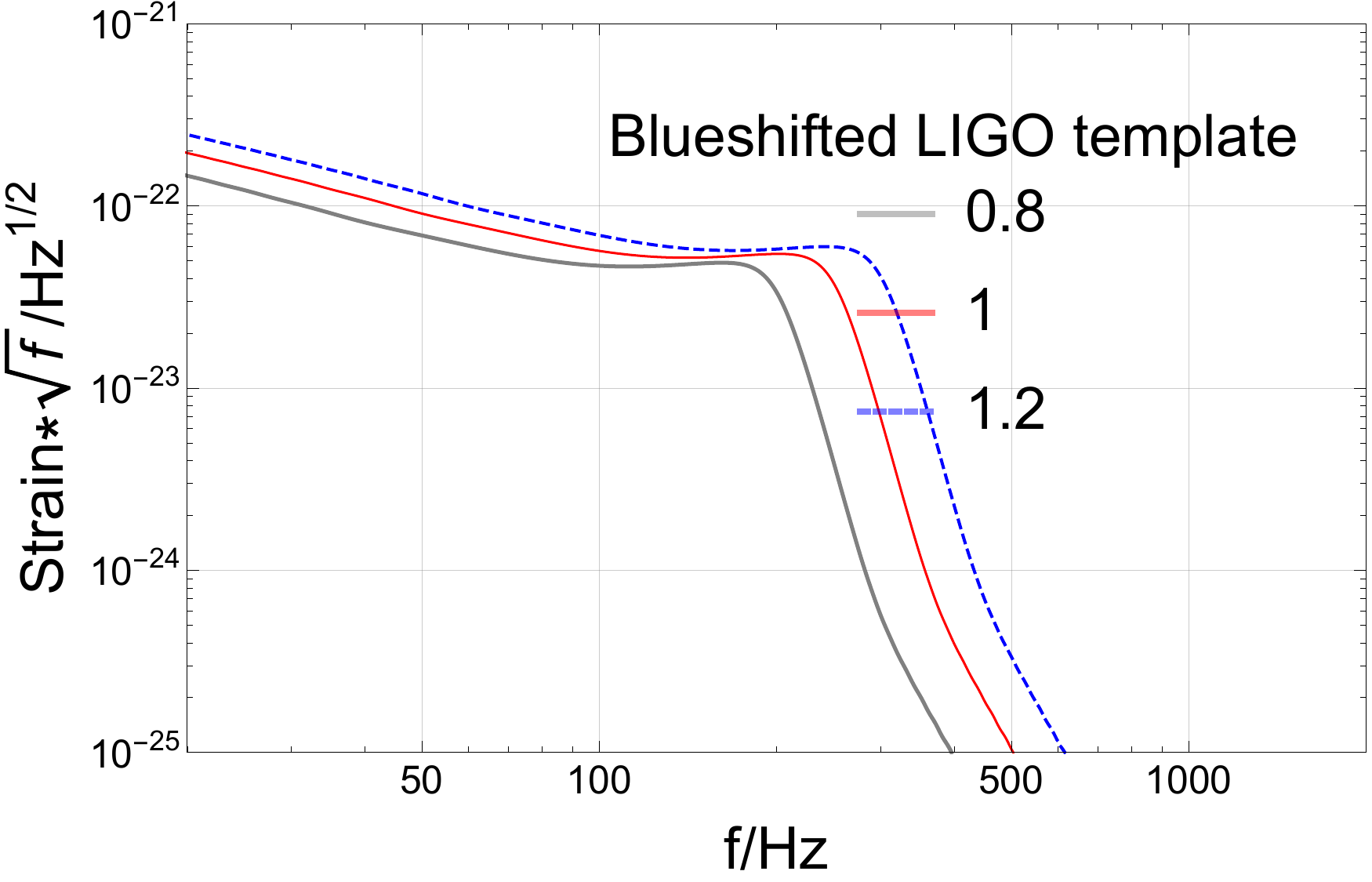}
\includegraphics[width=0.3\linewidth]{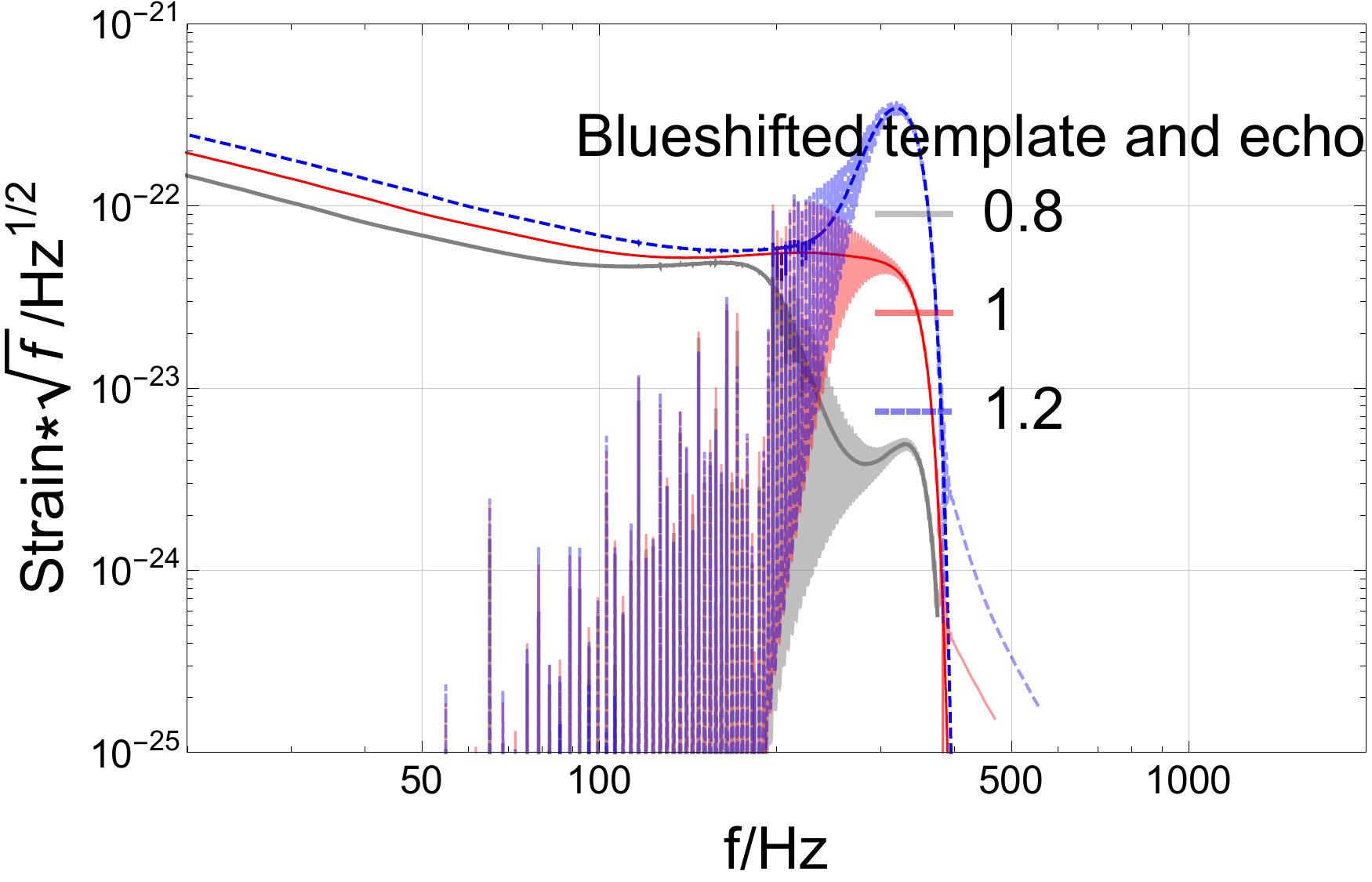}
\caption{\label{7} Echoes predicted for GW150914, expected for redshifted (blueshifted) initial conditions with respect to our minimal model. We see that lower frequency initial conditions lead to lower amplitude, but more persistent, echoes as they cannot penetrate the angular momentum barrier efficiently.}
\end{figure*}

As shown in Fig. \ref{7},  redshifted (blueshifted) initial conditions give echoes which damp more slowly (quickly), since low frequencies leak more slowly through the angular momentum barrier. This also dramatically changes the amplitude of first few echoes. Blueshift parameter $s$ can be a free parameters for data fitting purposes.

\begin{figure*}
    \includegraphics[width=0.45\linewidth]{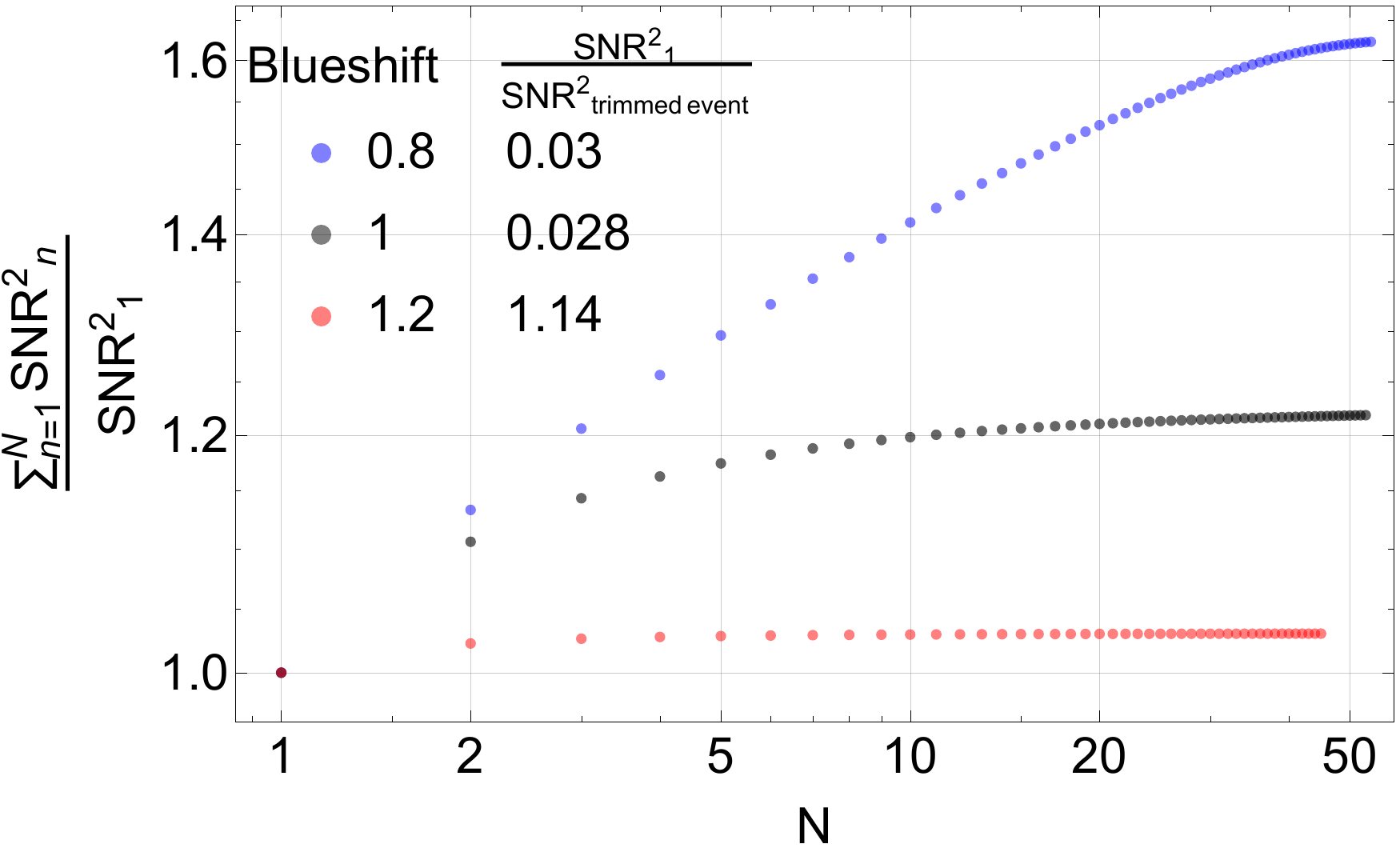}
    \includegraphics[width=0.45\linewidth]{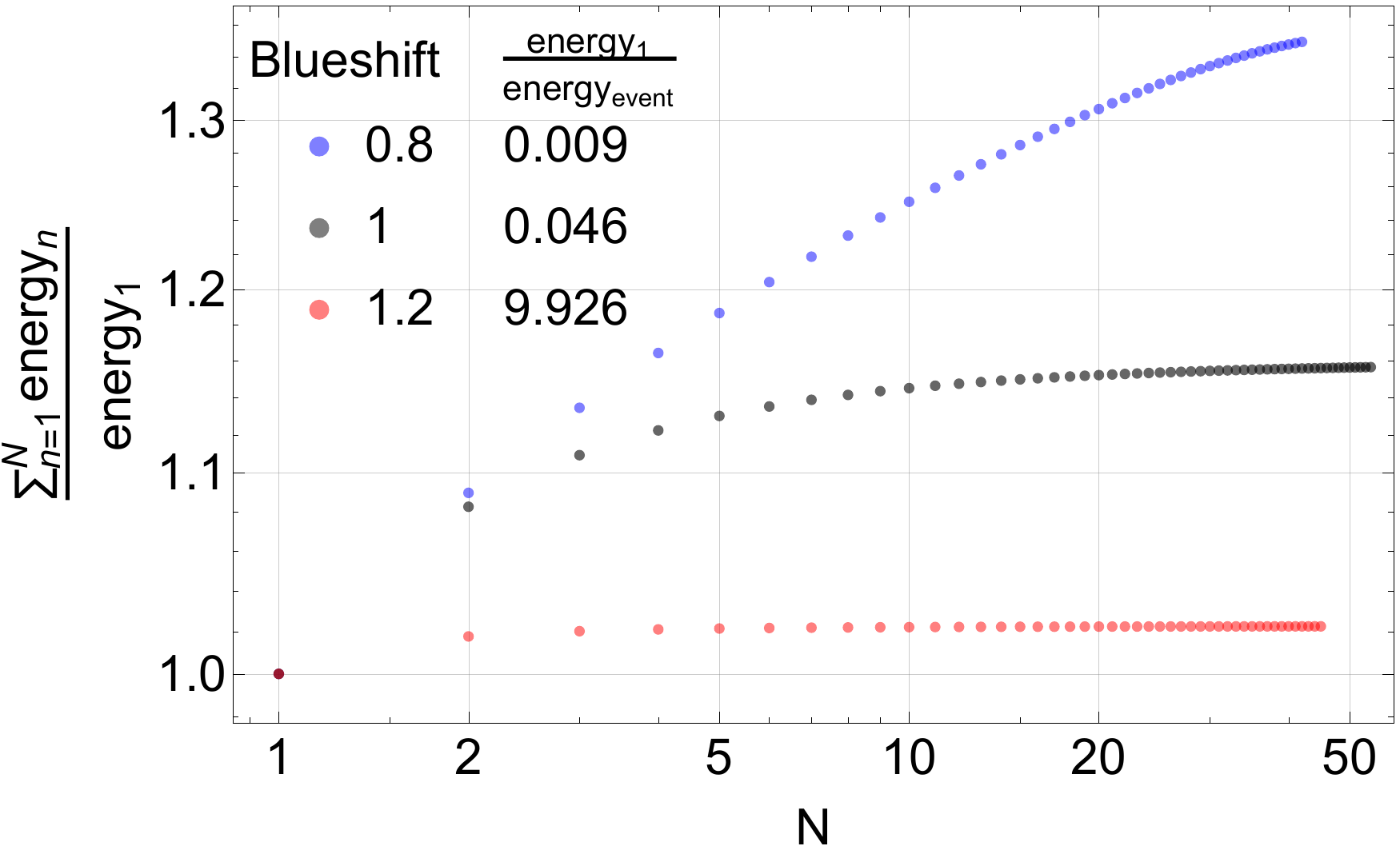}
\caption{\label{8} signal-to-noise ratios(SNR) and energy for blueshifted echoes compared with the first echo. We see that there is more (less) information in subsequent echoes for lower (higher) frequency initial conditions. Furthermore, the amplitude of first echo is hard to predict and can change by more than 1.5 orders of magnitude.We also list SNRs and energy for blueshifted first echoes compared with the event. Since we assume white noise to calculate the SNR in time domain, we trim the merger template at around 0.076 seconds before the peak (similar to the LIGO noise whitening for GW150914 template). }
\end{figure*}

The effect is clearer if we compared SNR of echoes to first echo, as shown in Fig. \ref{8}.  $\rm SNR^2_{\rm n}$ is $\rm SNR^2$ of our numerical solution of $\rm n^{th}$ echo and we trimmed a single echo with $\ln\left[|h(t)|/|h|_{\rm max, n}\right] > -1.5$. We assume white gaussian noise ${\sigma_{\omega}}=1$ so that 

\begin{eqnarray}\label{SNR}
\rm SNR^2_{\rm n}=  \sum_{\omega} \frac{ | \hat{h}_{\rm n, \omega}| ^2}{{\sigma_{\omega}}^2} =  \sum_{t}  | h_{\rm n}| ^2. \\
\end{eqnarray}

Fig \ref{8} (right panel) shows that later echoes contain more (less) information in redshifted (blueshifted) templates, since they decay more slowly (quickly). The left panel also shows the relative SNR of 1st echo compared to the trimmed main event in our model. The fact that this number can change by more than 1.5 orders of magnitude suggests that the amplitude of 1st echo is very sensitive to the nonlinear merger physics and cannot be reliably predicted. \cite{Gupta:2018znn} simulates a binary black hole merger and finds the ratio of the energy falling into the black hole to the energy out is around 1:1, which can be used as a normalization of amplitude of echoes.

\begin{figure*}
\minipage{0.33\textwidth}
  \includegraphics[width=1.151\linewidth]{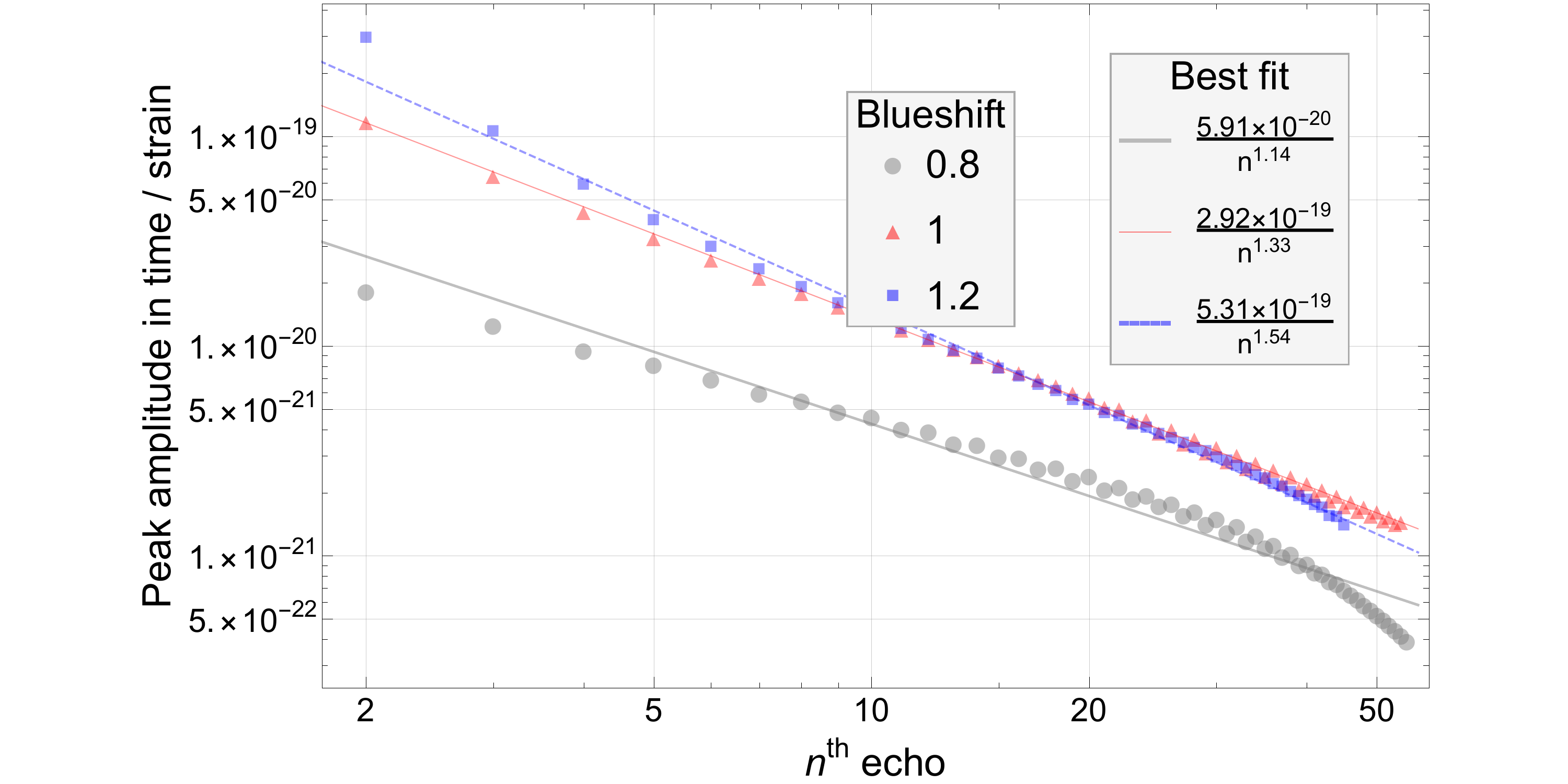}
    \includegraphics[width=1.151\linewidth]{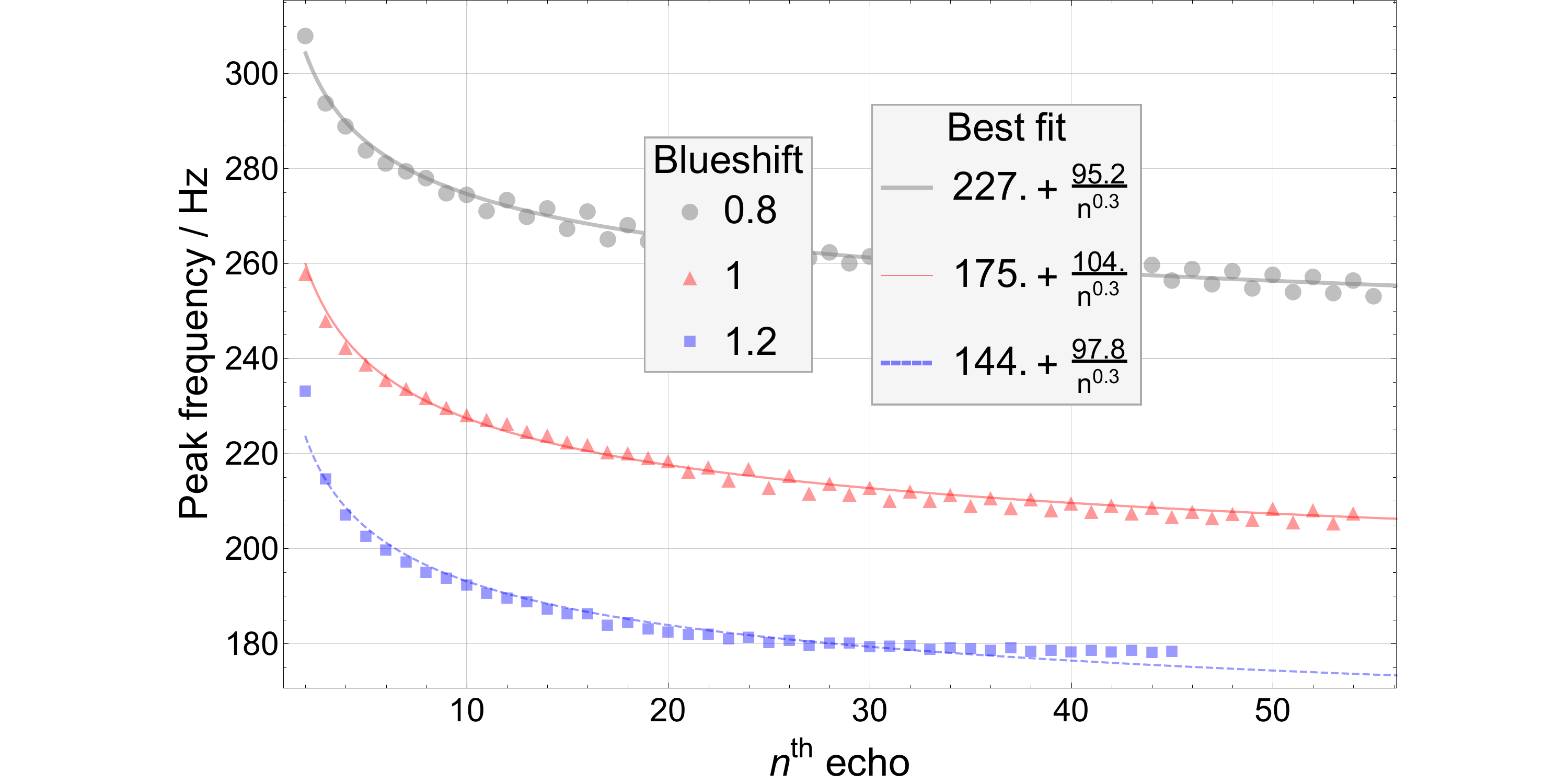}
\endminipage\hfill
\minipage{0.32\textwidth}
   \includegraphics[width=1.16\linewidth]{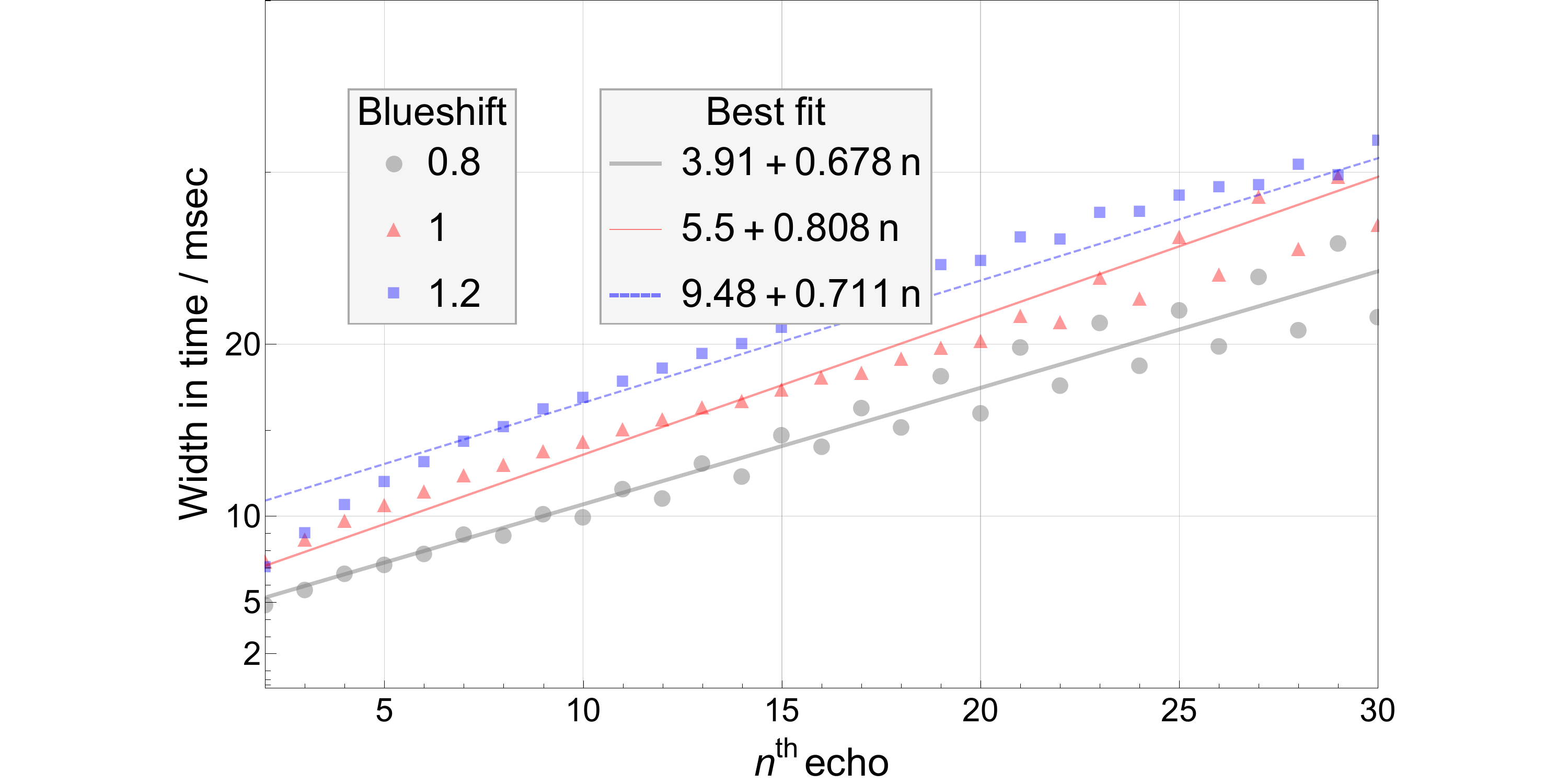}
    \includegraphics[width=1.17\linewidth]{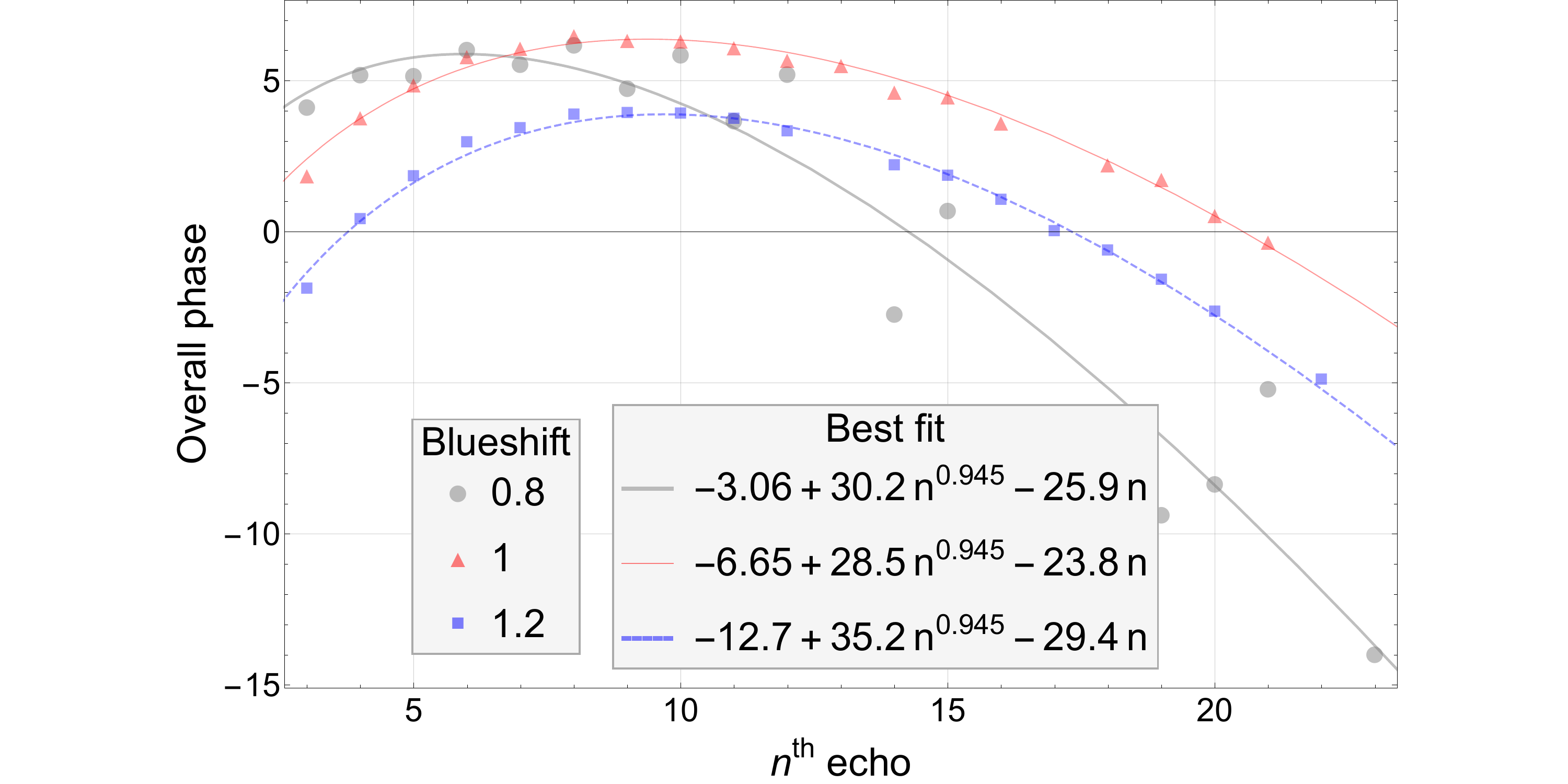}
\endminipage\hfill
\minipage{0.32\textwidth}%
   \includegraphics[width=1.18\linewidth]{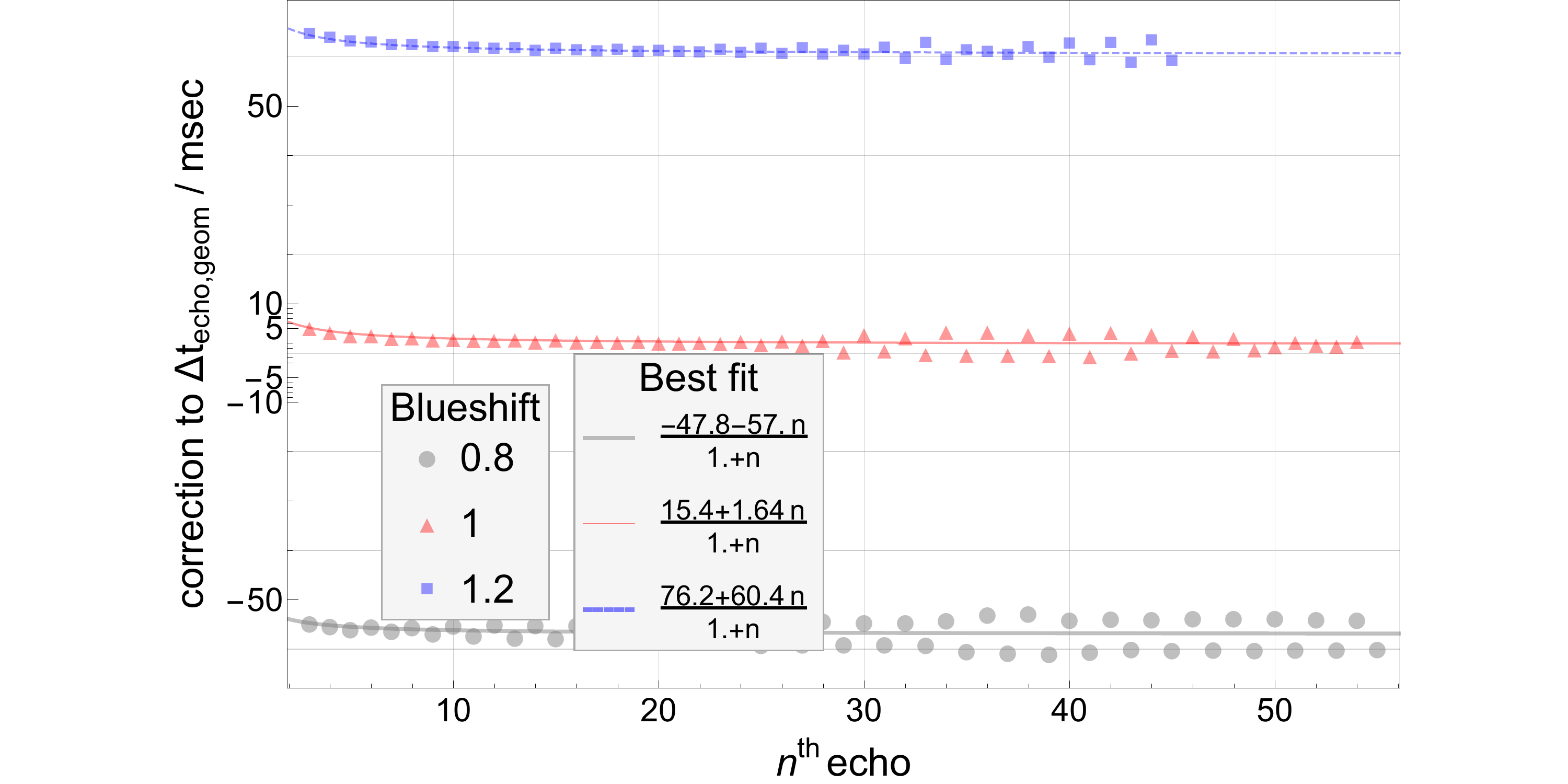}
    \includegraphics[width=1.18\linewidth]{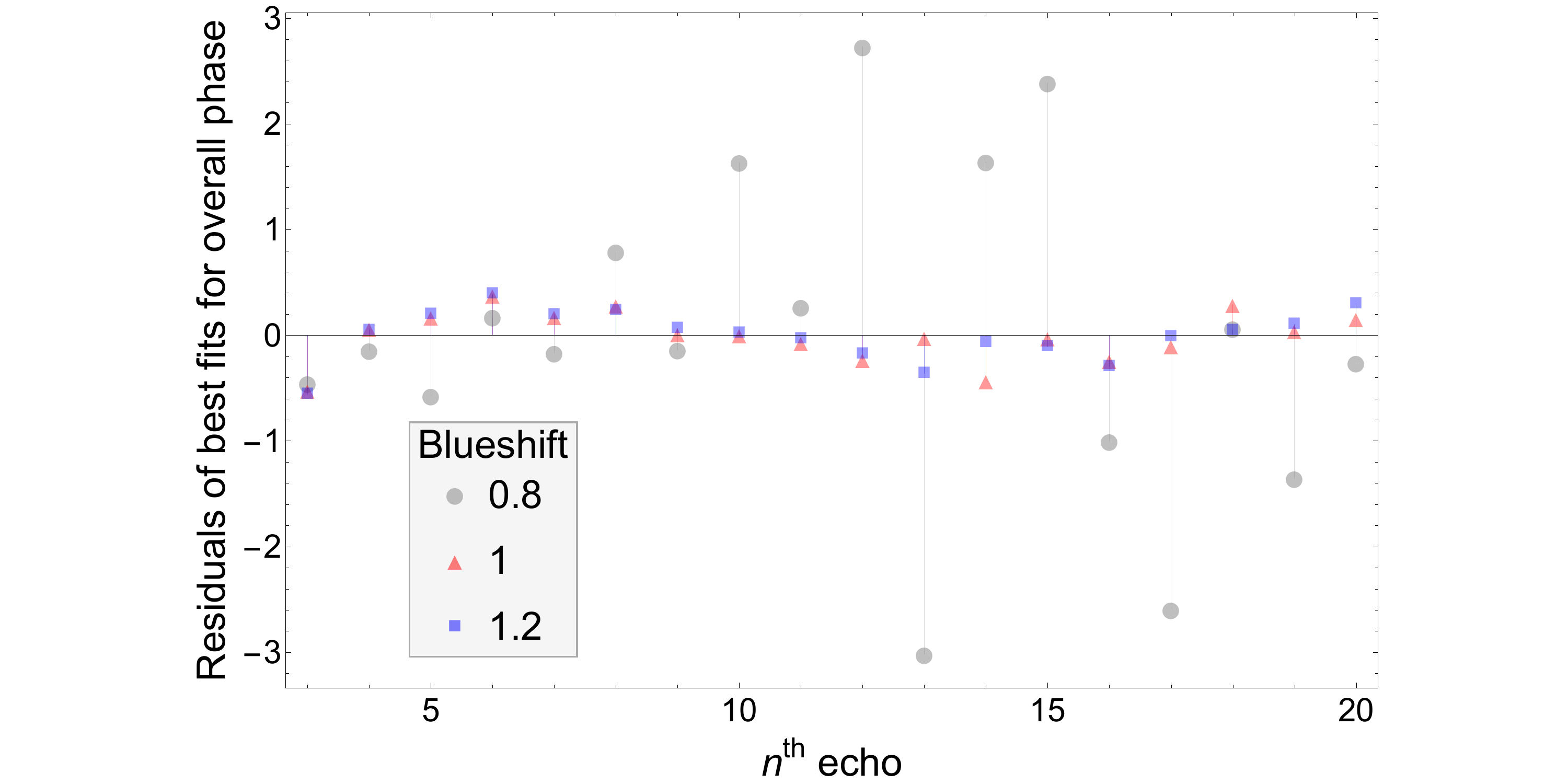}
\endminipage
\caption{\label{72} Same is Fig. (\ref{52}), but using the different blueshift factors $s$ (Eq. \ref{s_shift}) for echo initial conditions (fitted for $\ln\left[|h(t)|/|h|_{\rm max, n}\right] > -1.5$). We see that redshifted initial conditions yield weaker, but more persistent echoes (see text for details).  %Redshifted echo template of LIGO event GW150914 for second and later echoes. The top three is in the time domain while  the below is the frequency. Starting from left, amplitude of echo and of main frequency all fit the power law. With redshifted LIGO event, power law could be extended to first echo. And the more redshifted, the lower damping rate since high frequencies leak fast. Middle two is the width of the echoes, which become wider evolving along time in the time domain and opposite int the frequency domain since the high frequencies leaks at the early time. The right up plots gives correction to $\Delta t_{echo,geom}$ and they are different for different shifted LIGO event since different frequencies bounce back at different position and travel in different speed. And the below one is the main frequency which damps along time. The redshifted (blueshifted) has lower(higher) frequencies.
}
\end{figure*}

Table \ref{t5} and Fig. \ref{72} compare the best fit echo parameters for different blueshift factors. We see in the left panels that the blueshifted initial condition ($s=1.2$) has a transient excess in amplitude that decays quickly and falls in line the minimal model. In contrast, the redshifted model ($s=0.8$) has a significantly smaller but more persistent amplitude. Surprisingly, the middle panels show that the redshifted echoes remain narrower in time.  Even more puzzling is that the redshifted initial conditions have higher frequency echoes as shown in Fig. \ref{72} the bottom left panel. This is due to the fact that the echo peak frequency depends on the slope (and not the amplitude) of the spectral density $\hat{h}_{\rm out}(\omega)=R_{\rm ECO}(\omega) \frac{\hat{h}_{\rm LIGO}(\omega)}{R_{\rm BH}(\omega)} f_{\text{cutoff}}(\omega)$ from Eqn. \ref{eq:hout}, which involve several complicated components. As we see in the middle panel of Fig. (\ref{7}), this slope is not monotonic which leads to the counterintuitive behavior, even though the amplitude of the redshifted model is smaller compared to the blueshifted.

%$\hat{h}_{\rm LIGO}(\omega)$, which should  cancel the slope of $R_{\rm echo} f_\text{cutoff}/R_{\rm BH}$ (Eq. \ref{eq:hout}). As we see in the middle panel of Fig. (\ref{7}), this slope is not monotonic which leads to the counterintuitive behavior.   

%  The left and up plot of time-delay correction to $\Delta t_{echo,geom}$ changes, because different frequencies travel at slight different velocities and bounce back at different positions other than pure geometric picture. The last plot presents that redshifted template has bigger main frequency, which seems counterintuitive. But it only indicates the main frequency but not the amplitude of the frequency. The main frequency is decided by combination of $\hat{h}_{\rm LIGO}$, $R_{\rm BH}$ and $f_\text{cutoff}$ of a specific frequency, so once we redshift $\hat{h}_{\rm LIGO}$, the related $R_{\rm BH}$ and $f_\text{cutoff}$ change so that give different main frequency. But for sure, blueshifted one has large amplitude of the main frequency compared with redshifted one shown as Fig. \ref{7} and left and below plot in Fig. \ref{72}, which is also the reason why it has larger first echo.

\begin{widetext}

\begin{table}%The best place to locate the table environment is directly after its first reference in text
\caption{\label{t5} Same as Table \ref{t3}, but contrasting with redshifted/blueshifted initial conditions, fitted within $\ln\left[|h(t)|/|h|_{\rm max, n}\right] > -1.5$. }
\begin{tabular}{|l|l|l|l|}
\cline{1-4}
\textrm{blueshift factor $s$}&\textrm{ 0.8}&\textrm{$1$}&\textrm{$1.2$}\\
\colrule
\textrm{peak amplitude in time / strain}&$5.91 \times 10^{-20}/n^{1.14}$&
$2.92\times 10^{-19}/n^{1.33}$&
$5.31\times 10^{-19}/n^{1.54}$\\
\colrule
\textrm{width in time / msec} & $3.91+0.678 n$& $5.5+0.808 n$ & $9.48+0.711 n$ \\
\colrule
\textrm{correction to} $\Delta t_{\rm echo,geom} \textrm{/ msec}$  & $-47.8-57.0/(1+n)$& $15.4+1.64/(1+n)$& $76.2+60.4/(1+n)$ \\
\colrule
\textrm{peak frequency / Hz}& $227+95.2/ n^{0.3}$ & $175+104/ n^{0.3}$& $144+97.8/ n^{0.3}$ \\
\colrule
\textrm{Overall phase}  & $-3.06+30.2n^{0.945}-25.9n$ & $-6.65+28.5n^{0.945}-23.8n$ & $-12.7+35.2n^{0.945}-29.4n$ \\
\cline{1-4}
\end{tabular}
\end{table}

\end{widetext}

\subsection{Soft Wall}

Motivated by quantum models of black holes, the wall must at least partially absorb the energy incident on the wall \cite{abedi2016echoes}. For example, in fuzzball models \cite{Mathur:2012jk} high energy particles (with $\hbar \omega  \gg kT_{\rm H}$, where $T_{\rm H}$ is the Hawking temperature) excite the fuzzball microstates and thus will be absorbed by the wall. On the other hand, particles with $\hbar\omega \leq kT_{\rm H} $ may be (at least partially) reflected (but see \cite{Guo:2017jmi} for recent counter-arguments). Ringdown phase of mergers of two BHs is in the intermediate range ($\sim 100$  Hz for GW150914). Therefore, a realistic quantum gravity model for the echoes is expected to involve a {\it soft} wall. For example, frequency of electromagnetic emissions from accretion into BHs is much higher, which is expected to be absorbed by the wall \cite{Broderick:2009ph, Broderick:2015tda}. However, possible loopholes that could lead to astrophysical observables from quantum effects have been exploited in \cite{Pen:2013qva, Afshordi:2015foa}.

A wall that absorbs high frequency modes will dramatically decrease the amplitude of the first echo, since these modes leak out quickly every time the wavepacket hits the angular momentum barrier. Therefore, the first echo contains most of the high frequency modes which, as shown in the top left panel in Fig \ref{43}, would be absorbed for a soft wall.

\begin{figure}
\includegraphics[width=0.5\textwidth]{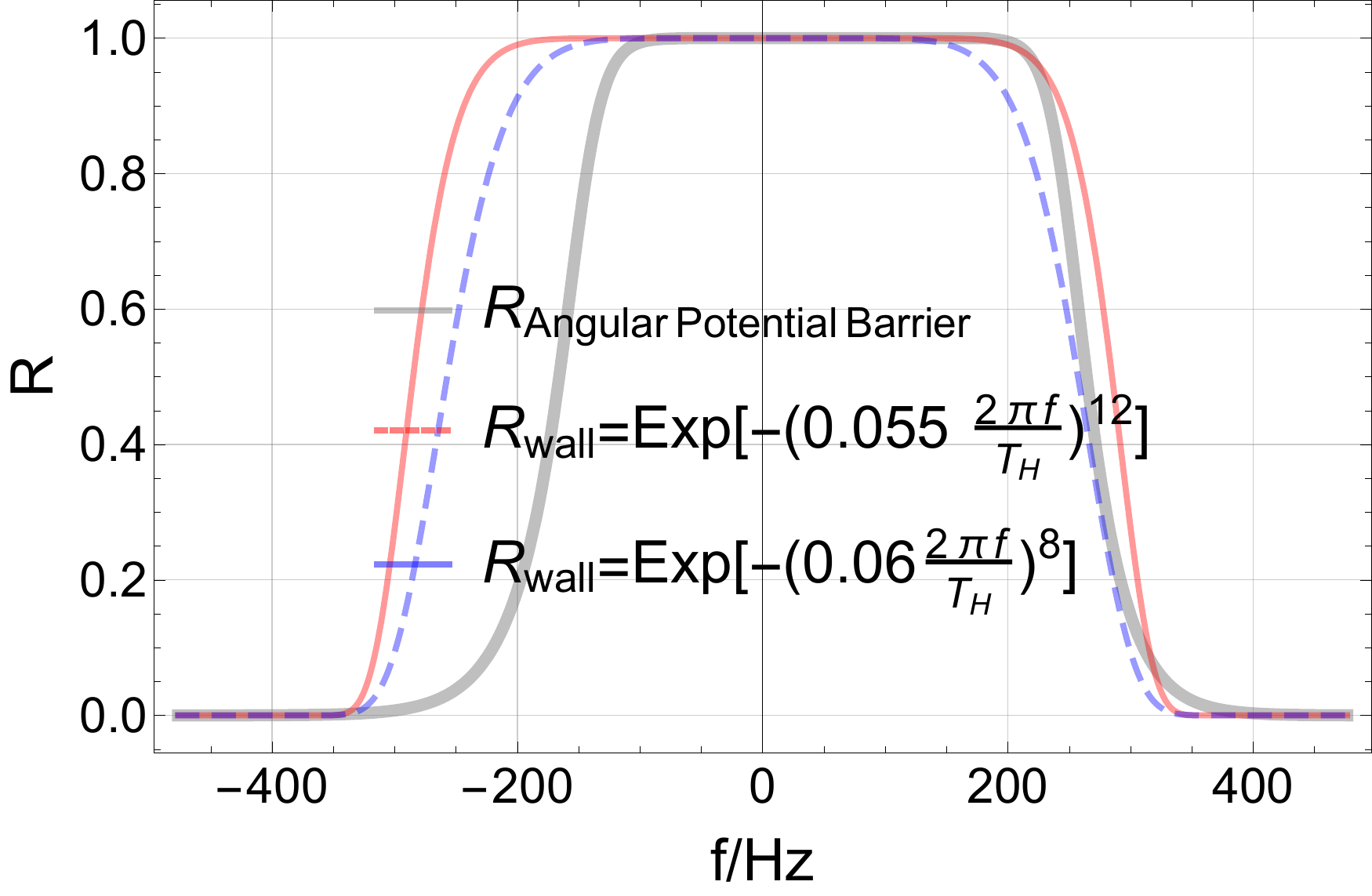}
\caption{\label{44} Comparison of soft wall reflectivity coefficients that we use, with that of the Kerr angular momentum barrier  \cite{Nakano:2017fvh}. The thin and dashed lines are the two reflectivity rates used in Fig \ref{42}.}
\end{figure}

\begin{figure}
\includegraphics[width=0.5\textwidth]{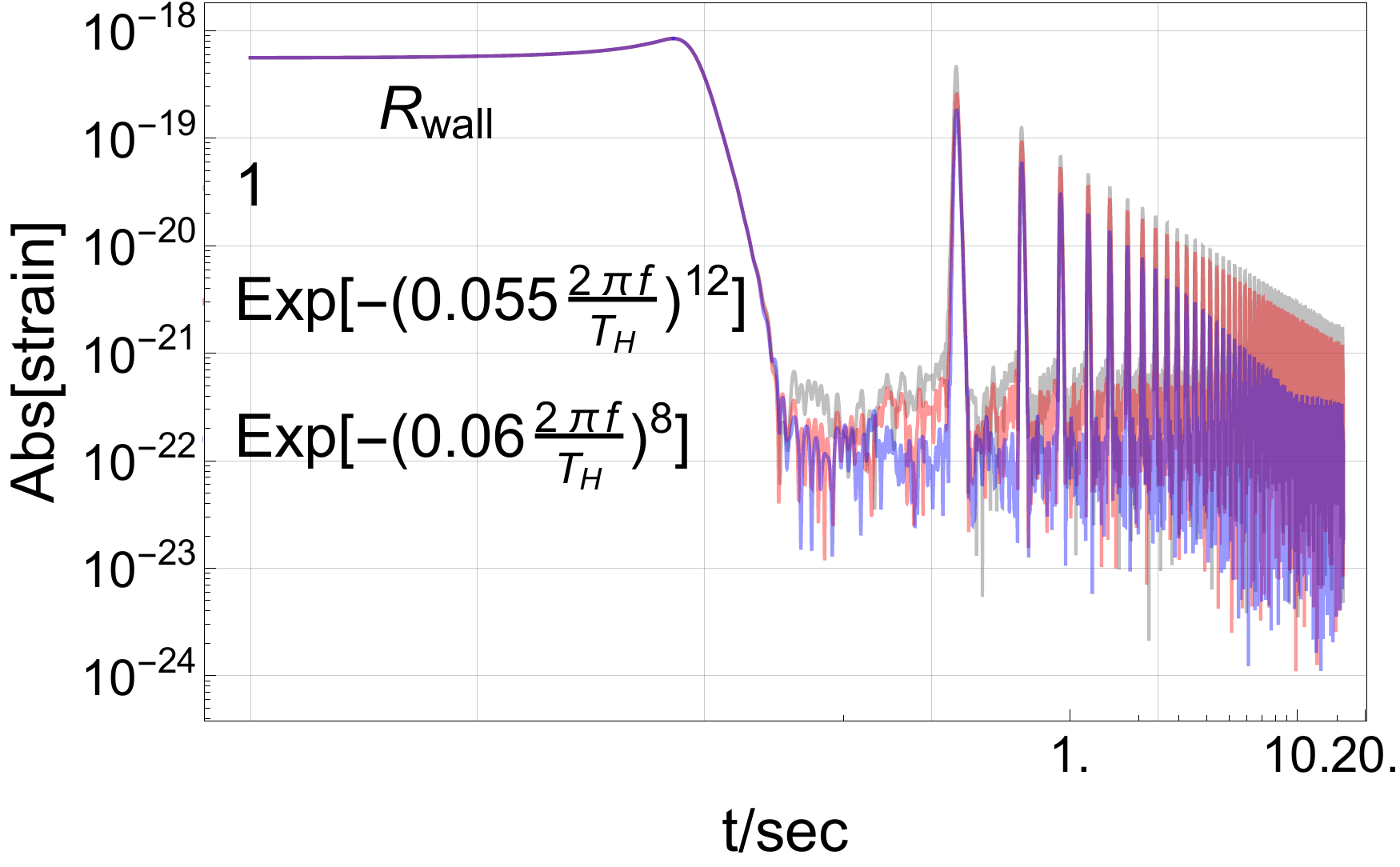}
\caption{\label{42} Echoes for GW15014, for soft vs. perfect walls. The top (gray) curve assumes a perfect wall/mirror, while the lower curves show soft walls with different energy reflectivity coefficients. %Dashed one is a Gaussian reflection rate which absorb high frequency and also lots of low ones so that we have small amplitude for all echoes. Thin one is power of Gaussian reflection rate so it absorb high frequency and reflect most low frequency. Hence, first echo become smaller but later echoes changes a little.
}
\end{figure}

\begin{figure*}
\minipage{0.33\textwidth}
  \includegraphics[width=1.151\linewidth]{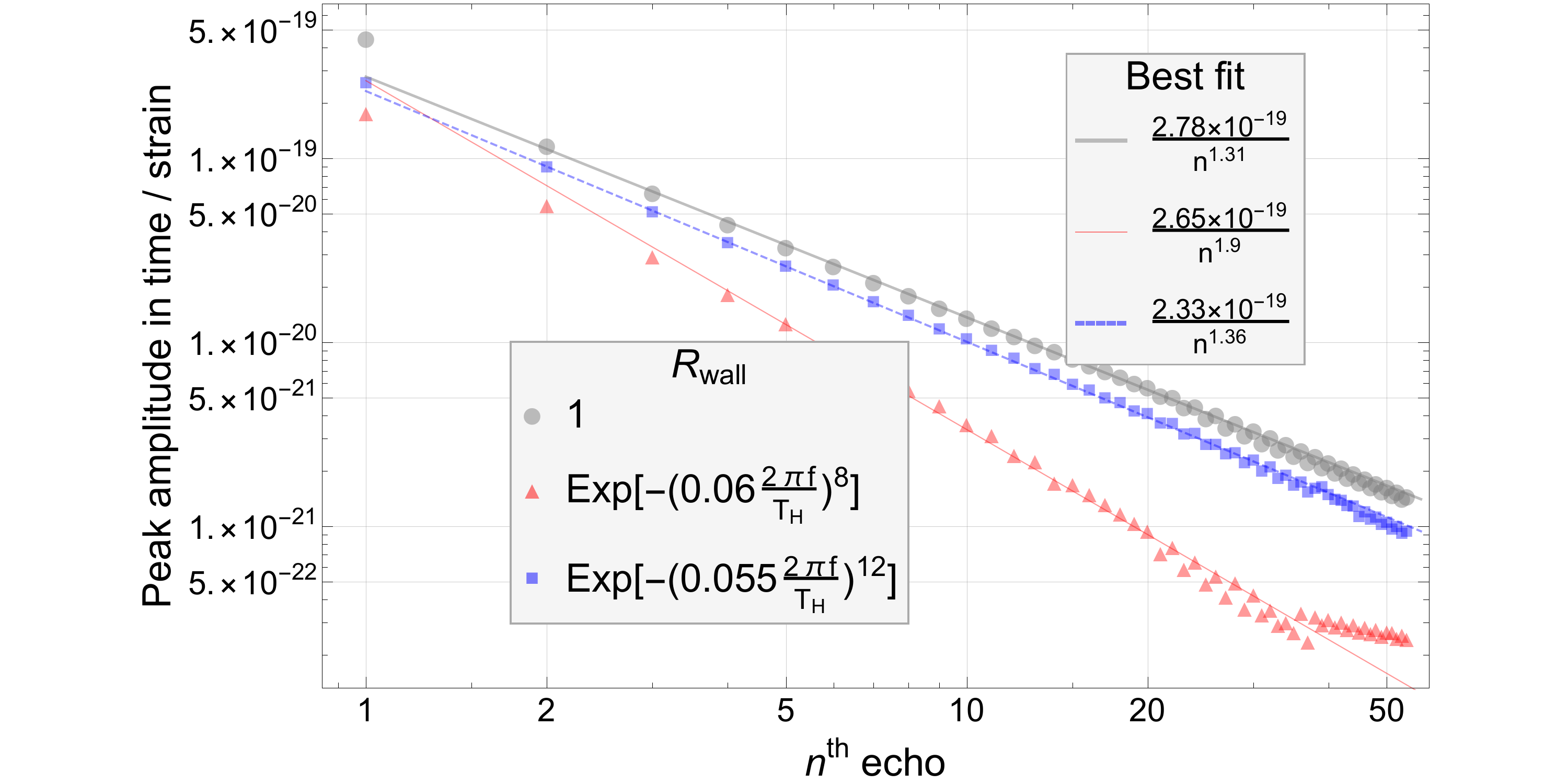}
    \includegraphics[width=1.151\linewidth]{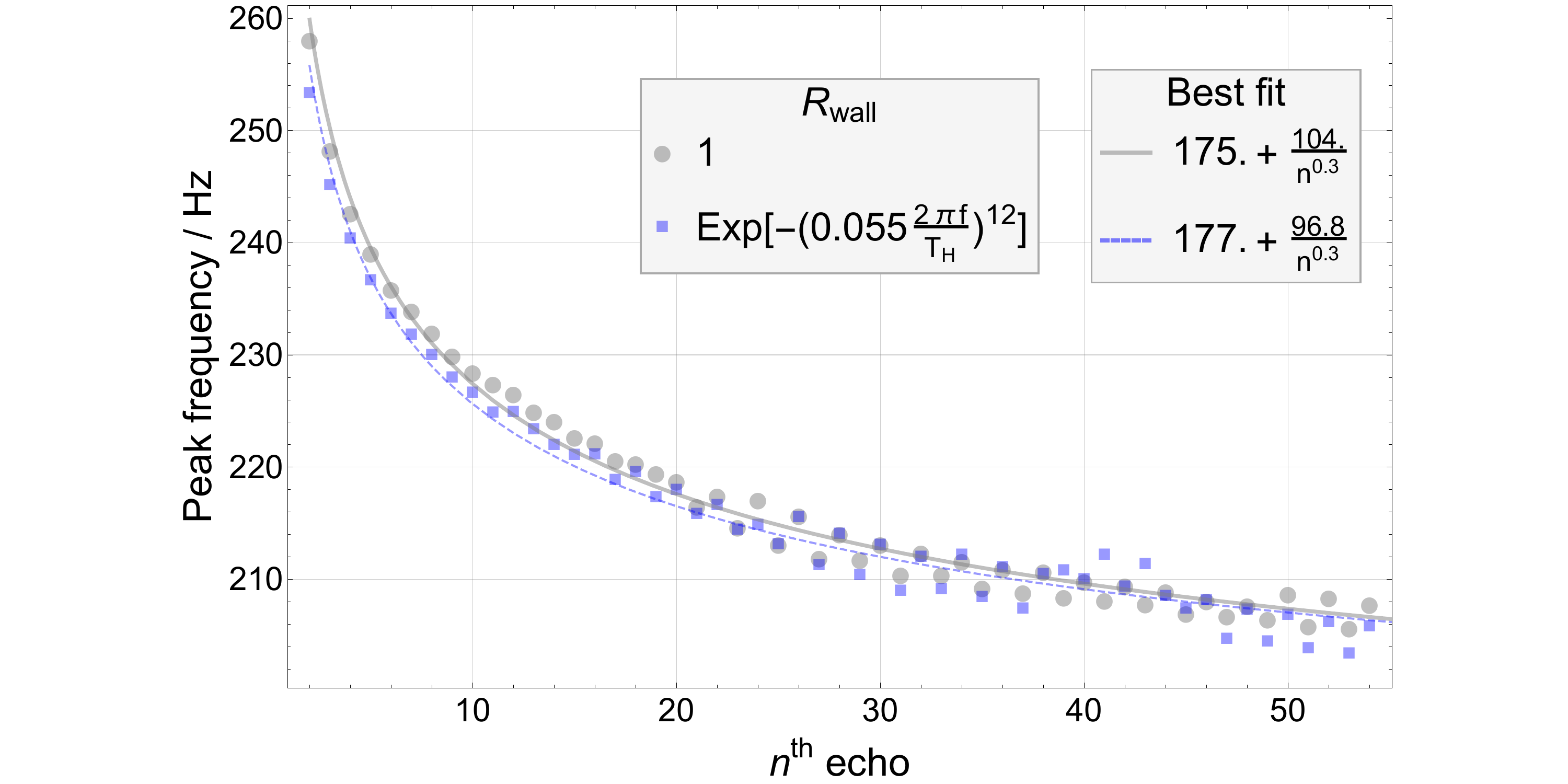}
\endminipage\hfill
\minipage{0.32\textwidth}
   \includegraphics[width=1.16\linewidth]{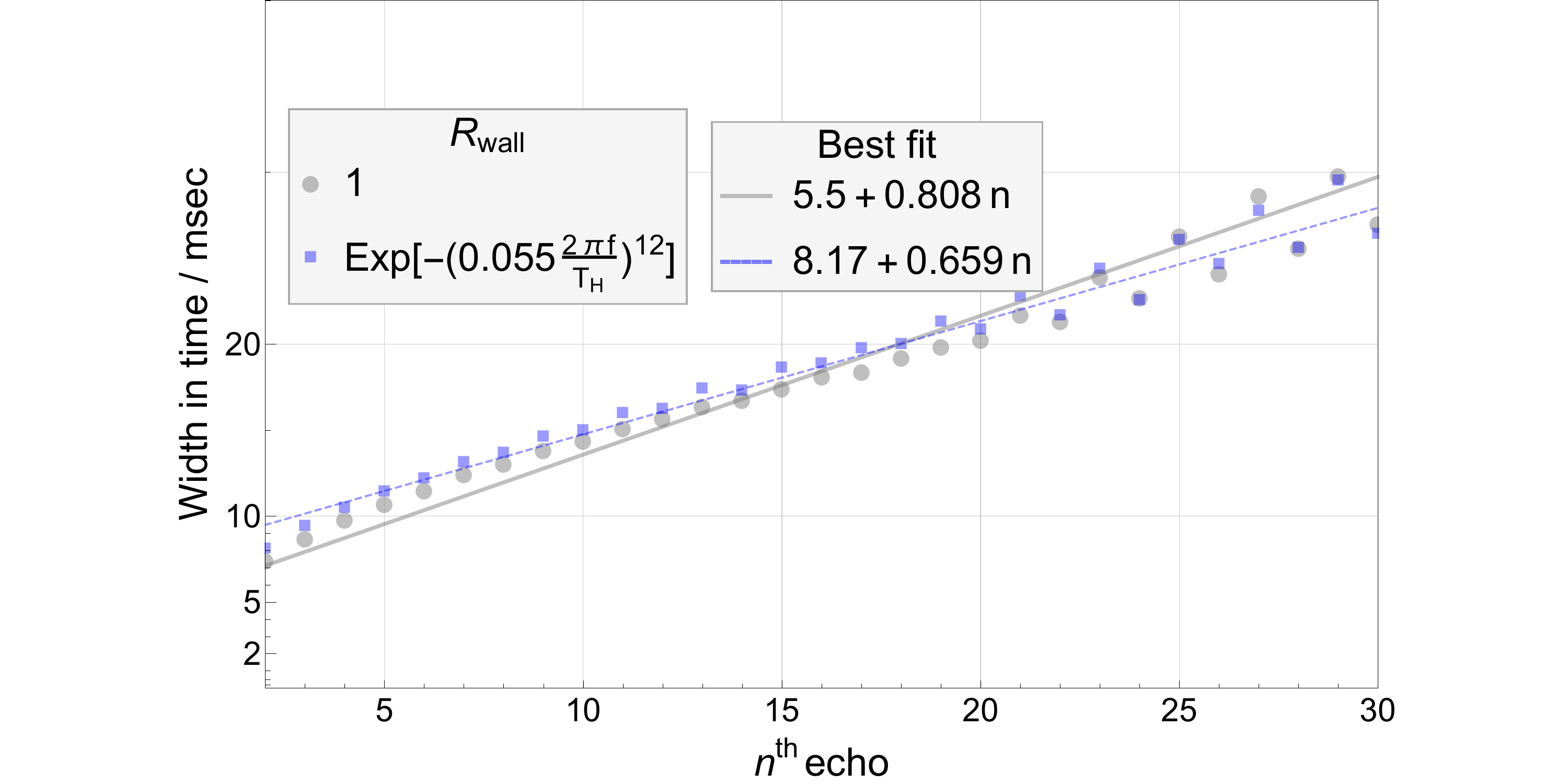}
    \includegraphics[width=1.17\linewidth]{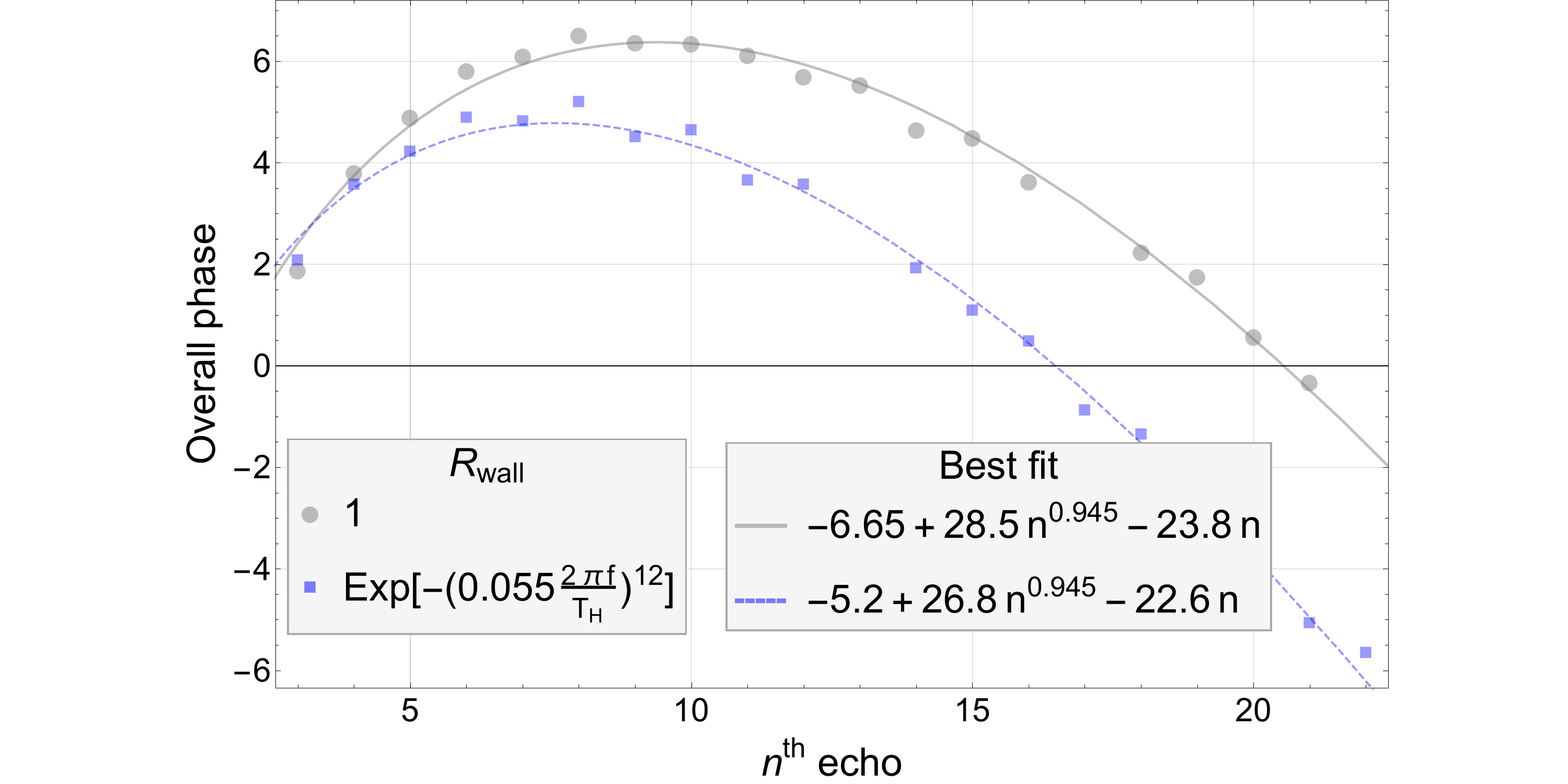}
\endminipage\hfill
\minipage{0.32\textwidth}%
   \includegraphics[width=1.18\linewidth]{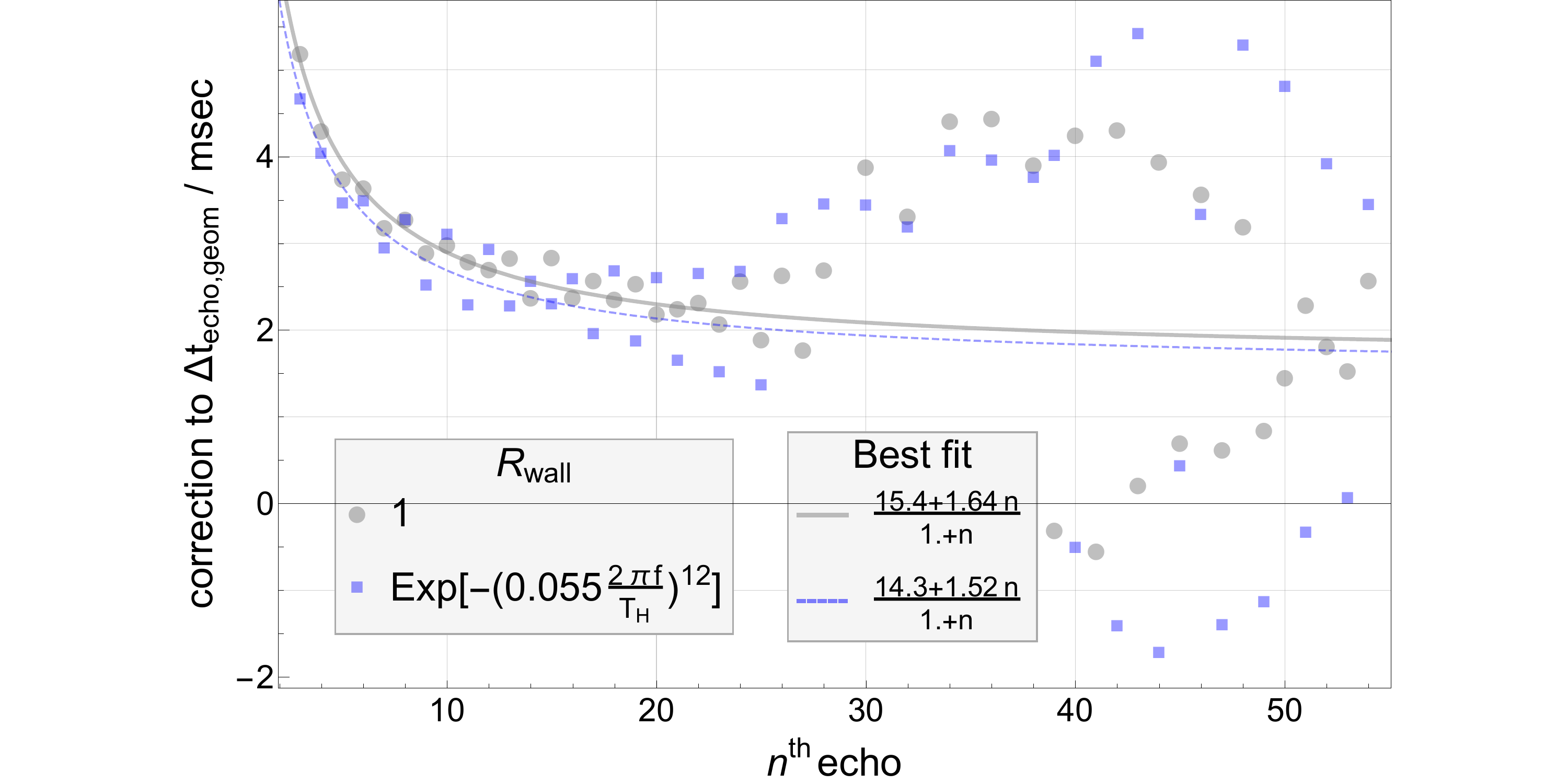}
    \includegraphics[width=1.18\linewidth]{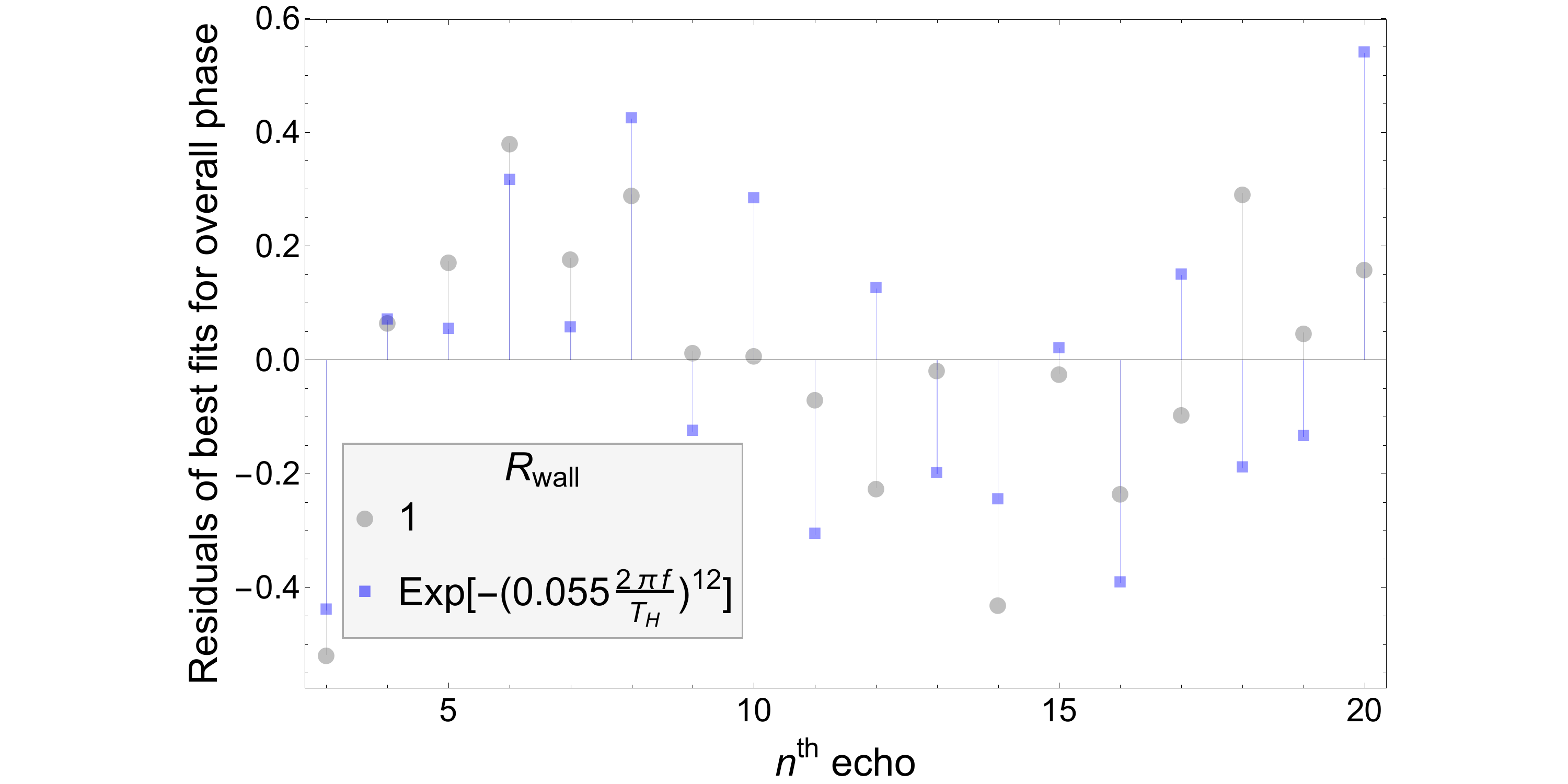}
\endminipage
\caption{\label{43} Same is Fig. (\ref{52}), comparing walls with different energy reflectivity coefficients (fitted for $\ln\left[|h(t)|/|h|_{\rm max, n}\right] > -1.5$). We see that echo amplitudes and peak frequencies decay more quickly for softer walls (see text for details). %Echo template of LIGO event GW150914 for second and later echoes with soft wall. The top three are in the time domain while  the below are the frequency. Starting from left, amplitude of echo and of main frequency all fit the power law. Power law could be extended to first echo with some reflectivity rate of wall. And with wall absorption, the echoes damp fast. As presented in the first plot, Wall with$R_{\rm wall}= \exp[-0.06 \frac{2 \pi \omega}{T_{\rm H}}^8]$ absorb too much so that there is few signals for detecting long-term echo, thus, we don't fit the template for this one. Middle two is the width of the echoes, which become wider evolving along time in the time domain and opposite int the frequency domain since the high frequencies leaks at the early time. The right up plots gives correction to $\Delta t_{echo,geom}$ and they are similar with soft wall. And the below one is the main frequency which damps along time. The soft wall absorb high frequencies, thus, main frequencies damp faster.
}
\end{figure*}

Of course, the actual frequency-dependent reflection of the wall depends on the specific quantum theory of black holes. We explore a phenomenological model for the wall with a Gaussian-like energy reflection rate 
\begin{equation}
R_{\rm wall}(\omega) \simeq \exp\left[-\left( \alpha\frac{\omega}{T_{\rm H}} \right)^q\right],
\end{equation} 
where $T_{\rm H}=\frac{r_+^2-a^2}{4 \pi r_+(r_+^2+a^2)}$ is the Hawking temperature for Kerr BH. While smooth $R_{\rm wall}$'s, such as gaussian or Boltzmann reflectivity ($q=2$ or $1$, respectively) may appear natural, they do tend to essentially wipe out the echoes, unless $\alpha \ll 1$, which is inconsistent with the tentative echoes found in \cite{abedi2016echoes}. In contrast, a sharper function with, e.g., $q=12$ then can damp the first echo, but not significantly influence later echoes, as shown in Fig \ref{42} \footnote{Fig \ref{42} also shows that if the wall absorbs too much, the late echoes will stop decaying. This is due to superradiant instability which we shall discuss in the next section.}. We can also compare these reflectivity functions with that of the angular momentum barrier of the Kerr BH,  for the same spin and mass, as shown in Fig \ref{44}, which provides another motivation for sharper $R_{\rm wall}$'s.   %This sharp structure of wall absorbs most of high frequency but remains low frequency for long-time signals. While the true quantum BH and wall structure remained unknown, this reasonable guess of soft wall does give us later echo signals since low frequencies survived longer.

\begin{widetext}

\begin{table}%The best place to locate the table environment is directly after its first reference in text
\caption{\label{t4} Same as Table \ref{t3}, but contrasting perfect ($R_{\rm wall}=1$) and soft ($R_{\rm wall}= \exp[-(0.055 \frac{\omega}{T_{\rm H}})^{12}]$) walls, fitted within $\ln\left[|h(t)|/|h|_{\rm max, n}\right] > -1.5$. }
\begin{tabular}{|l|l|l|}
\cline{1-3}
\textrm{wall type}&\textrm{perfect}&\textrm{soft}\\
\colrule
\textrm{peak amplitude in time / strain}&
$2.78 \times10^{-19}/n^{1.31}$&
$2.33 \times 10^{-19}/n^{1.36}$\\
\colrule
\textrm{width in time / msec} & $5.5+0.808 n$ & $8.17+0.659 n$ \\
\colrule
\textrm{correction to} $\Delta t_{\rm echo,geom}$ \textrm{msec} & $15.4+1.64/(1+n)$& $14.3+1.52/(1+n)$ \\
\colrule
\textrm{peak frequency / Hz} & $175+104/ n^{0.3}$& $177+96.8/ n^{0.3}$ \\
\colrule
\textrm{Overall phase} & $-6.65+28.5 n^{0.945} -23.8 n$& $-5.2+26.8 n^{0.945} -22.6 n$ \\
\cline{1-3}
\end{tabular}
\end{table}

\end{widetext}

Table \ref{t4} and Fig. \ref{43} compare the template for the perfect and soft walls, similar to Figs. (\ref{52}) and (\ref{72}). We see in the top left panel that, due to absorption of high frequency modes, the power law fit to the amplitudes could be extended to first echo for the soft walls. More generally, echoes decay faster for a softer wall. 
As echoes for a wall with $R_{\rm wall}= \exp[- (0.06\frac{ \omega}{T_{\rm H}})^8]$ decay too fast, we only focus on the $R_{\rm wall}= \exp[-(0.055 \frac{\omega}{T_{\rm H}})^{12}]$ case in subsequent panels of Fig. \ref{43}, and provide numerical fits for echo properties in Table \ref{t4}. With this choice, the evolution of echo properties is similar to those in Figs. (\ref{52}) and (\ref{72}), with the notable difference that peak frequency decays more rapidly as the soft wall absorbs high frequencies.

\section{\label{sec7}Conclusions}

We have provided realistic templates for echoes of BH mergers by numerically solving the linearized Einstein equation (or Teukolsky equation) in Kerr spacetime with boundary conditions at a Planck length proper distance outside the (would-be) event horizon. We obtain analytic approximations for the echo waveforms and time-delays, and explore their dependence on the softness of the wall (or frequency-dependence of the reflection rate), as well as nonlinear effects during merger event. These analytic templates should be useful in echo searches in current and future gravitational wave data. Finally, we studied the occurrence of superradiant instability and showed that it has negligible effect, for the first few dozen echoes of in typical BH mergers such as GW150914. 

Let us close with some open questions and future directions: 
\begin{itemize}
\item The strain is dominated by mode $l=2, m=\pm{2}$. We only show mode $m=2$ here and solution of $m=-2$ can easily be found by $R_{slm}[\omega]=R^*_{sl-m}[-\omega]$. More realistic templates should combine all other modes by appropriate weight. 

\item We cannot provide a reliable waveform for the first echo as it is too sensitive to the {\it ad hoc} cutoff function (\ref{cutoff}) that we use to set up our initial conditions. This highlights the need for a covariant numerical implementation of ECOs within a dynamical spacetime, which could provide realistic nonlinear initial conditions for echoes. 

\item Another big uncertainty is the expected softness of the wall. While this is ultimately a question for the quantum models of black holes, it highlights the need for a covariant and causal description of the wall dynamics. It might be possible to describe this dynamics in terms of the properties of a surface (2+1d) fluid and Israel junction conditions (e.g., see  \cite{Saravani:2012is}).  

\item The computation of the echo phase beyond $\sim 20$ echoes is limited by numerical precision and frequency resolution. This can be improved in the future, by either brute force or novel numerical/analytic methods. 
\end{itemize}
\nocite{*}

\begin{acknowledgements}
We thank Vitor Cardoso, Rafael Sorkin, Huan Yang,  Aaron Zimmerman, Bob Holdom, Randy S. Conklin, Ren Jing, Ofek Birnholtz, William East and Connor Adair  for helpful comments and discussions. We also thank all the participants in our weekly group meetings for their patience during our discussions. This work was supported by the University of Waterloo, Natural Sciences and Engineering Research Council of Canada (NSERC), and the Perimeter Institute for Theoretical Physics. Research at the Perimeter Institute is supported by the Government of Canada through Industry Canada, and by the Province of Ontario through the Ministry of Research and Innovation.
\end{acknowledgements}

\bibliography{main}% Produces the bibliography via BibTeX.

\appendix

\section{\label{a1}Ergoregion Instability }

%Scattering off Kerr BH can lead to superradiant amplification of modes with frequency $0<\omega<m \Omega_{\rm H}$, which can extract energy from a spinning background \cite{1972Natur.238..211P}. Adding a (partially) reflective wall near horizon could turn this amplification to an instability, since modes trapped between the wall and the angular momentum barrier can extract the spin energy repeatedly \cite{1978CMaPh..63..243F, Cardoso:2007az}. In this section, we study this effect for echoes, both in frequency and time domains. 

In this appendix, we further discuss the emergence of superradiance and ergoregion instability in ECOs. 

\begin{figure}
\includegraphics[width=0.5\textwidth]{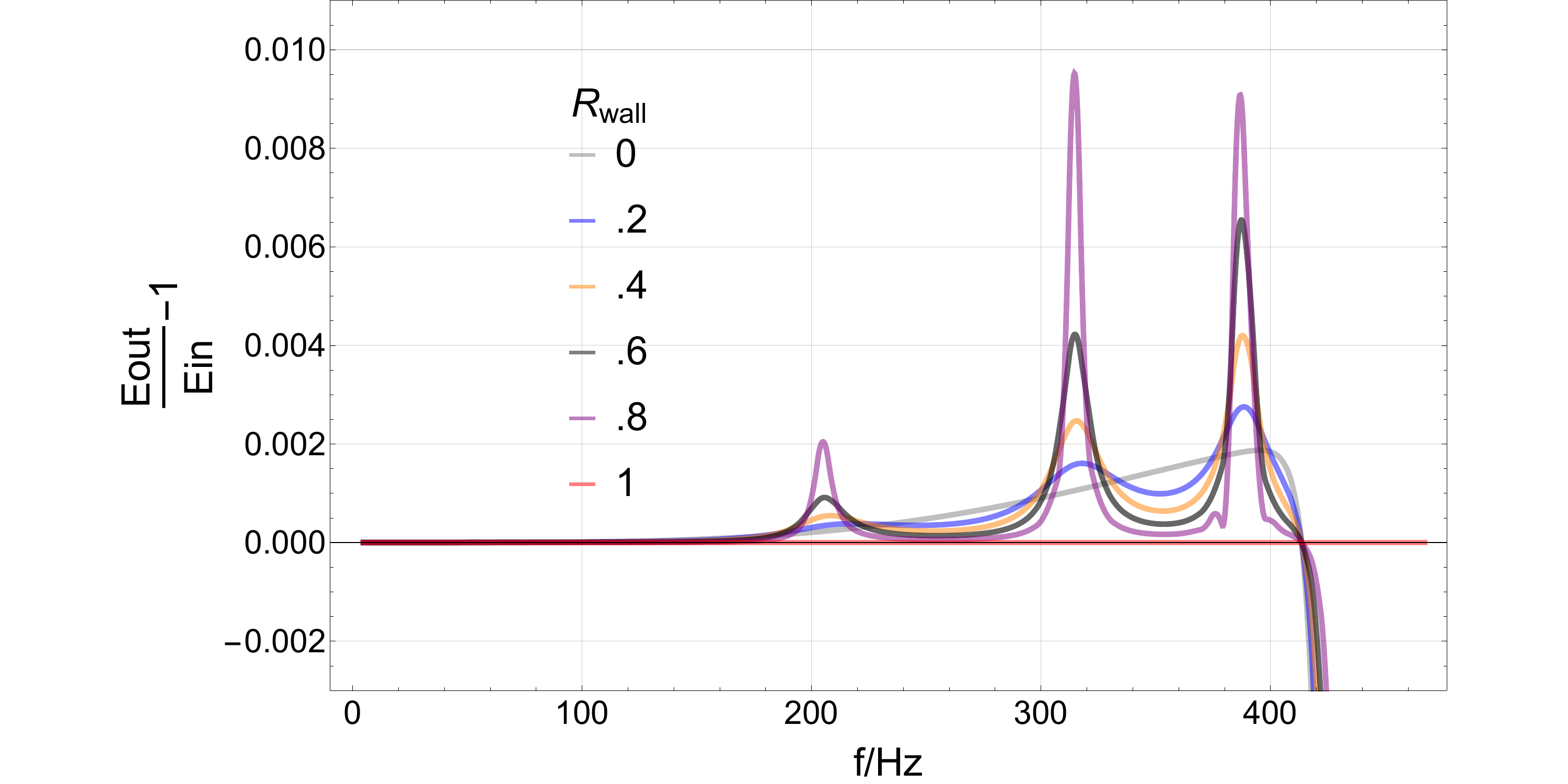}
\caption{\label{61} Superradiance by a spinning ECO/BH with $a=0.99$ and $M = 67.6 M_{\odot}$, assuming different wall reflection coefficients $R_{\rm wall}$ \footnote{For simplicity, we present results with scalar mode $(l,m,s)=(2,2,0)$, but we have confirmed the same results for gravitational waves.}, with wall positition$r_{\rm wall}=r_{\rm h} (1+\delta), \delta = 0.05$. The horizontal axis is frequency, while the vertical axis is the relative energy extracted from ECO/BH. $R_{\rm wall} =0$ is the classical BH (with no reflection on horizon), showing a smooth response with superradiance at low frequencies. A soft wall with $0< R_{\rm wall}<1$ shows several peaks,  corresponding to the resonance frequencies of the cavity formed by the wall and the angular momentum barrier, which amplify superradiance. A perfect reflective wall kills superradiance  by definition, as all the energy that goes in, comes out eventually.}
\end{figure}

Fig \ref{61} shows the superradiance in the frequency domain. We notice that the amplification develops resonance peaks in the presence of an imperfect wall  \footnote{For simplicity, we present results with scalar mode $(l,m,s)=(2,2,0)$, but we have confirmed the same results for gravitational waves.}, and these peaks become sharper with increasing the reflectivity of wall. However, perfect wall seems to kill the superradiance for scattered waves. However, it is much more plausible that we in fact develop infinitely sharp resonance peaks, associated with trapped ergoregion w-modes, which cannot be captured by finite resolution in the frequency domain. We integrate the plots over superradiance range for different reflectivity of the wall, shown as Table \ref{ta}. The area is roughly conserved for all soft walls and equal to the area of a classical BH. Hence,we conclude that when approaching $R_{\rm wall}=1$, we still have the superradiance, but only at discrete frequencies. The zero of  both Fig. \ref{61} and Table \ref{ta} are just because of finite resolution in the frequency domain, and we are not able to see the infinitely sharp resonances when approaching $R_{\rm wall} \rightarrow1$.

\begin{table}%The best place to locate the table environment is directly after its first reference in text
\caption{\label{ta} Integrals of superradince profiles (curves in Fig.(\ref{61}, up to the superradiance threshold) for different wall positions $\delta$, or absorptions $R_{\rm wall}$. We see that the integrated superradiance appears to have a universal value, independent of wall presence or properties. The same is likely to be the case of the perfect wall (last row), but cannot be resolved numerically due to infinitely sharp resonance structure. }
\begin{tabular}{|r|l|l|l|l|}
\cline{1-4}
$\delta=$ & 0.05 & 0.015 & 0.0005\\
\colrule
$R_{\rm wall}=0$&$0.2202$&$0.2202$&$0.2202$\\
\colrule
0.2&$0.2226$&$0.2226$&$0.2225$\\
\colrule
0.4&$0.2246$&$0.2246$&$0.2243$\\
\colrule
0.6&$0.2260$&$0.226r$&$0.2256$\\
\colrule
0.8&$0.2223$&$0.2276$&$0.2190$\\
\colrule
1&$0.0000$&$0.0000$&$0.0000$\\
\cline{1-4}
\end{tabular}
\end{table}

\begin{figure}
\includegraphics[width=0.42\textwidth]{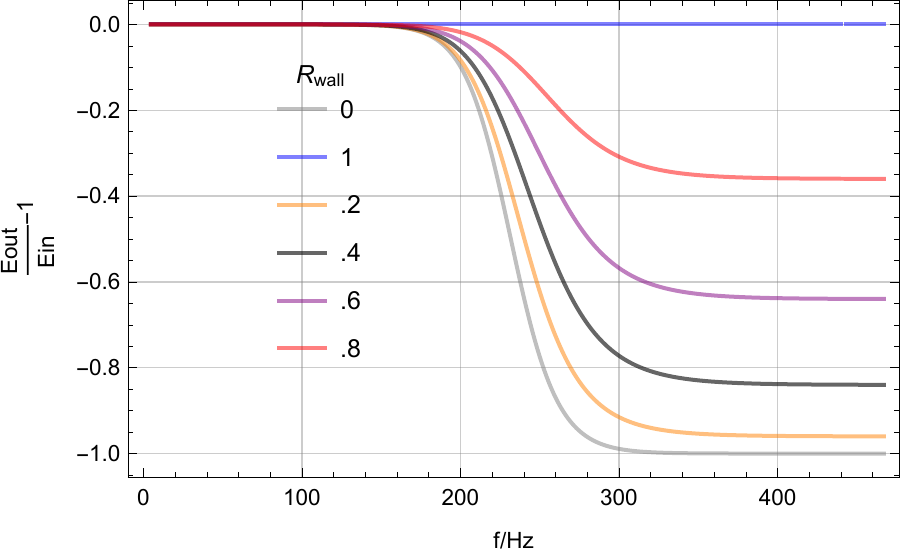}
\caption{\label{621} Same as Fig. \ref{61}, but with $a=0$. We see that superradiance, and superradiant resonance peaks disappear as spin goes to zero. }
\end{figure}

\begin{figure}
\includegraphics[width=0.42\textwidth]{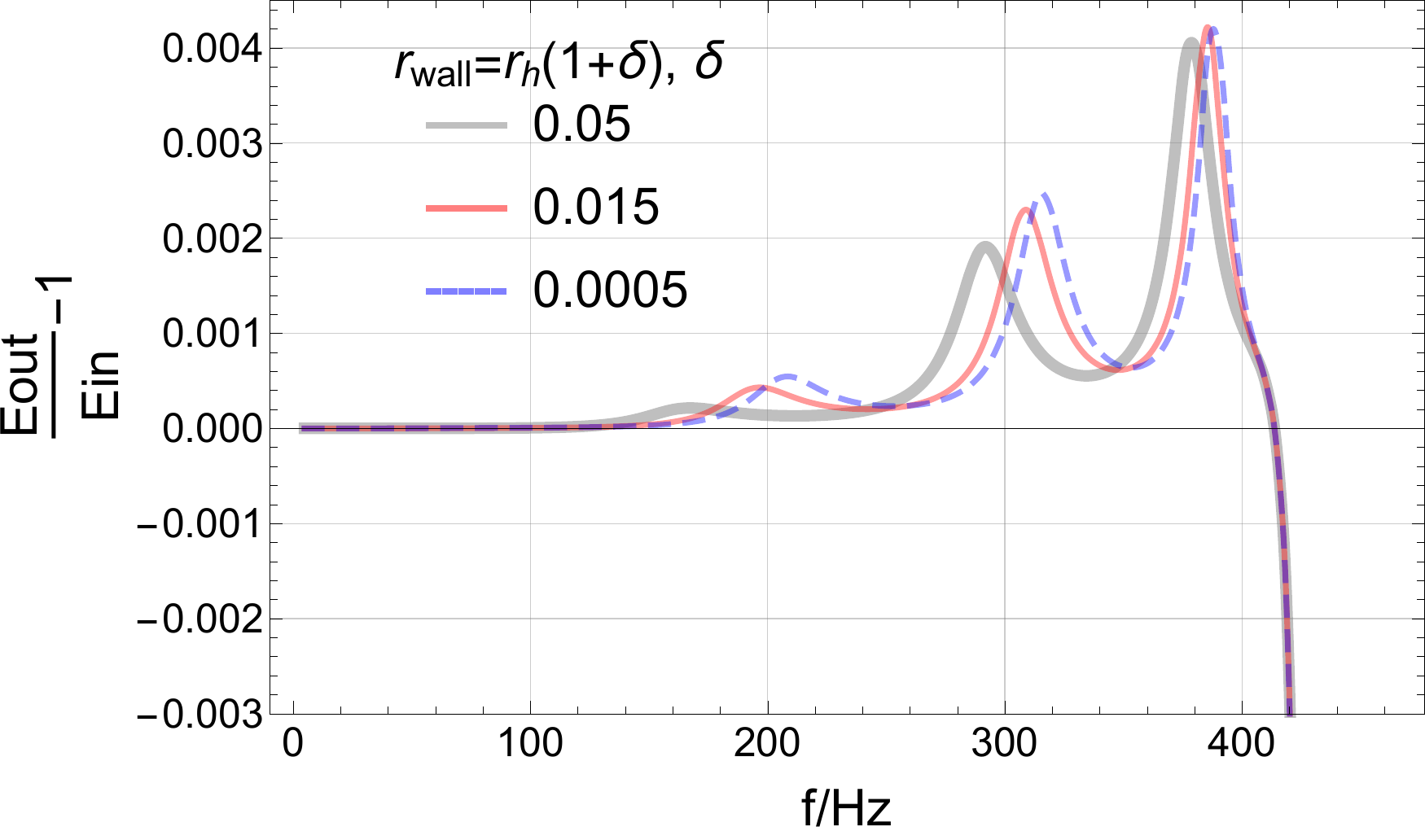}
\caption{\label{622} Same as Fig. \ref{61}, but with different wall positions (fixing $R_{\rm wall}$=0.4). We see that this shifts the resonance frequencies. }
\end{figure}

Fig \ref{621} shows that resonance superradiant peaks disappear as $a \rightarrow 0$, while Fig \ref{622} shows that they shift when we shift the position of the wall, as expected. 
%We use two ways to confirm that the peaks are really from superradiant with wall structure. First, we plots spinless case and don't see superradiant or peaks, as shown in Fig \ref{621}. Another evidence is that since these resonance frequencies depend on the distance between wall and barrier. Second, we change the wall positions $r_{\rm wall}=r_{\rm horizon}(1+\delta)$ and it does give different resonance frequencies, as shown in Fig \ref{622}. 

Note that, even in the absence of superradiant scattering, there could still be instabilities that manifest themselves as the poles of the amplification in the upper complex plane of frequency space.  Indeed, ergoregion instability was predicted in  \cite{1978CMaPh..63..243F} in the absence of horizons and/or dissipation. 
However, we do not see any significant growth, at least in the first 50 echoes we predict for GW150914 in Fig. \ref{51}. This can be understood by noticing that, as shown in Fig \ref{52}, the echoes are dominated by frequencies  $f \gtrsim 210$ Hz, but superradiance happens at $f< m \Omega_H \simeq 180$ Hz for this event. Hence, the instability does not take over until peak frequency drops below this limit. 

\begin{figure}
\includegraphics[width=0.4\textwidth]{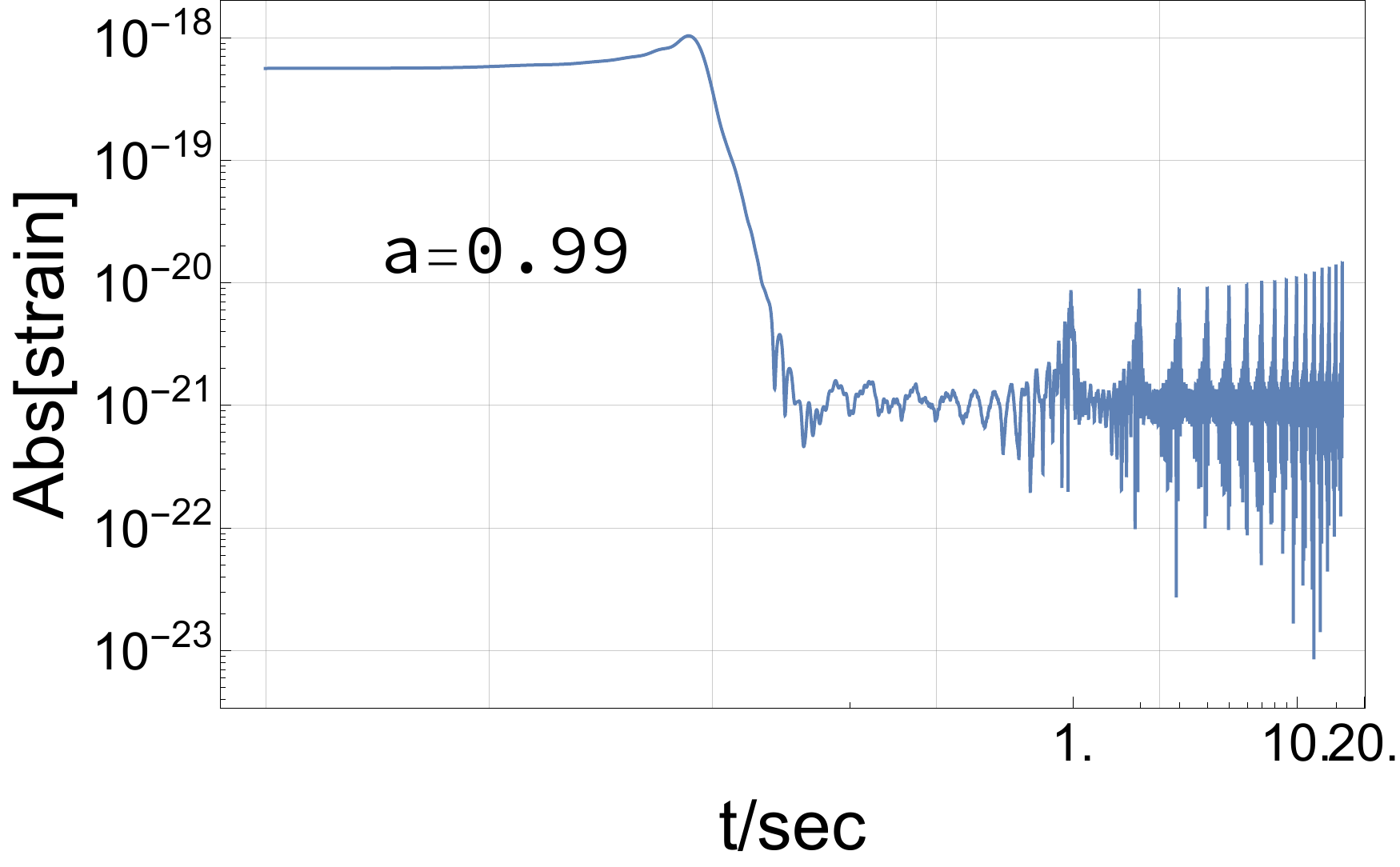}
\caption{\label{63} Occurrence of ergoregion instability in later echoes in ECOs with higher spin (near extremal Kerr). }
\end{figure}

We can increase the spin to see the superradiant instability develop faster, as shown in Fig \ref{63}. For spin $a=0.80$, the echoes stop decaying at some point and for $a=0.99$, they start to increase. While this example does demonstrate the appearance of ergoregion instability, it should be treated as a toy model, as the initial wavepacket was designed to reproduce the merger/ringdown template LIGO event GW150914 with $a=0.67$. 
%We should stressed that we still use LIGO event GW150914 template to set up incoming wavepacket but actually since we change the spin, the real event should have different ringdown waveform, which indicate different initial condition. That's the reason why echoes are very small. When we increase the spin, more high frequency will be reflected because of the superradiant so that we have more high frequency in initial ringdown form compared with lower spin one to set up the initial incoming wavepacket. But it's enough for our purpose to show that wall does add up the superradiant.

As we mentioned in Sec. \ref{sec1}, whether or not ergoregion instability acts in nature depends on the wall absorption properties \cite{Maggio:2017ivp}. Indeed, observation  of astrophysical black holes at significant spins  \cite{Narayan:2013gca} does suggest that the instability must be suppressed.

\end{document}